%% file: main.tex
\pdfoutput=1

\documentclass[12pt,a4paper]{article}

\usepackage{ifthen} 
\newboolean{pdflatex}
\setboolean{pdflatex}{true} 

\newboolean{articletitles}
\setboolean{articletitles}{true} 

\newboolean{uprightparticles}
\setboolean{uprightparticles}{true} 

\newboolean{inbibliography}
\setboolean{inbibliography}{false} 

\def\paperauthors{LHCb collaboration} 
\def\paperasciititle{
Study of bquark bquarkbar correlations in high energy 
proton-proton collisions} 
\def\papertitle{
Study of $\bquark\bquarkbar$~correlations in high energy 
proton\nobreakdash-proton collisions} 
\def\paperkeywords{{High Energy Physics}, {LHCb}} 
\def\papercopyright{CERN on behalf of the LHCb collaboration}
\def\paperlicence{CC-BY-4.0}
\def\paperlicenceurl{https://creativecommons.org/licenses/by/4.0/}

\input{preamble}

\usepackage{rotating}
\usepackage{pifont}
\input{local-symbols}

\usepackage{rotating}
\usepackage{multirow}
\usepackage{graphpap}
\usepackage{subfigure} 
\usepackage{wasysym} 
\usepackage{array} 
\usepackage{pifont} 
\usepackage{dashrule} 
\usepackage{verbatim} 
\usepackage{tikz} 
\usetikzlibrary{patterns}

\definecolor{RootOne}  {rgb}{0,0,0}
\definecolor{RootTwo}  {rgb}{1,0,0}
\definecolor{RootThree}{rgb}{0,1,0}
\definecolor{RootFour} {rgb}{0,0,1}
\definecolor{RootFive} {rgb}{1,1,0}
\definecolor{RootSix}  {rgb}{1,0,1}
\definecolor{RootSeven}{rgb}{0,1,1}

\definecolor{Root1}  {rgb}{0,0,0}
\definecolor{Root2}  {rgb}{1,0,0}
\definecolor{Root3}  {rgb}{0,1,0}
\definecolor{Root4}  {rgb}{0,0,1}
\definecolor{Root5}  {rgb}{1,1,0}
\definecolor{Root6}  {rgb}{1,0,1}
\definecolor{Root7}  {rgb}{0,1,1}
\definecolor{Root8}  {rgb}{0.35,0.83 ,0.33}
\definecolor{Root92} {rgb}{1   ,0.747,0   }

\begin{document}

\renewcommand{\thefootnote}{\fnsymbol{footnote}}
\setcounter{footnote}{1}

\input{title-LHCb-PAPER}


\renewcommand{\thefootnote}{\arabic{footnote}}
\setcounter{footnote}{0}



\pagestyle{plain} 
\setcounter{page}{1}
\pagenumbering{arabic}


%

\input{intro}
\input{detector}

\input{signal}

\input{correlations}
\input{summary}

\input{acknowledgements}

\input{appendix}
\clearpage



\clearpage

\addcontentsline{toc}{section}{References}
\setboolean{inbibliography}{true}
\bibliographystyle{LHCb}
\bibliography{main,LHCb-PAPER,LHCb-CONF,LHCb-DP,LHCb-TDR,local}

\newpage


 
\newpage
\input{LHCb_Authorship_flat_06-Jun-2017}

\end{document}

%% file: preamble.tex
\usepackage[top=1in, bottom=1.25in, left=1in, right=1in]{geometry}

\usepackage{microtype}
\usepackage{lineno}  
\usepackage{xspace} 
\usepackage{caption} 

\usepackage{graphicx}  
\usepackage{color}
\usepackage{colortbl}
\graphicspath{{./figs/}} 

\usepackage{amsmath} 
\usepackage{amssymb}
\usepackage{amsfonts}
\usepackage{upgreek} 

\newcommand*\patchAmsMathEnvironmentForLineno[1]{%
\expandafter\let\csname old#1\expandafter\endcsname\csname #1\endcsname
\expandafter\let\csname oldend#1\expandafter\endcsname\csname
end#1\endcsname
 \renewenvironment{#1}%
   {\linenomath\csname old#1\endcsname}%
   {\csname oldend#1\endcsname\endlinenomath}%
}
\newcommand*\patchBothAmsMathEnvironmentsForLineno[1]{%
  \patchAmsMathEnvironmentForLineno{#1}%
  \patchAmsMathEnvironmentForLineno{#1*}%
}
\AtBeginDocument{%
\patchBothAmsMathEnvironmentsForLineno{equation}%
\patchBothAmsMathEnvironmentsForLineno{align}%
\patchBothAmsMathEnvironmentsForLineno{flalign}%
\patchBothAmsMathEnvironmentsForLineno{alignat}%
\patchBothAmsMathEnvironmentsForLineno{gather}%
\patchBothAmsMathEnvironmentsForLineno{multline}%
\patchBothAmsMathEnvironmentsForLineno{eqnarray}%
}

\usepackage{hyperxmp}

\usepackage[pdftex,
            pdfauthor={\paperauthors},
            pdftitle={\paperasciititle},
            pdfkeywords={\paperkeywords},
            pdfcopyright={Copyright (C) \papercopyright},
            pdflicenseurl={\paperlicenceurl}]{hyperref}

\usepackage[all]{hypcap} 

\input{lhcb-symbols-def} 

\usepackage{cite} 
\usepackage{mciteplus}

%% file: lhcb-symbols-def.tex
\usepackage{xspace} 
\usepackage{upgreek}

\def\lhcb {\mbox{LHCb}\xspace}

\def\MagUp {\mbox{\em Mag\kern -0.05em Up}\xspace}

\ifthenelse{\boolean{uprightparticles}}%
{

 \def\Peta        {\ensuremath{\upeta}\xspace}

 \def\Pmu         {\ensuremath{\upmu}\xspace}

 \def\Ppi         {\ensuremath{\uppi}\xspace}

 \def\Pphi        {\ensuremath{\upphi}\xspace}

 \def\Ppsi        {\ensuremath{\uppsi}\xspace}

 \def\PDelta      {\ensuremath{\Delta}\xspace}                 
 \def\PXi      {\ensuremath{\Xi}\xspace}                 
 \def\PLambda      {\ensuremath{\Lambda}\xspace}                 
 \def\PSigma      {\ensuremath{\Sigma}\xspace}                 
 \def\POmega      {\ensuremath{\Omega}\xspace}                 
 \def\PUpsilon      {\ensuremath{\Upsilon}\xspace}                 
 
 \def\PB      {\ensuremath{\mathrm{B}}\xspace}                 
                  
 \def\PD      {\ensuremath{\mathrm{D}}\xspace}

 \def\PJ      {\ensuremath{\mathrm{J}}\xspace}                 
 \def\PK      {\ensuremath{\mathrm{K}}\xspace}

 \def\PX      {\ensuremath{\mathrm{X}}\xspace}

 \def\Pb      {\ensuremath{\mathrm{b}}\xspace}                 
 \def\Pc      {\ensuremath{\mathrm{c}}\xspace}

 \def\Pi      {\ensuremath{\mathrm{i}}\xspace}

 \def\Pp      {\ensuremath{\mathrm{p}}\xspace}

 \def\Ps      {\ensuremath{\mathrm{s}}\xspace}

}
{

 \def\Peta        {\ensuremath{\eta}\xspace}

 \def\Pmu         {\ensuremath{\mu}\xspace}

 \def\Ppi         {\ensuremath{\pi}\xspace}

 \def\Pphi        {\ensuremath{\phi}\xspace}

 \def\Ppsi        {\ensuremath{\psi}\xspace}                 
                  
 \mathchardef\PDelta="7101
 \mathchardef\PXi="7104
 \mathchardef\PLambda="7103
 \mathchardef\PSigma="7106
 \mathchardef\POmega="710A
 \mathchardef\PUpsilon="7107
                  
 \def\PB      {\ensuremath{B}\xspace}                 
                  
 \def\PD      {\ensuremath{D}\xspace}

 \def\PJ      {\ensuremath{J}\xspace}                 
 \def\PK      {\ensuremath{K}\xspace}

 \def\PX      {\ensuremath{X}\xspace}

 \def\Pb      {\ensuremath{b}\xspace}                 
 \def\Pc      {\ensuremath{c}\xspace}

 \def\Pi      {\ensuremath{i}\xspace}

 \def\Pp      {\ensuremath{p}\xspace}

 \def\Ps      {\ensuremath{s}\xspace}

}

\makeatletter
\ifcase \@ptsize \relax
  \newcommand{\miniscule}{\@setfontsize\miniscule{4}{5}}
\or
  \newcommand{\miniscule}{\@setfontsize\miniscule{5}{6}}
\or
  \newcommand{\miniscule}{\@setfontsize\miniscule{5}{6}}
\fi
\makeatother

\DeclareRobustCommand{\optbar}[1]{\shortstack{{\miniscule (\rule[.5ex]{1.25em}{.18mm})}
  \\ [-.7ex] $#1$}}

\def\mup        {{\ensuremath{\Pmu^+}}\xspace}
\def\mun        {{\ensuremath{\Pmu^-}}\xspace} 
\def\mumu       {{\ensuremath{\Pmu^+\Pmu^-}}\xspace}

\def\squark    {{\ensuremath{\Ps}}\xspace}

\def\cquark    {{\ensuremath{\Pc}}\xspace}
\def\cquarkbar {{\ensuremath{\overline \cquark}}\xspace}

\def\bquark    {{\ensuremath{\Pb}}\xspace}
\def\bquarkbar {{\ensuremath{\overline \bquark}}\xspace}

\def\pion   {{\ensuremath{\Ppi}}\xspace}

\def\pip    {{\ensuremath{\pion^+}}\xspace}

\def\kaon    {{\ensuremath{\PK}}\xspace}
  \def\Kbar    {{\kern 0.2em\overline{\kern -0.2em \PK}{}}\xspace}

\def\KorKbar    {\kern 0.18em\optbar{\kern -0.18em K}{}\xspace}

\def\Kp      {{\ensuremath{\kaon^+}}\xspace}
\def\Km      {{\ensuremath{\kaon^-}}\xspace}

  \def\Dbar    {{\kern 0.2em\overline{\kern -0.2em \PD}{}}\xspace}

\def\DorDbar    {\kern 0.18em\optbar{\kern -0.18em D}{}\xspace}

\def\B       {{\ensuremath{\PB}}\xspace}
\def\Bbar    {{\ensuremath{\kern 0.18em\overline{\kern -0.18em \PB}{}}}\xspace}

\def\BorBbar    {\kern 0.18em\optbar{\kern -0.18em B}{}\xspace}

\def\Bu      {{\ensuremath{\B^+}}\xspace}

\def\Bd      {{\ensuremath{\B^0}}\xspace}
\def\Bs      {{\ensuremath{\B^0_\squark}}\xspace}

\def\Bc      {{\ensuremath{\B_\cquark^+}}\xspace}

\def\jpsi     {{\ensuremath{{\PJ\mskip -3mu/\mskip -2mu\Ppsi\mskip 2mu}}}\xspace}

  \def\Y#1S{\ensuremath{\PUpsilon{(#1S)}}\xspace}

\def\proton      {{\ensuremath{\Pp}}\xspace}
\def\antiproton  {{\ensuremath{\overline \proton}}\xspace}

\def\Lz          {{\ensuremath{\PLambda}}\xspace}
\def\Lbar        {{\ensuremath{\kern 0.1em\overline{\kern -0.1em\PLambda}}}\xspace}
\def\LorLbar    {\kern 0.18em\optbar{\kern -0.18em \PLambda}{}\xspace}

\def\Lb      {{\ensuremath{\Lz^0_\bquark}}\xspace}

\def\to                 {\ensuremath{\rightarrow}\xspace}

\def\AT#1     {\ensuremath{A_{\mathrm{T}}^{#1}}\xspace}           

\def\C#1      {\ensuremath{\mathcal{C}_{#1}}\xspace}                       
\def\Cp#1     {\ensuremath{\mathcal{C}_{#1}^{'}}\xspace}                    
\def\Ceff#1   {\ensuremath{\mathcal{C}_{#1}^{\mathrm{(eff)}}}\xspace}        
\def\Cpeff#1  {\ensuremath{\mathcal{C}_{#1}^{'\mathrm{(eff)}}}\xspace}       
\def\Ope#1    {\ensuremath{\mathcal{O}_{#1}}\xspace}                       
\def\Opep#1   {\ensuremath{\mathcal{O}_{#1}^{'}}\xspace}                    

\newcommand{\tev}{\ifthenelse{\boolean{inbibliography}}{\ensuremath{~T\kern -0.05em eV}}{\ensuremath{\mathrm{\,Te\kern -0.1em V}}}\xspace}
\newcommand{\gevbib}{\ifthenelse{\boolean{inbibliography}}{\ensuremath{~G\kern -0.05em eV}}{\ensuremath{\mathrm{\,Ge\kern -0.1em V}}}\xspace}
\newcommand{\gev}{\ensuremath{\mathrm{\,Ge\kern -0.1em V}}\xspace}
\newcommand{\mev}{\ensuremath{\mathrm{\,Me\kern -0.1em V}}\xspace}
\newcommand{\kev}{\ensuremath{\mathrm{\,ke\kern -0.1em V}}\xspace}
\newcommand{\ev}{\ensuremath{\mathrm{\,e\kern -0.1em V}}\xspace}
\newcommand{\gevc}{\ensuremath{{\mathrm{\,Ge\kern -0.1em V\!/}c}}\xspace}
\newcommand{\mevc}{\ensuremath{{\mathrm{\,Me\kern -0.1em V\!/}c}}\xspace}
\newcommand{\gevcc}{\ensuremath{{\mathrm{\,Ge\kern -0.1em V\!/}c^2}}\xspace}
\newcommand{\gevgevcccc}{\ensuremath{{\mathrm{\,Ge\kern -0.1em V^2\!/}c^4}}\xspace}
\newcommand{\mevcc}{\ensuremath{{\mathrm{\,Me\kern -0.1em V\!/}c^2}}\xspace}

\def\mum  {\ensuremath{{\,\upmu\mathrm{m}}}\xspace}

\def\invfb   {\ensuremath{\mbox{\,fb}^{-1}}\xspace}

\def\deriv {\ensuremath{\mathrm{d}}}

\def\gsim{{~\raise.15em\hbox{$>$}\kern-.85em
          \lower.35em\hbox{$\sim$}~}\xspace}
\def\lsim{{~\raise.15em\hbox{$<$}\kern-.85em
          \lower.35em\hbox{$\sim$}~}\xspace}

\def\sqs   {\ensuremath{\protect\sqrt{s}}\xspace}

\def\ptot       {\mbox{$p$}\xspace}
\def\pt         {\mbox{$p_{\mathrm{ T}}$}\xspace}

\def\evtgen     {\mbox{\textsc{EvtGen}}\xspace}

\def\geant      {\mbox{\textsc{Geant4}}\xspace}

\def\photos     {\mbox{\textsc{Photos}}\xspace}

\def\pythia     {\mbox{\textsc{Pythia}}\xspace}

\def\tell1  {TELL1\xspace}
\def\ukl1   {UKL1\xspace}

\newcommand{\ie}{\mbox{\itshape i.e.}\xspace}

%% file: local-symbols.tex
\def\X      {{\ensuremath{\PX}}\xspace}

\newcommand{\ptpsi}{\ensuremath{p_{\mathrm{T}}^{\jpsi}}\xspace}
\newcommand{\ypsi} {\ensuremath{y^{\jpsi}}\xspace}
\def\jpsione{{\ensuremath{{\PJ\mskip -3mu/\mskip -2mu\Ppsi_1\mskip 2mu}}}\xspace}
\def\jpsitwo{{\ensuremath{{\PJ\mskip -3mu/\mskip -2mu\Ppsi_2\mskip 2mu}}}\xspace}

%% file: title-LHCb-PAPER.tex

\begin{titlepage}
\pagenumbering{roman}

\vspace*{-1.5cm}
\centerline{\large EUROPEAN ORGANIZATION FOR NUCLEAR RESEARCH (CERN)}
\vspace*{1.5cm}
\noindent
\begin{tabular*}{\linewidth}{lc@{\extracolsep{\fill}}r@{\extracolsep{0pt}}}
\ifthenelse{\boolean{pdflatex}}
{\vspace*{-2.7cm}\mbox{\!\!\!\includegraphics[width=.14\textwidth]{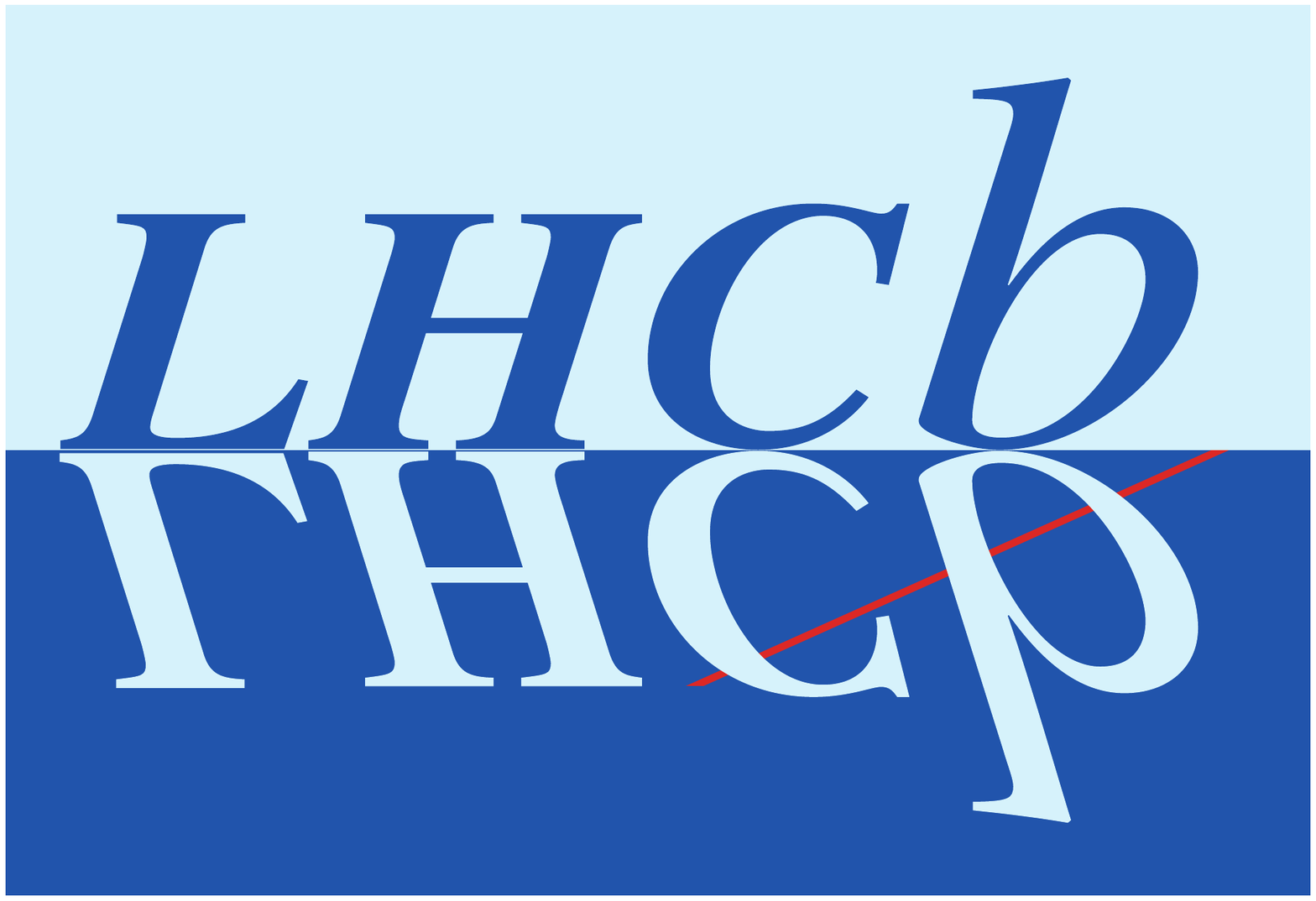}} & &}%
{\vspace*{-1.2cm}\mbox{\!\!\!\includegraphics[width=.12\textwidth]{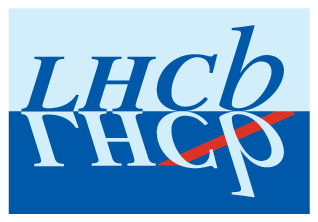}} & &}%
\\
 & & CERN-EP-2017-178    \\  
 & & LHCb-PAPER-2017-020 \\  
 & & August  14, 2017    \\ 
 & & \\
\end{tabular*}

\vspace*{4.0cm}

{\normalfont\bfseries\boldmath\huge
\begin{center}
  \papertitle 
\end{center}
}

\vspace*{2.0cm}

\begin{center}
\paperauthors\footnote{Authors are listed at the end of this paper.}
\end{center}

\vspace{\fill}

\begin{abstract}
  \noindent
  Kinematic correlations for pairs of beauty hadrons, 
  produced in high energy proton\nobreakdash-proton collisions, 
  are studied. 
  The~data sample used 
  was collected with the~\lhcb
  experiment at centre\nobreakdash-of\nobreakdash-mass energies of 
  7~and~8\tev and corresponds to an~integrated luminosity of 3\invfb.
  The~measurement is performed using 
  inclusive $\bquark\to\jpsi\mathrm{X}$~decays
  in the~rapidity range~\mbox{$2<y^{\jpsi}<4.5$}.
  The~observed correlations are in good agreement with theoretical predictions.
\end{abstract}

\vspace*{2.0cm}

\begin{center}
  Pubished in JHEP  1711 (2017) 030
\end{center}

\vspace{\fill}

{\footnotesize 
\centerline{\copyright~\papercopyright, licence \href{\paperlicenceurl}{\paperlicence}.}}
\vspace*{2mm}

\end{titlepage}


\newpage
\setcounter{page}{2}
\mbox{~}

\cleardoublepage

%% file: intro.tex
\section{Introduction}\label{sec:intro}

The~production of heavy\nobreakdash-flavour hadrons in high energy collisions provides  
important tests for the predictions of quantum chromodynamics\,(QCD).
Open\nobreakdash-charm hadron production 
has been studied in 
$\proton\proton$~collisions 
at the~Large Hadron Collider\,(LHC)
by the~LHCb collaboration at
centre\nobreakdash-of\nobreakdash-mass 
energies~\mbox{$\sqs=5$}, 
7~and~13\tev~\cite{LHCb-PAPER-2016-042,LHCb-PAPER-2012-041,LHCb-PAPER-2015-041},
by the~ATLAS collaboration at~\mbox{$\sqs=7\tev$}~\cite{Aad:2015zix}
and 
by the~ALICE collaboration 
at $\sqs=2.76$~and~\mbox{$7\tev$}~\cite{Abelev:2012vra,ALICE:2011aa,Abelev:2012tca,Adam:2016ich}.
In~addition, the~CDF collaboration has studied 
the~production of open\nobreakdash-charm hadrons 
in $\proton\antiproton$~collisions 
at the~Tevatron
at~\mbox{$\sqs=1.96\tev$}~\cite{Acosta:2003ax,Aaltonen:2016wdo}.
For~beauty hadrons,
the~production cross\nobreakdash-sections 
in high energy 
$\proton\proton$~and
$\proton\antiproton$~collisions 
have been studied 
by a~number of collaborations~\cite{Albajar:1988th,Abachi:1994kj,Abe:1995dv,Abulencia:2006ps}.
Most~recently, 
at the~LHC,
the~LHCb~collaboration at~\mbox{$\sqs=7$}, 
$8$ and $13\tev$ and 
the~CMS~collaboration at~\mbox{$\sqs=8\tev$}
studied beauty hadron production 
using semileptonic decays~\cite{LHCb-PAPER-2010-002,Khachatryan:2011hf},
inclusive decays of beauty hadrons 
into $\jpsi$~mesons~\cite{LHCb-PAPER-2011-003,LHCb-PAPER-2013-016,LHCb-PAPER-2015-037},
and 
exclusive  
\mbox{$\Bd\to\jpsi\kaon{\mathrm{(892)^{*0}}}$},  
\mbox{$\Bu\to\jpsi\Kp$}, 
\mbox{$\Bs\to\jpsi\Kp\Km$}~\cite{LHCb-PAPER-2011-043,LHCb-PAPER-2013-004,Khachatryan:2011mk,Chatrchyan:2011pw}, 
\mbox{$\Lb\to\jpsi\proton\Km$}~\cite{LHCb-PAPER-2015-032,Chatrchyan:2012xg}
and 
\mbox{$\Bc\to\jpsi\pip$}~\cite{LHCb-PAPER-2012-028,LHCb-PAPER-2014-050}
decays.
The~transverse momentum, $\pt$,
and rapidity, $y$, spectra are found to be 
in agreement with calculations at 
next\nobreakdash-to\nobreakdash-leading 
order\,(NLO). 
These~calculations are made using the~general\nobreakdash-mass 
variable\nobreakdash-flavour\nobreakdash-number 
scheme\,(GMVFNS)~\cite{Kniehl:2012ti,Kniehl:2004fy,
  Kniehl:2005ej,Kneesch:2007ey,Kniehl:2009ar}, {\sc{Powheg}}~\cite{Gauld:2015yia} and 
fixed\nobreakdash-order with next\nobreakdash-to\nobreakdash-leading\nobreakdash-log
resummation\,(FONLL)~\cite{Cacciari:1998it,Cacciari:2001td,Cacciari:2003zu,Cacciari:2005uk,
Cacciari:2012ny,Cacciari:2015fta}. 
For~\Bc~mesons, a~good agreement 
in the~shapes of the~\pt and 
$y$~spectra is found~\cite{LHCb-PAPER-2014-050} 
with calculations based on a~complete 
order\nobreakdash-$\alpha_{\mathrm{s}}^4$~approach~\cite{Chang:1992jb,Chang:1994aw,Berezhnoy:1994ba,Kolodziej:1995nv}.
However, the~inclusive single\nobreakdash-heavy\nobreakdash-flavour hadron 
transverse momentum and rapidity spectra have limited sensitivity 
to the~subprocesses of the~production mechanism
and the~size of higher\nobreakdash-order QCD corrections.

The~kinematic correlations between the~heavy quark and antiquark
provide additional information and can enable a~better understanding 
of the~production mechanism, such as the~contribution of 
the~gluon\nobreakdash-splitting, 
flavour\nobreakdash-creation and  
flavour\nobreakdash-excitation processes,
as well as the~role of higher\nobreakdash-order corrections.
Such~correlations have been studied for pairs of open\nobreakdash-charm 
mesons 
by the~CDF collaboration in the~central rapidity 
region~\mbox{$\left|y\right|<1$}~\cite{Reisert:2007zza,Reisert:2007zz} 
and by the~LHCb collaboration in the~forward rapidity region~\mbox{$2<y<4$}~\cite{LHCb-PAPER-2012-003}.
The~difference in the~azimuthal angle, $\Pphi$, between 
two reconstructed open\nobreakdash-charm mesons
shows a~strong correlation, which demonstrates
the~importance of the~gluon\nobreakdash-splitting mechanism for 
the~production of~$\cquark\cquarkbar$~events.
For~charm production in the~central rapidity region, the~contributions from 
flavour\nobreakdash-creation and flavour\nobreakdash-excitation processes 
have been identified, in addition to that from 
gluon splitting~\cite{Reisert:2007zza,Reisert:2007zz}. 

The~azimuthal and rapidity correlations 
in $\bquark\bquarkbar$ production 
have been studied 
by the~UA1~\cite{Albajar:1993be},  
D0~\cite{Abbott:1999se} 
and 
CDF~\cite{Abe:1996zt,Abe:1998ac,Acosta:2004nj,Aaltonen:2007zza}~collaborations
in $\proton\antiproton$~collisions at~\mbox{$\sqs=0.63$},
$1.8$ and~\mbox{$1.96\tev$}.
At~the~LHC, the~first study of \bquark\bquarkbar~correlations 
in high energy $\proton\proton$~collisions
in the~central rapidity region has been  performed 
by the~CMS collaboration~\cite{Khachatryan:2011wq}.
The~collaboration found that none of the~available
calculations describe 
the~shapes of the~differential cross\nobreakdash-section 
well~\cite{Frixione:2002ik,*Frixione:2003ei,*Frixione:2008ym,
Jung:2000hk,*Catani:1993ww}.
In~particular, 
the~region where the~contributions of gluon\nobreakdash-splitting 
processes are expected to be large
is not adequately 
described by any of the~predictions
from 
{\sc{MC@NLO}}~\cite{Frixione:2002ik,*Frixione:2003ei,*Frixione:2008ym},
{\sc{Cascade}}~\cite{Jung:2000hk,*Catani:1993ww},
\mbox{$\pythia\,8$}~\cite{Sjostrand:2007gs},
or {\sc{MadGraph}}~\cite{Maltoni:2002qb,*Alwall:2007st}.
Recently, a~study of \bquark\bquarkbar~correlations 
in $\proton\proton$~collisions in 
the~central rapidity region has been  performed 
by the~ATLAS collaboration~\cite{Aaboud:2017vqt}
and a~good agreement with calculations was obtained.
The~four\nobreakdash-flavour 
{\sc{MadGraph5}} prediction~\cite{Alwall:2014hca}
provides the~best overall agreement with data,
and performs  better than the~\mbox{$\pythia\,8$}
and {\sc{Herwig++}}~\cite{Bahr:2008pv} generators. 

This paper reports the~study of \bquark\bquarkbar~correlations 
in high energy hadron collisions in 
the~forward rapidity region. The~data sample used was collected with the~\lhcb
experiment at centre\nobreakdash-of\nobreakdash-mass energies of 
7~and~8\tev and corresponds to integrated luminosities of 
1~and 2\invfb,  respectively.
The~beauty hadrons are reconstructed via 
their inclusive decays into $\jpsi$ mesons, 
denoted here as $\bquark\to\jpsi\mathrm{X}$~decays,
using \jpsi~mesons decaying into the~\mumu~final state.
The~results are  compared 
with the~leading\nobreakdash-order\,(LO) and NLO expectations 
from 
\pythia~\cite{Sjostrand:2006za,Sjostrand:2007gs}
and 
{\sc{Powheg}}~\cite{Nason:2004rx,Frixione:2007vw,Frixione:2007nw,Alioli:2010xd}, 
respectively. 

%% file: detector.tex
\section{Detector and simulation}
\label{sec:Detector}

The~\lhcb detector~\cite{Alves:2008zz,LHCb-DP-2014-002} is 
a~single\nobreakdash-arm forward
spectrometer covering the~\mbox{pseudorapidity} range~\mbox{$2<\Peta <5$},
designed for the~study of particles containing \bquark or \cquark~quarks. 
The~detector includes 
a~high\nobreakdash-precision tracking system
consisting of a~silicon\nobreakdash-strip vertex detector surrounding 
the~$\proton\proton$~interaction  region~\cite{LHCb-DP-2014-001}, 
a~large\nobreakdash-area silicon\nobreakdash-strip detector located
upstream of a~dipole magnet with a bending power of about~$4{\mathrm{\,Tm}}$, 
and three stations of silicon\nobreakdash-strip detectors and straw
drift tubes placed downstream of the~magnet.
The~tracking system provides a~measurement of momentum, \ptot, of charged particles with
a~relative uncertainty that varies from 0.5\%~at low momentum to 1.0\% at~200\gevc.
The~minimum distance of a~track to a~primary vertex\,(PV), 
the~impact parameter\,(IP), 
is measured with a~resolution of~$(15+29/\pt)\mum$,
where \pt~is the~component of the~momentum transverse 
to the~beam, in\,\gevc.
Different types of charged hadrons are distinguished using information
from two ring\nobreakdash-imaging Cherenkov detectors. 
Photons, electrons and hadrons are identified by a~calorimeter system consisting of
scintillating\nobreakdash-pad and preshower detectors, 
an~electromagnetic
calorimeter and a~hadronic calorimeter. 
Muons are identified by a~system composed of alternating 
layers of iron and multiwire proportional chambers~\cite{LHCb-DP-2012-002}.

The~online event selection is performed by a trigger~\cite{LHCb-DP-2012-004}, 
which consists of a~hardware stage, based on information from 
the~calorimeter and muon systems; followed by a~software stage, 
which applies a~full event reconstruction.
The~hardware trigger selects pairs of opposite\nobreakdash-sign muon candidates
with a~requirement that the~product of the~muon transverse momenta is larger than
\mbox{$1.7\,(2.6)\,\mathrm{GeV}^2/c^2$} for data 
collected at~$\sqs=7\,(8)\tev$.
The~subsequent software trigger is composed of two stages, 
the~first of which performs a~partial event reconstruction.
A~full event reconstruction is then made at the~second stage.
In~the~software trigger, 
the~invariant mass of well\nobreakdash-reconstructed 
pairs of oppositely charged muons that form 
a~vertex with good reconstruction quality
is required to exceed 2.7\gevcc
and the~vertex 
is required to be significantly displaced from all~PVs.

Simulated samples are used to determine 
the~reconstruction and trigger efficiencies.
%
Proton\nobreakdash-proton
collisions are generated using
\pythia~\cite{Sjostrand:2006za,Sjostrand:2007gs} 
with a~specific \lhcb configuration~\cite{LHCb-PROC-2010-056}.  
Decays of hadronic particles
are described by \evtgen~\cite{Lange:2001uf}, 
in which final\nobreakdash-state
radiation is generated using \photos~\cite{Golonka:2005pn}. 
The~interaction of the~generated particles with the~detector, 
and its response, are implemented using the~\geant toolkit~\cite{Allison:2006ve, *Agostinelli:2002hh} 
as described in Ref.~\cite{LHCb-PROC-2011-006}.


%% file: signal.tex
\section {Signal selection and efficiency determination}

Selected events are required to have two reconstructed 
\mbox{$\jpsi\to\mumu$}~candidates.
In~the~following these two candidates 
are marked with subscripts 1~and~2, 
which are randomly assigned.
The~muon candidates must be identified as muons, 
have good reconstruction quality, 
\mbox{$\pt>500\mevc$} 
and~\mbox{$2 < \Peta< 5$}~\cite{LHCb-DP-2013-001,LHCb-DP-2012-002}.
Both~reconstructed \jpsi~candidates are required to have 
a~good\nobreakdash-quality vertex,
a~reconstructed mass in the~range~\mbox{$3.00<m^{\mumu}<3.18\gevcc$},
\mbox{$2< \ptpsi < 25\gevc$} and  \mbox{$2< \ypsi < 4.5$}. 
These~criteria 
ensure a~good reconstruction and trigger efficiency.
Only~events triggered by at least one of the~\jpsi~candidates are retained.
The~two \jpsi~candidates are required to be associated with the~same PV
and, in order to suppress background 
from promptly produced \jpsi~mesons,
both dimuon vertices are required to be significantly displaced from that~PV.

\begin{figure}[t]
  \setlength{\unitlength}{1mm}
  \centering
  \begin{picture}(150,115)
    \put(  0,  55){ 
      \includegraphics*[width=75mm,height=60mm,%
      ]{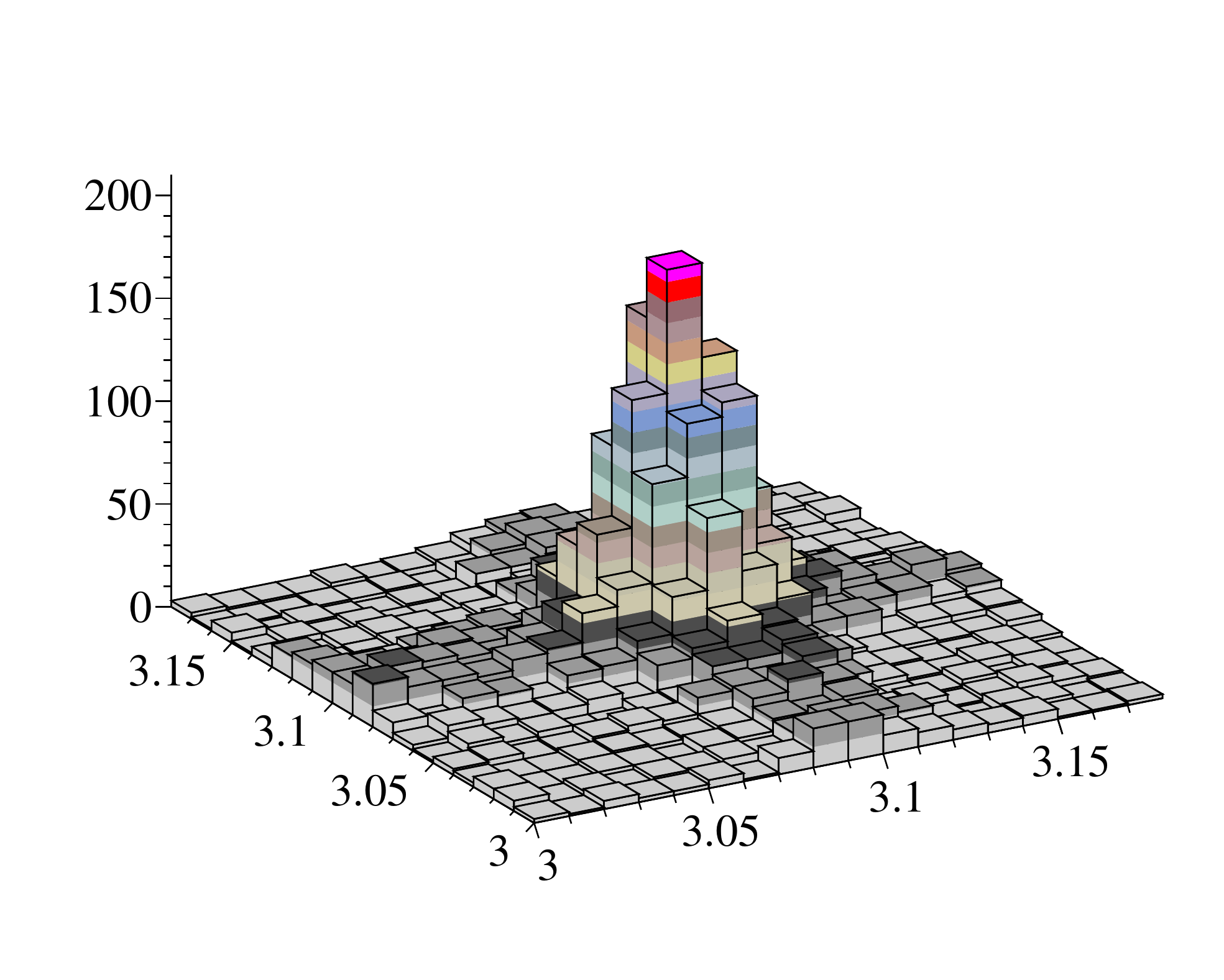}
    }
    \put( 75,  55){ 
      \includegraphics*[width=75mm,height=60mm,%
      ]{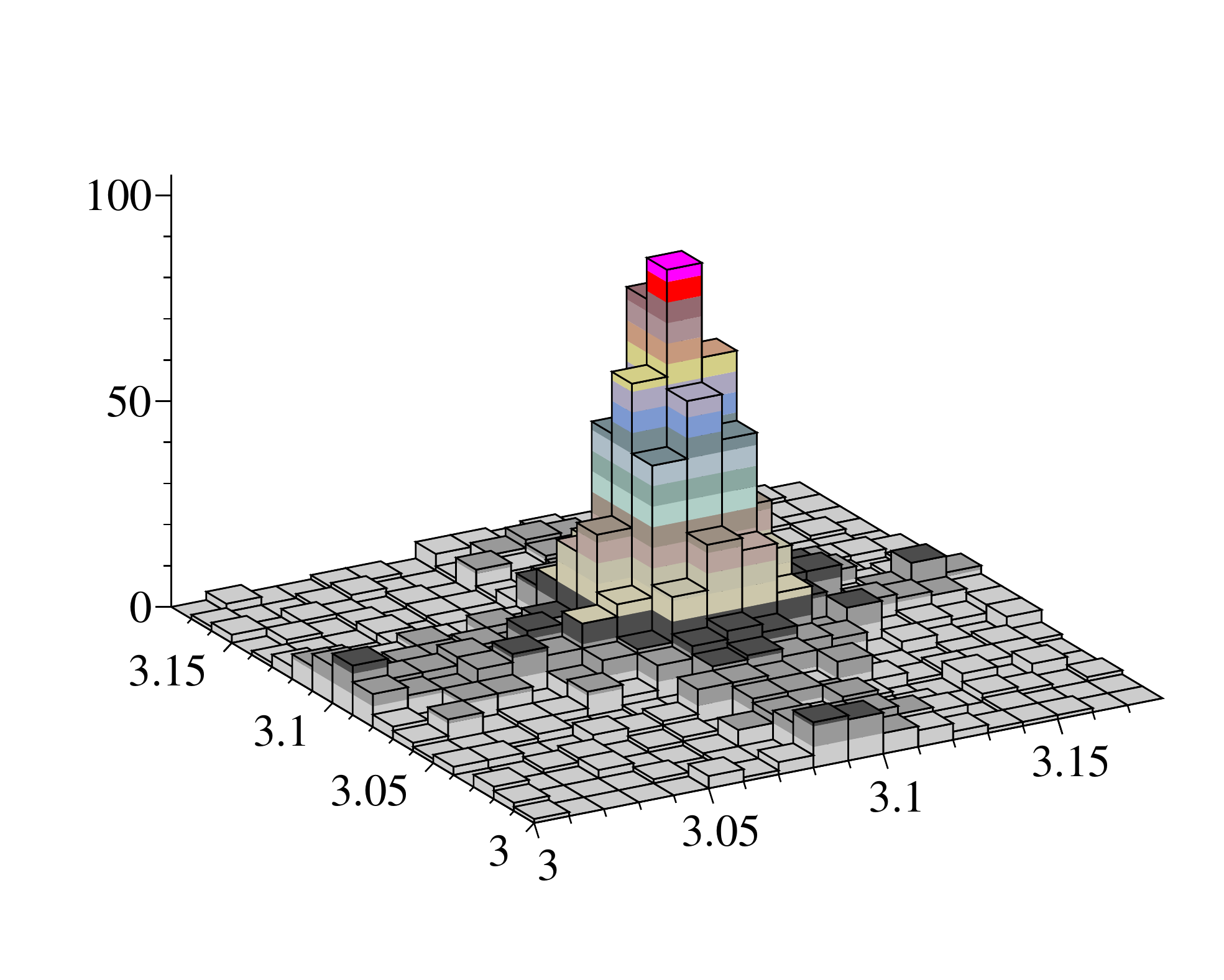}
    }
    \put(  0,  0){ 
      \includegraphics*[width=75mm,height=60mm,%
      ]{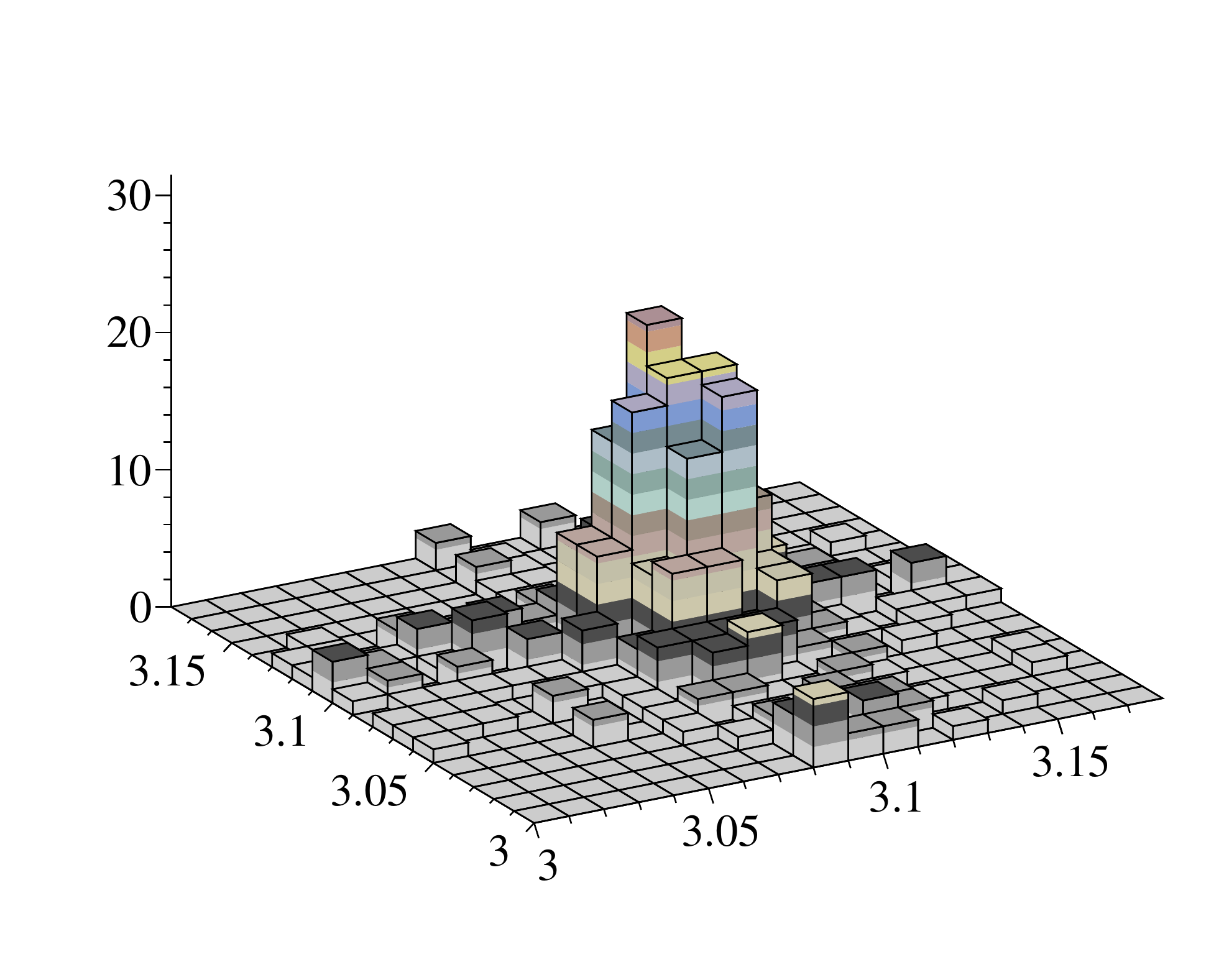}
    }
    \put( 75,  0){ 
      \includegraphics*[width=75mm,height=60mm,%
      ]{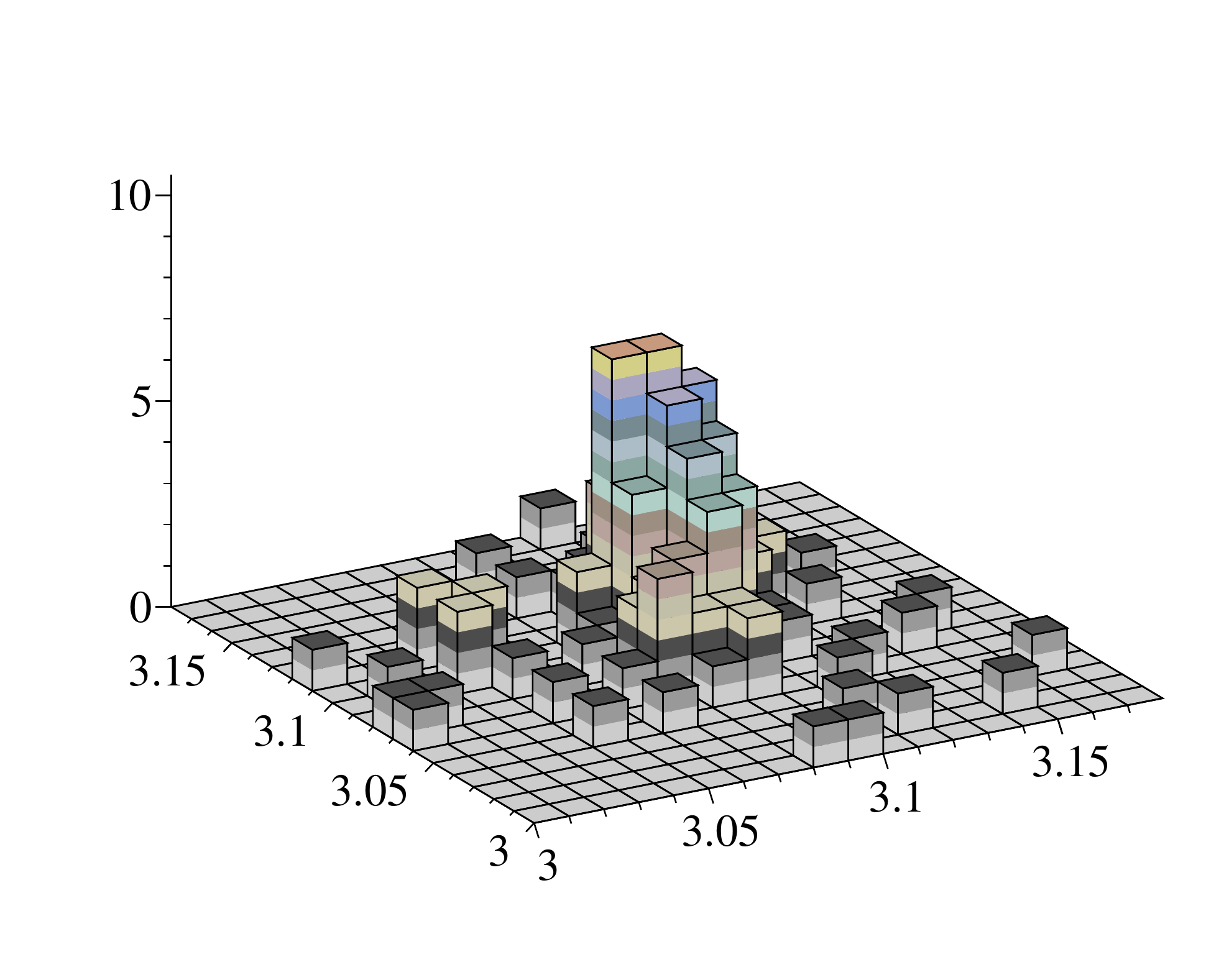}
    }
    \put(  5, 71)  { 
      \begin{rotate}{-33} \small 
        $m_1^{\mumu}~\left[\!\gevcc\right]$
      \end{rotate}
    }
    \put( 45, 59)  { 
      \begin{rotate}{10} \small 
        $m_2^{\mumu}~\left[\!\gevcc\right]$
      \end{rotate}
    }
    \put( 80, 71)  { 
      \begin{rotate}{-33} \small 
        $m_1^{\mumu}~\left[\!\gevcc\right]$
      \end{rotate}
    }
    \put(120, 59)  { 
      \begin{rotate}{10} \small 
        $m_2^{\mumu}~\left[\!\gevcc\right]$
      \end{rotate}
    }
    \put(  5, 16)  { 
      \begin{rotate}{-33} \small 
        $m_1^{\mumu}~\left[\!\gevcc\right]$
      \end{rotate}
    }
    \put( 45,  4)  { 
      \begin{rotate}{10} \small 
        $m_2^{\mumu}~\left[\!\gevcc\right]$
      \end{rotate}
    }
    \put( 80, 16)  { 
      \begin{rotate}{-33} \small 
        $m_1^{\mumu}~\left[\!\gevcc\right]$
      \end{rotate}
    }
    \put(120,  4)  { 
      \begin{rotate}{10} \small 
        $m_2^{\mumu}~\left[\!\gevcc\right]$
      \end{rotate}
    }
    \put(  2, 75) {\scriptsize\begin{sideways}Candidates/$(10\mevcc)^2$\end{sideways}} 
    \put( 77, 75) {\scriptsize\begin{sideways}Candidates/$(10\mevcc)^2$\end{sideways}} 
    \put(  2, 20) {\scriptsize\begin{sideways}Candidates/$(10\mevcc)^2$\end{sideways}} 
    \put( 77, 20) {\scriptsize\begin{sideways}Candidates/$(20\mevcc)^2$\end{sideways}} 
    \put(120,105) { \small $\begin{array}{r}\mathrm{LHCb} \\ \sqs=7,8\tev\end{array}$} 
    \put( 15,104) {a)~{\small{$\ptpsi>2\gevc$}}}
    \put( 90,104) {b)~{\small{$\ptpsi>3\gevc$}}}
    \put( 15, 49) {c)~{\small{$\ptpsi>5\gevc$}}}
    \put( 90, 49) {d)~{\small{$\ptpsi>7\gevc$}}}
  \end{picture}
  \caption { \small
    Distribution of $m_1^{\mumu}$ vs $m_2^{\mumu}$ 
    for selected pairs of $\jpsi\to\mumu$~candidates 
    in different $\ptpsi$~regions.
  }
  \label{fig:lego}
\end{figure}

The two\nobreakdash-dimensional distribution
of the~$\mumu$~masses,
$m^{\mumu}_1$ and $m^{\mumu}_2$,
for the~selected pairs of \mbox{$\jpsi\to\mumu$}~candidates is presented 
in Fig.~\ref{fig:lego} for several requirements on \ptpsi.
A~clear signal peak, corresponding to events with 
two \jpsi~mesons detached from the~PV,
is visible.

\begin{figure}[t]
  \setlength{\unitlength}{1mm}
  \centering
  \begin{picture}(150,60)
    %
    \put(  0,  0){ 
      \includegraphics*[width=75mm,height=60mm,%
      ]{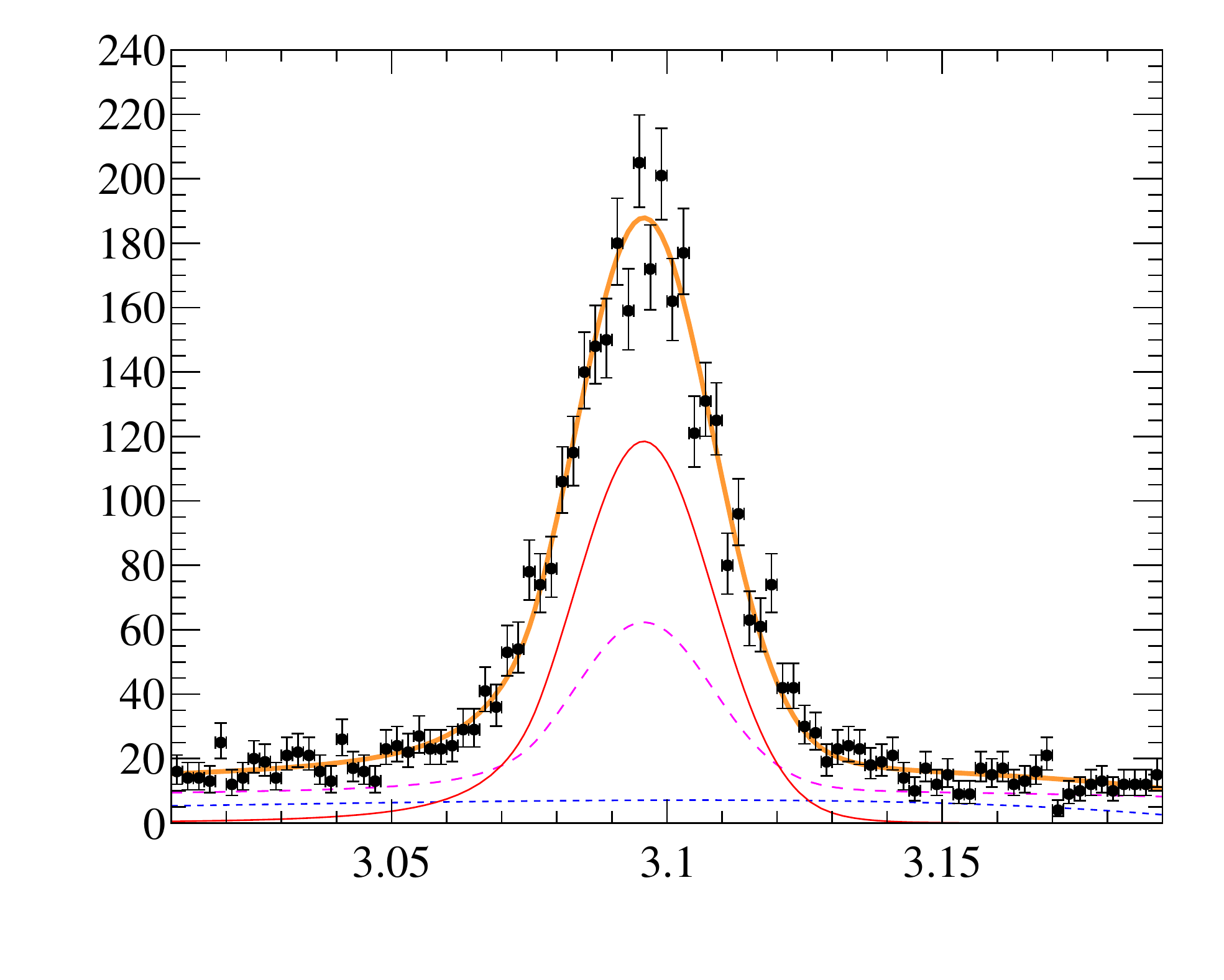}
    }
    \put( 75,  0){ 
      \includegraphics*[width=75mm,height=60mm,%
      ]{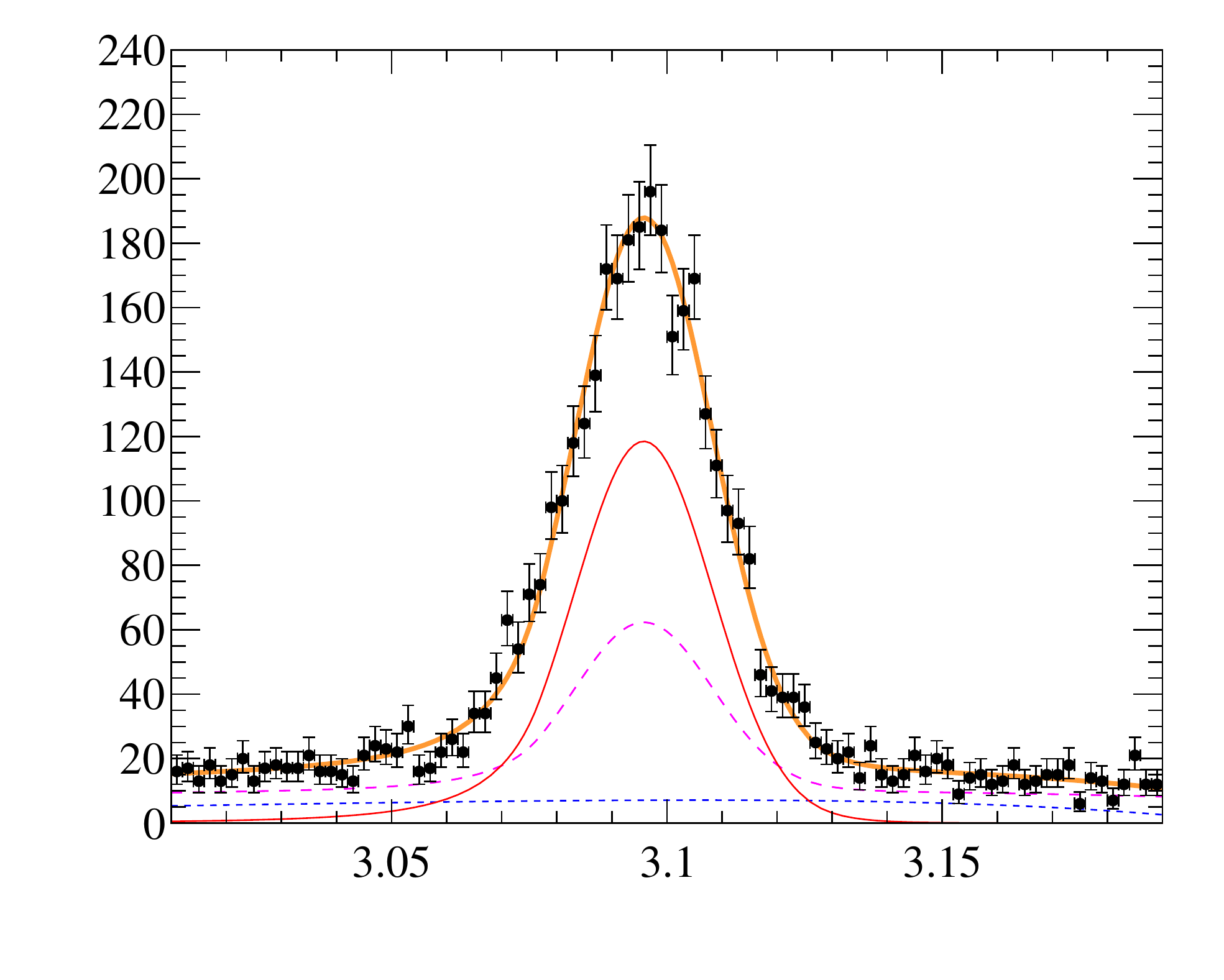}
    }
    \put(-0.5, 15) { \begin{sideways}Candidates/$(2\mevcc)$\end{sideways}} 
    \put(74.5, 15) { \begin{sideways}Candidates/$(2\mevcc)$\end{sideways}} 
    \put( 35,  1) { $m_1^{\mumu}$ } \put( 58,  1) { $\left[\!\gevcc\right]$ }  
    \put(110,  1) { $m_2^{\mumu}$ } \put(133,  1) { $\left[\!\gevcc\right]$ }  
    %
    \put(118,49) { \small $\begin{array}{r}\mathrm{LHCb} \\ \sqs=7,8\tev\end{array}$} 
    \put(44,51.7) { \color[rgb]{1,0,0}  {\rule{4mm}{0.5pt}}} 
    \put(44,47)   { \color[rgb]{1,0,1}  {\hdashrule[0.5ex][x]{4mm}{0.8pt}{1.2mm 0.3mm} } } 
    \put(44,43)   { \color[rgb]{0,0,1}  {\hdashrule[0.5ex][x]{4mm}{0.8pt}{0.5mm 0.5mm} } } 
    \put(44,39.7) { \color[rgb]{1,0.747,0} {\rule{4mm}{2.0pt}}} 
    \put(49,51){ \small $\jpsi\jpsi$ } 
    \put(49,47){ \small $\jpsi+{\mathrm{comb.}}$ } 
    \put(49,43){ \small ${\mathrm{comb.}}$ } 
    \put(49,39){ \small ${\mathrm{total~fit}}$ } 
    \put(89,51){ \small $\ptpsi>2\gevc$ } 
  \end{picture}
  \caption { \small
    Projections of the~extended unbinned maximum likelihood fit
    to 
    (left)\,$m_1^{\mumu}$ and 
    (right)\,$m_2^{\mumu}$ for 
    $\ptpsi>2\gev$.
    The~total fit function is shown as a~solid thick orange line.
    The~solid~thin~red curve shows the~signal component,
    while  
    the~background with one true \jpsi~candidate is shown by 
    the~dashed magenta line
    and the~pure combinatorial background 
    is shown with a~dotted thin blue line.
  }
  \label{fig:fit}
\end{figure}

The~signal yield is determined by performing an~extended unbinned 
maximum likelihood fit to the~two\nobreakdash-dimensional mass distribution.
The~distribution is fitted with the~function
\begin{subequations}
  \begin{eqnarray*} \label{eq:fit}
    \mathfrak{F}(m_1, m_2 ) 
    &  =   &        N_{SS}     \,\,     S\left(m_1\right) \,  S\left(m_2\right)           \label{eq:fit_ss}\\ 
    &  +   & \dfrac{N_{SB}}{2} \,\,    
    \Big( S\left(m_1\right)  B^{\prime}\left( m_2 \right) +  B^{\prime}\left(m_1\right) S\left( m_2 \right) \Big) \label{eq:fit_sb} \\
    &  +   &        N_{BB}     \,\,  B^{\prime\prime}\left(m_1, m_2\right),   \label{eq:fit_bb}
  \end{eqnarray*}
\end{subequations}
where the~first term corresponds to a~signal of two \jpsi~mesons,  
the~second term corresponds to a~combination of 
one \jpsi~meson and combinatorial background; and 
the~last term describes pure combinatorial background.
The~coefficients 
$N_{SS}$,
$N_{SB}$ and 
$N_{BB}$ are the~yields for these three components.
The~signal component, denoted as $S(m)$, is modelled by 
a~double\nobreakdash-sided Crystal~Ball 
function~\cite{Skwarnicki:1986xj,LHCb-PAPER-2011-013}.
The~background component, $B^{\prime}(m)$,
is parameterized as
the~product of an~exponential and a~first\nobreakdash-order 
polynomial function
and 
the~background component $B^{\prime\prime}(m_1,m_2)$ is parameterized as 
the~product of two exponential functions 
$\mathrm{e}^{-\uptau m_1}$ and 
$\mathrm{e}^{-\uptau m_2}$,
with the~same slope parameter, $\uptau$, 
and a~symmetric second\nobreakdash-order polynomial. 
With these parameterizations
the~overall function is symmetric,
$\mathfrak{F}(m_2, m_1 )\equiv\mathfrak{F}(m_1, m_2 )$. 
The~power\nobreakdash-law tail parameters of
the~double\nobreakdash-sided Crystal~Ball 
function are fixed to the~values obtained from simulation, 
leaving the~mean and the~core width as free parameters. 
Results of the~extended unbinned maximum likelihood fit for 
the~different requirements on \ptpsi are presented in Table~\ref{tab:fits}.
Figure~\ref{fig:fit} 
shows the~projections of the fit for~\mbox{$\ptpsi>2\gevc$}.

\begin{table}[t]
  \centering
  \caption{ \small
    Signal and background yields from 
    the~extended unbinned maximum likelihood fit 
    for different requirements on \ptpsi.
    The~uncertainties are statistical only.
  } \label{tab:fits}
  \vspace*{3mm}
  \begin{tabular*}{0.95\textwidth}{@{\hspace{1mm}}l@{\extracolsep{\fill}}cccc@{\hspace{1mm}}}
    &   $p^{\jpsi}_{\mathrm{T}}>2\gevc$ & $p^{\jpsi}_{\mathrm{T}}>3\gevc$ & $p^{\jpsi}_{\mathrm{T}}>5\gevc$ & $p^{\jpsi}_{\mathrm{T}}>7\gevc$ \\
    \hline 
    $N_{SS}$  & $2066          \pm72$  & $1092          \pm50$ & $302\pm17$           &  $98\pm13$           \\
    $N_{SB}$  & $2066          \pm88 $ & $\phantom{0}949\pm58$ & $217\pm17$           &  $40\pm13$           \\
    $N_{BB}$  & $\phantom{0}945\pm73$  & $\phantom{0}343\pm50$ & $\phantom{0}39\pm12$ &  $11\pm9\phantom{0}$ 
  \end{tabular*}
\end{table}

Several background sources potentially contribute to the~observed 
$\jpsi$\nobreakdash-pair signal.
The~first group of sources involves events where 
two \jpsi~mesons originate from different $\proton\proton$~collision vertices:
it includes events with two \jpsi~mesons from decays of beauty hadrons,
events with one \jpsi~meson originating from a~beauty hadron decay and 
another \jpsi~meson produced promptly 
and, finally, events with two prompt \jpsi~mesons. 
The~second group of sources consists of 
events where both \jpsi~mesons
originate from the~same $\proton\proton$~collision, 
namely prompt $\jpsi$\nobreakdash-pair production~\cite{LHCb-PAPER-2011-013,LHCb-PAPER-2016-057},
and associated production of a~prompt $\jpsi$~meson 
and a~$\bquark\bquarkbar$~pair,
where one of the~\bquark~hadrons decays into a~\jpsi~meson.

The~contribution from the~first group of background sources 
is estimated from 
the~measured production cross\nobreakdash-sections 
for $\bquark\to\jpsi\X$~and prompt \jpsi~events~\cite{LHCb-PAPER-2011-003,LHCb-PAPER-2013-016}, 
the~multiplicity of $\proton\proton$~collision vertices and the~size 
of the~beam collision region. 
Taking from   simulation an~estimate for the~probability of 
reconstructing two spatially close PVs as a~single PV,
the~total relative contribution 
from these sources is found to be less than~0.1\%. 

For the second group of background sources, 
the~contribution from prompt $\jpsi$\nobreakdash-pair production 
is significantly suppressed by the~requirement that 
both dimuon vertices are displaced from the~PV.
Using the~production cross\nobreakdash-section 
for prompt \jpsi~pairs\footnote{ The~production 
  cross\nobreakdash-section of \jpsi~pairs 
  is measured at~\mbox{$\sqs=7\tev$}~\cite{LHCb-PAPER-2011-013}.
  The~cross\nobreakdash-section at~\mbox{$\sqs=8\tev$}
  is estimated using a~linear interpolation 
  between the measurements 
  at~\mbox{$\sqs=7\tev$}  
  and~\mbox{$\sqs=13\tev$}~\cite{LHCb-PAPER-2016-057}.},
the~relative contribution from this source is estimated 
to be less than~0.05\%. 
The~background from associated production of $\bquark\bquarkbar$~and 
a~prompt \jpsi~meson in 
the~same $\proton\proton$~collision is calculated 
assuming double parton scattering is the~dominant
production mechanism, 
following Ref.~\cite{LHCb-PAPER-2015-046}.
The~relative contribution from this source is estimated 
to be less than~0.05\%. 

Normalized differential 
cross\nobreakdash-sections~\cite{LHCb-PAPER-2012-003,LHCb-PAPER-2015-046}
are  presented 
as a~function of kinematic variables, 
defined below, and here generically denoted as~$v$,
\begin{equation}\label{eq:shape}
  \dfrac{1}{\upsigma}
  \dfrac{\deriv\upsigma}{\deriv v} \equiv 
  \dfrac{1}{N^{\mathrm{cor}}}
  \dfrac{\Delta N^{\mathrm{cor}}_i}{\Delta v_i},
\end{equation}
where 
$N^{\mathrm{cor}}$ 
is the~total number of efficiency\nobreakdash-corrected signal candidates, 
$\Delta N^{\mathrm{cor}}_i$ 
is the~number of efficiency\nobreakdash-corrected signal candidates in bin~$i$, 
and $\Delta v_i$ is the~corresponding bin width. 
The~efficiency\nobreakdash-corrected yields 
$N^{\mathrm{cor}}$ and $\Delta N^{\mathrm{cor}}_i$ 
are calculated as in Refs.~\cite{LHCb-PAPER-2012-003,LHCb-PAPER-2013-062}
\begin{eqnarray*}
         N^{\mathrm{cor}}   & = & \sum_j \dfrac{ \upomega_j}{\epsilon_{\mathrm{tot},j}^{\jpsi\jpsi}},   \\ 
  \Delta N^{\mathrm{cor}}_i & = & \sum_{j\subset i} \dfrac{ \upomega_j}{\epsilon_{\mathrm{tot},j}^{\jpsi\jpsi}},   
\end{eqnarray*}
where the~sum runs over all pairs of \jpsi~candidates in the~case of $N^{\mathrm{cor}}$ and 
all pairs of \jpsi~candidates in bin~$i$ in the~case of $\Delta N^{\mathrm{cor}}_i$.
Here~$\epsilon_{\mathrm{tot}}^{\jpsi\jpsi}$ is the~total efficiency 
for the~pair of \jpsi~candidates 
and the~weights $\upomega_j$ are determined using the~$sPlot$ technique~\cite{Pivk:2004ty}.

The~total efficiency of the~\jpsi pair is estimated 
on an~event\nobreakdash-by\nobreakdash-event basis 
as in Refs.~\mbox{\cite{LHCb-PAPER-2011-013,LHCb-PAPER-2012-003,LHCb-PAPER-2013-062,LHCb-PAPER-2015-046,LHCb-PAPER-2016-057}}
\begin{equation}\label{eq:eff}
  \epsilon_{\mathrm{tot}}^{\jpsi\jpsi} = 
  \epsilon_{\mathrm{acc}}^{\jpsi\jpsi}\, 
  \epsilon_{\mathrm{rec\&sel}}^{\jpsi\jpsi}\ 
  \epsilon_{\mathrm{\Pmu ID}}^{\jpsi\jpsi}\,  
  \epsilon_{\mathrm{trg}}^{\jpsi\jpsi},
\end{equation}
where 
$\epsilon_{\mathrm{acc}}$~is the~geometrical acceptance of 
the~LHCb~detector, 
$\epsilon_{\mathrm{rec\&sel}}$~is the~reconstruction and selection
efficiency for candidates with all final\nobreakdash-state muons 
inside the~geometrical acceptance, 
$\epsilon_{\mathrm{\Pmu ID}}$~is the~muon identification\,($\Pmu$ID) efficiency 
for the~selected candidates
and 
$\epsilon_{\mathrm{trg}}$~is the~trigger efficiency for 
the~selected candidates satisfying the~$\Pmu$ID requirement. 
The~efficiencies, 
$\epsilon_{\mathrm{acc}}$,
$ \epsilon_{\mathrm{rec\&sel}}$ and 
$ \epsilon_{\mathrm{\Pmu ID}}$, 
are factorized as 
\begin{equation}
  \epsilon^{\jpsi\jpsi} \equiv  
  \epsilon^{\jpsione}  
  \epsilon^{\jpsitwo}, 
\end{equation}
while the~trigger efficiency is decomposed 
as in Refs.~\cite{LHCb-PAPER-2011-013,LHCb-PAPER-2012-003,LHCb-PAPER-2016-057}
\begin{equation}
  \epsilon_{\mathrm{trg}}^{\jpsi\jpsi} \equiv 
  1 - 
  \left(   1 - \epsilon_{\mathrm{trg}}^{\jpsione} \right) 
  \left(   1 - \epsilon_{\mathrm{trg}}^{\jpsitwo} \right). 
\end{equation}
The~efficiencies 
$\epsilon_{\mathrm{acc}}^{\jpsi}$, 
$\epsilon_{\mathrm{rec\&sel}}^{\jpsi}$ and  
$\epsilon_{\mathrm{trg}}^{\jpsi}$ are estimated 
as functions 
of the~transverse momentum and rapidity of the~\jpsi~meson
using simulation. 
The~trigger efficiency for single \jpsi~mesons, 
$\epsilon_{\mathrm{trg}}^{\jpsi}$, has been validated using data. 
The~muon identification efficiency for \jpsi~mesons is factorized as 
\begin{equation}
  \epsilon_{\mathrm{\Pmu ID}}^{\jpsi} \equiv 
  \epsilon_{\mathrm{\Pmu ID}}^{\mup}\, 
  \epsilon_{\mathrm{\Pmu ID}}^{\mun},  
\end{equation}
where the~corresponding single\nobreakdash-muon identification efficiency, 
$\epsilon_{\mathrm{\Pmu ID}}^{\Pmu^{\pm}}$, is 
determined 
as a~function of muon momentum 
and pseudorapidity 
using large samples of prompt \jpsi~mesons.

\subsection {Systematic uncertainties}\label{sec:systematics}

The~systematic uncertainty due to the~imprecise determination of 
the~luminosity
does not enter in the~normalized differential 
cross\nobreakdash-sections. 
The~systematic uncertainties, related to the~evaluation of 
the~efficiency\nobreakdash-corrected signal yields 
$N^{\mathrm{cor}}$ and $\Delta N^{\mathrm{cor}}_i$ from 
Eq.~\eqref{eq:shape}
are summarized in Table~\ref{tab:syst} and are discussed in detail below.

\begin{table}[t]
  \centering
  \caption{ \small
    Summary of relative systematic uncertainties for 
    the~efficiency\nobreakdash-corrected signal yield.
  } \label{tab:syst}
  \vspace*{3mm}
  \begin{tabular*}{0.6\textwidth}{@{\hspace{5mm}}l@{\extracolsep{\fill}}c@{\hspace{5mm}}}
    Source                    &  Uncertainty $\left[\%\right]$\\
    \hline 
    Signal determination      &   $<1.0$                             \\
    Muon identification       &   $\phantom{<}\;0.4$                   \\ 
    Track reconstruction      &   $\phantom{<}\;1.7$                   \\ 
    Trigger                   &   $\phantom{<}\;1.2$                   \\
    Simulated sample size     &   $<0.1$ 
  \end{tabular*}
\end{table}

Systematic uncertainties associated with the~signal determination
are studied by varying the~signal and background shapes used for the~fit function.
For~the~signal parameterization, the~power\nobreakdash-law tail 
parameters of the~double\nobreakdash-sided Crystal~Ball function 
are varied according to the~results of fits to 
large samples of
low\nobreakdash-background \mbox{$\bquark\to\jpsi\X$} and \mbox{$\Bu\to\jpsi\Kp$}~candidates.
The~alternative signal shape parameterization from Ref.~\cite{Santos:2013gra} 
is also used in the~fits.
For~the~parameterization of the~background functions, 
$B^{\prime}(m)$ and $B^{\prime\prime}(m_1,m_2)$, 
the~order of the~polynomial functions is varied.
The~difference in the~fitted signal yields does not exceed 
1\% in all of the~above cases.

The~systematic uncertainty related to the~muon identification is estimated 
to be 0.4\%. It~is obtained from the~uncertainties 
for the~single\nobreakdash-particle identification efficiencies,
$\epsilon_{\mathrm{\Pmu ID}}^{\Pmu^{\pm}}$, using pseudoexperiments. 

The~efficiency $\epsilon_{\mathrm{rec\&sel}}^{\jpsi}$ 
is corrected on a~per\nobreakdash-track basis 
for small discrepancies between data 
and simulation using data\nobreakdash-driven 
techniques~\cite{LHCb-DP-2013-002,LHCb-DP-2013-001}.
The~uncertainty in the correction factor is propagated to 
the~determination of 
the~efficiency\nobreakdash-corrected signal yields 
using pseudoexperiments. This~results in a~systematic 
uncertainty of 0.6\%.
Added in quadrature to the~(correlated) uncertainty 
from the~track reconstruction of $0.4\%$  per track\,($1.6\%$ in total) 
these sources give an~overall systematic uncertainty 
associated with the~track reconstruction of~1.7\%.

The~trigger efficiency has been validated using large
low\nobreakdash-background samples of $\Bu\to\jpsi\Kp$~decays and 
inclusive samples of \jpsi~mesons.
Taking the~largest difference 
between simulation and data for $\epsilon^{\jpsi}_{\mathrm{trg}}$,
the~corresponding systematic uncertainty
for the~efficiency\nobreakdash-corrected yields is~1.2\%.

The~uncertainties in the~efficiencies $\epsilon^{\jpsi}_{\mathrm{acc}}$, 
$\epsilon^{\jpsi}_{\mathrm{rec\&sel}}$ and $\epsilon^{\jpsi}_{\mathrm{trg}}$,
which are due to 
the~limited size of the~simulation samples, 
are propagated to 
the~efficiency\nobreakdash-corrected signal yields 
using pseudoexperiments and are less than~0.1\%.

Part of the~uncertainties, 
summarized in Table~\ref{tab:syst}, cancel in
the~ratio $\tfrac{\Delta N^{\mathrm{cor}}_i}{N^{\mathrm{cor}}}$ 
and thus do not affect 
the~normalized differential cross\nobreakdash-sections. 
For~all bins
for  which the~normalized differential cross-sections 
are evaluated, 
the~systematic uncertainty is much smaller than 
the~corresponding statistical  uncertainty and 
is therefore neglected hereafter.

%% file: correlations.tex
{\boldmath{\section{Results}\label{sec:correlations}}}

The~normalized differential production cross\nobreakdash-sections 
defined by Eq.~\eqref{eq:shape} are presented
as a~function of the~following variables:
\vspace*{3mm}
\begin{itemize}
\item
$\left| \Delta \Pphi^{\ast} \right|$, 
  the~difference in the~azimuthal angle, $\Pphi^{\ast}$,
  between the~two beauty hadrons, 
  where $\Pphi^{\ast}$ is estimated 
  from the~direction of the~vector from the~PV to 
  the~decay vertex of the~\jpsi~meson;
\item
$\left| \Delta \Peta^{\ast} \right|$, 
  the~difference in the~pseudorapidity, $\Peta^{\ast}$,
  between 
  the~two beauty hadrons, 
  where   $\Peta^{\ast}$ is estimated from 
  the~direction of the~vector from the~PV to the~decay 
  vertex of the~\jpsi~meson;
\item
 $\mathcal{A}_{\mathrm{T}}\equiv\left| \frac
     {p^{\jpsione}_{\mathrm{T}}-p^{\jpsitwo}_{\mathrm{T}}}
         {p^{\jpsione}_{\mathrm{T}}+p^{\jpsitwo}_{\mathrm{T}}}\right|$, 
         the~asymmetry between the~transverse momenta of two~\jpsi~mesons;
\item
$m^{\jpsi\jpsi}$, the~mass of the~\jpsi~pair;
\item
$p^{\jpsi\jpsi}_{\mathrm{T}}$, the~transverse momentum 
  of the~\jpsi~pair;
\item
$y^{\jpsi\jpsi}$, the~rapidity of the~\jpsi~pair.
\end{itemize}
The~differential cross\nobreakdash-sections with respect to other 
variables are given in Appendix~\ref{app:more}.
The~shapes for the~differential production cross\nobreakdash-sections 
for
\mbox{$\left| \Delta \Pphi^{\ast} \right|$} and 
\mbox{$\left| \Delta \Peta^{\ast} \right|$} variables are 
independent of the~decay of the~long\nobreakdash-lived beauty hadrons
and directly probe the~production properties of pairs of beauty hadrons.
The~other variables have a~minor dependence both on the~branching 
fractions of different beauty hadrons, as well as on 
the~\mbox{$\bquark\to\jpsi\mathrm{X}$}~decay kinematics. 

\begin{figure}[t]
  \setlength{\unitlength}{1mm}
  \centering
  \begin{picture}(150,165)
    \put(  0,110){ 
      \includegraphics*[width=75mm,height=55mm,%
      ]{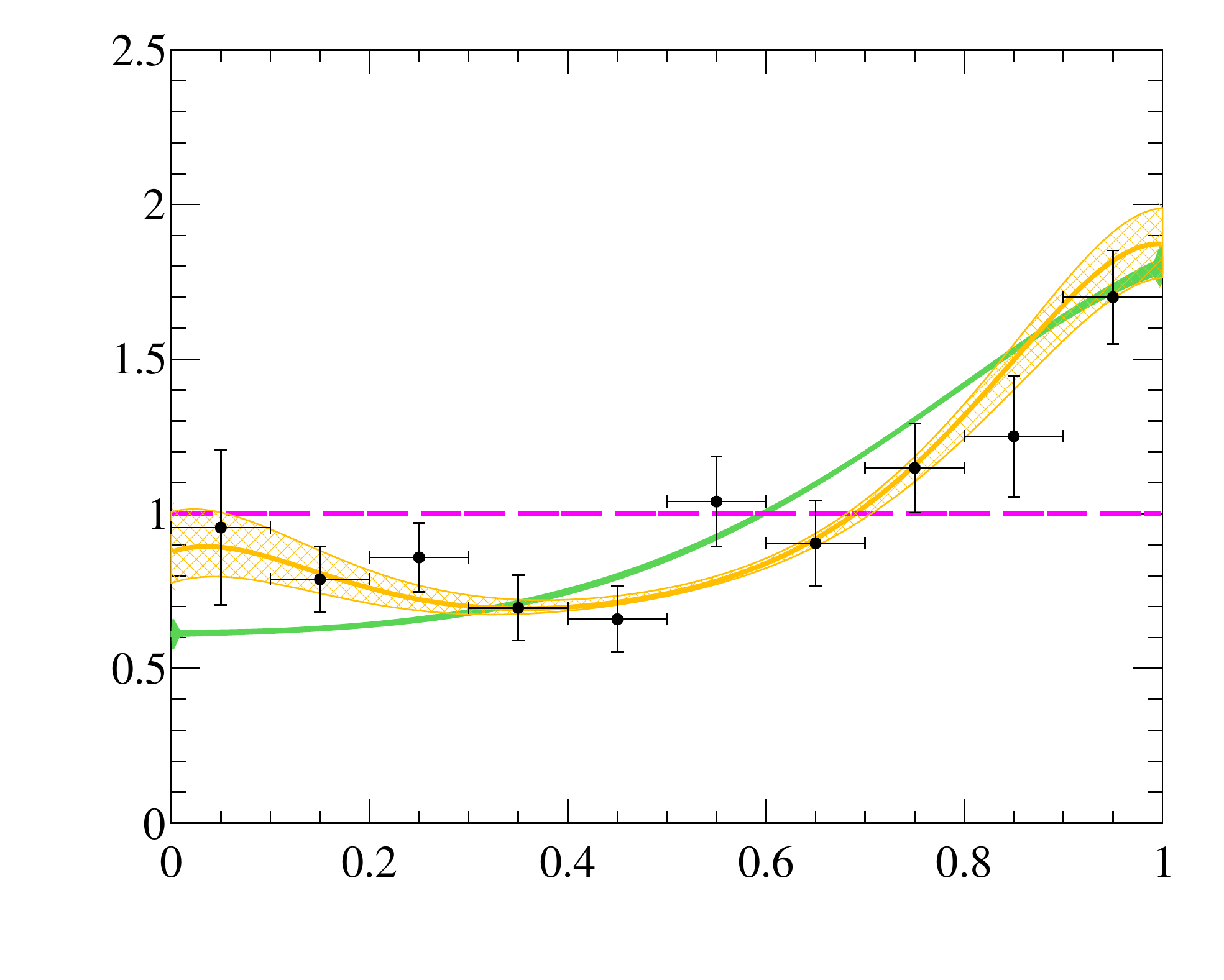}
    }
    \put( 75,110){ 
      \includegraphics*[width=75mm,height=55mm,%
      ]{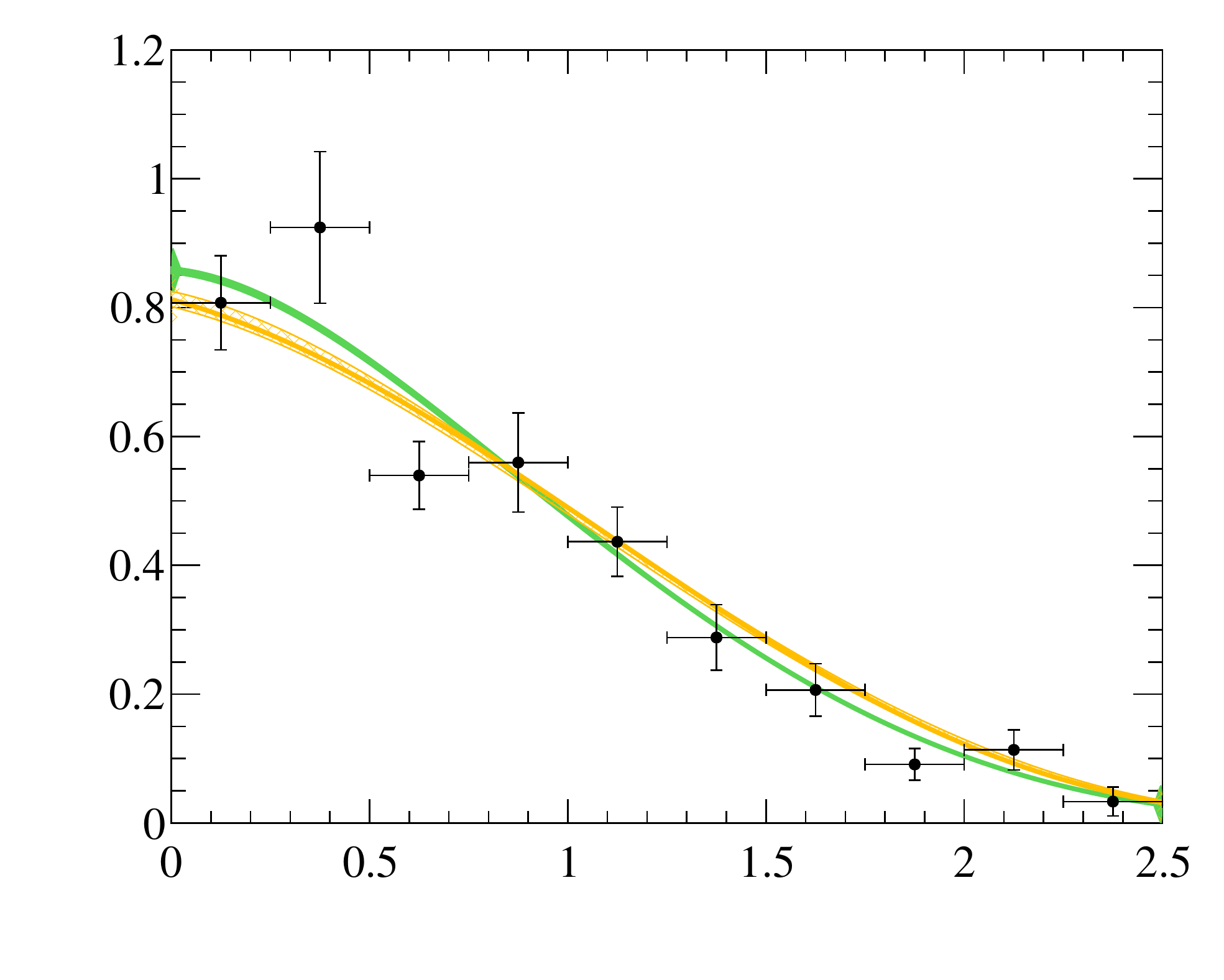}
    }
    \put(  0, 55){ 
      \includegraphics*[width=75mm,height=55mm,%
      ]{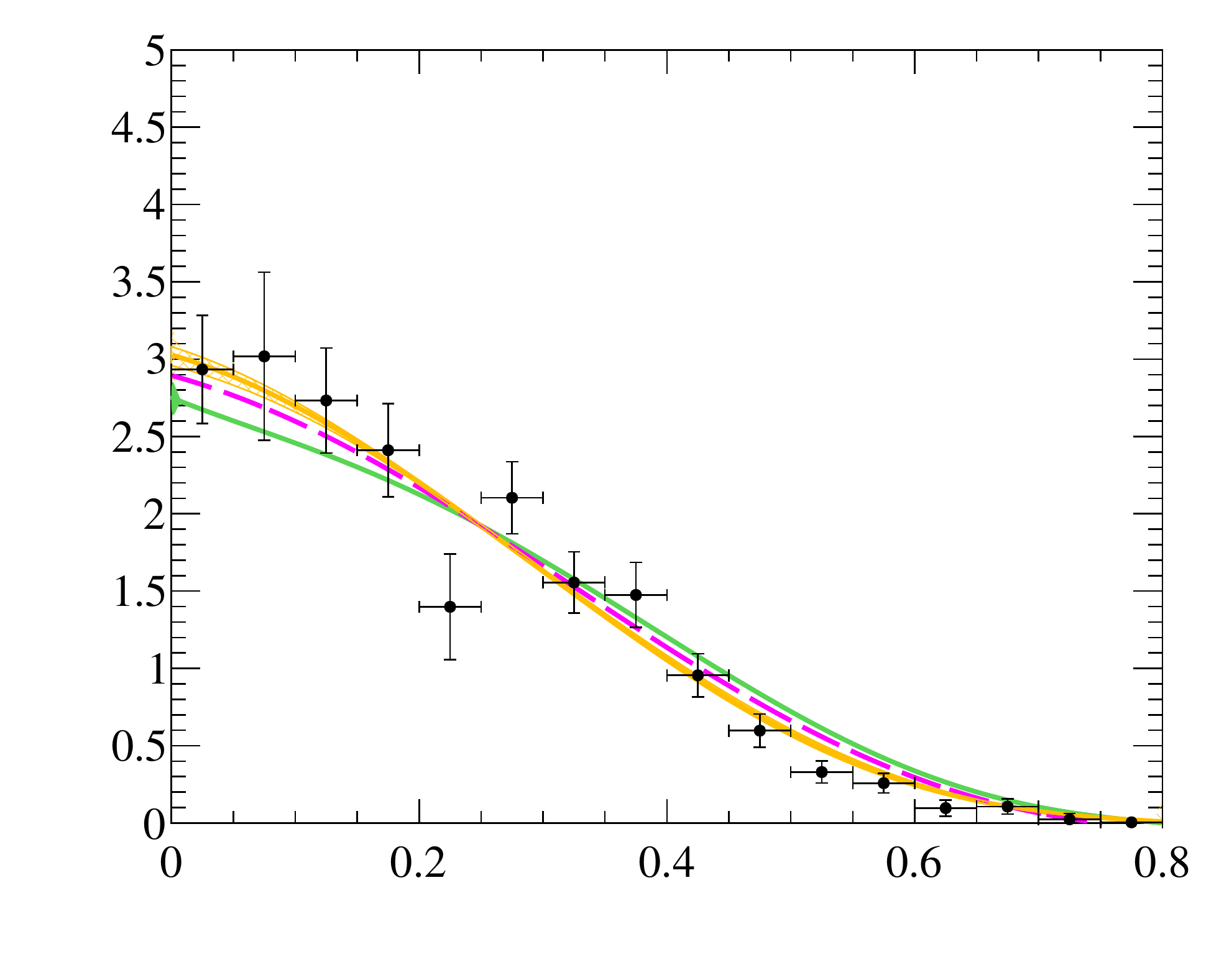}
    }
    \put( 75, 55){ 
      \includegraphics*[width=75mm,height=55mm,%
      ]{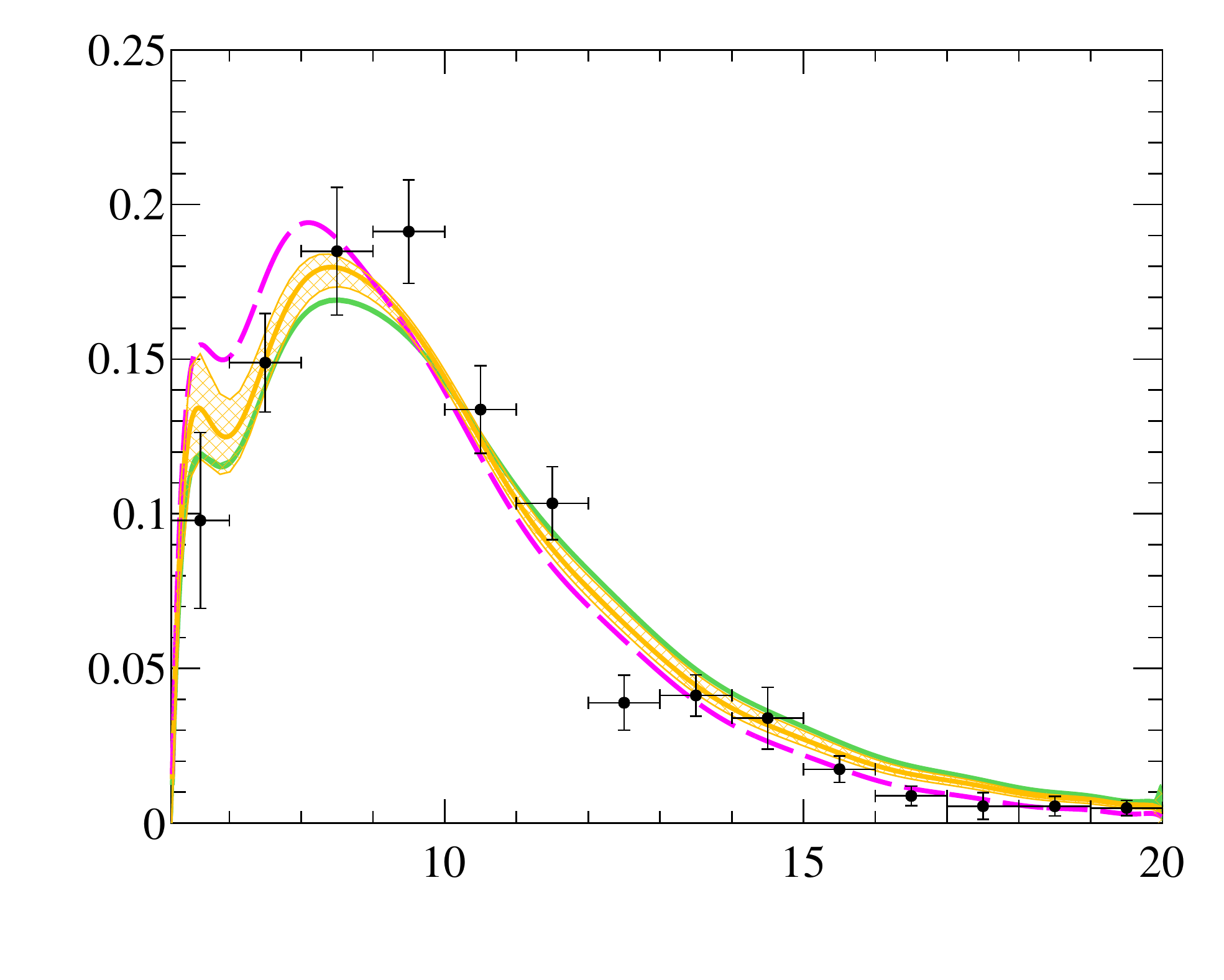}
    }
    \put(  0,  0){ 
      \includegraphics*[width=75mm,height=55mm,%
      ]{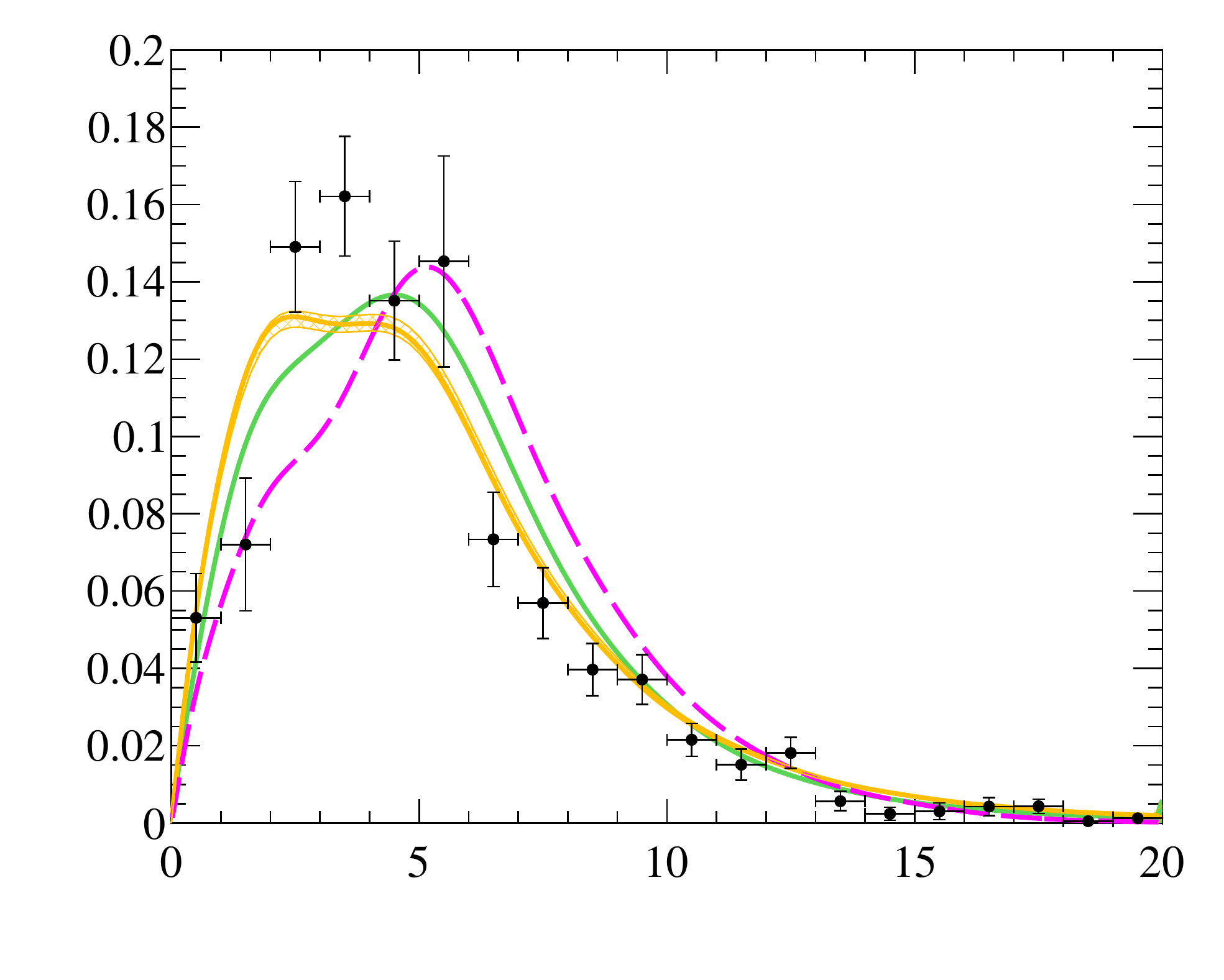}
    }
    \put( 75,  0){ 
      \includegraphics*[width=75mm,height=55mm,%
      ]{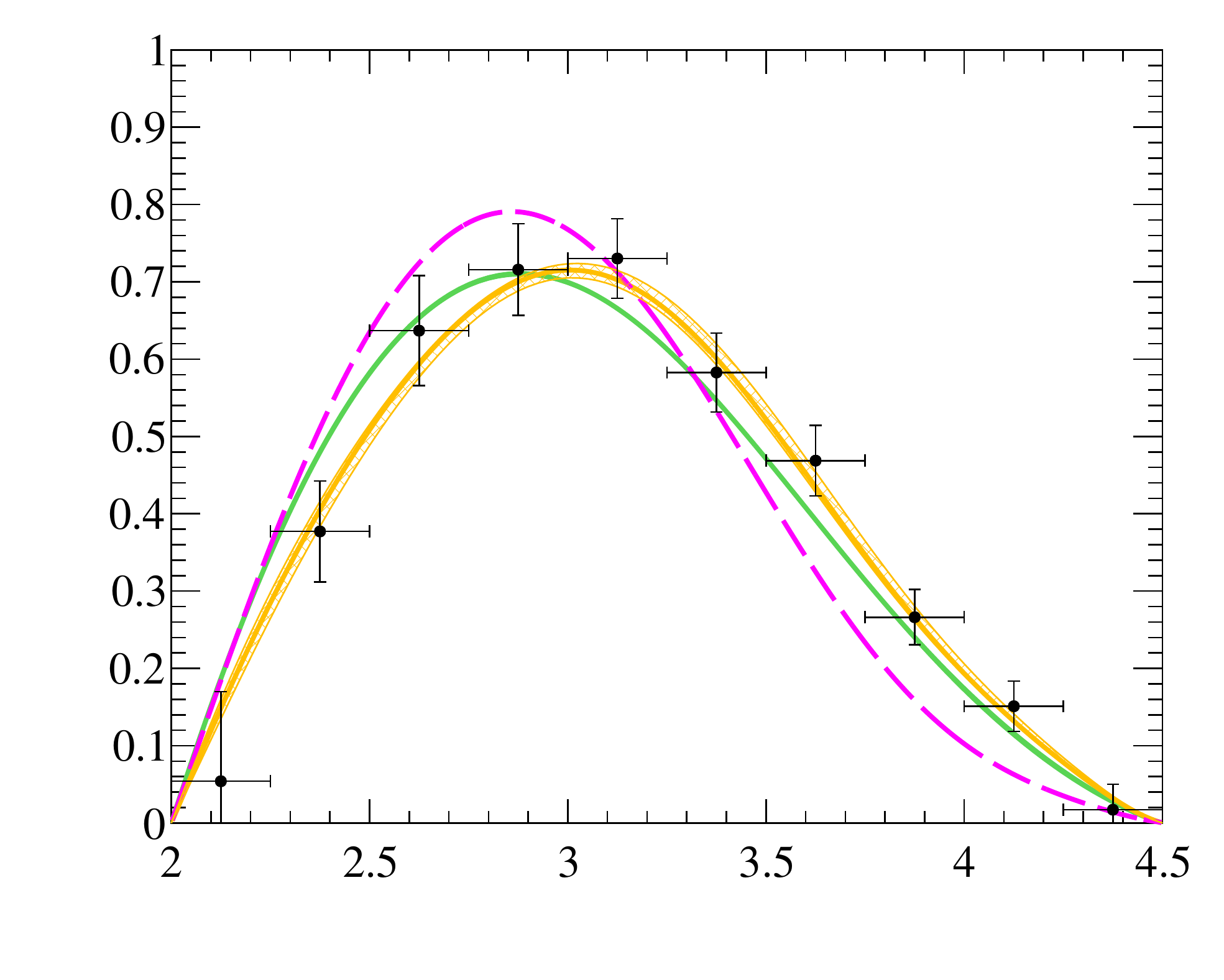}
    }
    \put( -1,150) { \small \begin{sideways} 
        $\frac{\pi}{\upsigma}\frac{\mathrm{d}\upsigma}{\mathrm{d}\left|\Delta\Pphi^{\ast}\right|}$
      \end{sideways}} 
    \put( 74,150) { \small \begin{sideways} 
        $\frac{1}{\upsigma}\frac{\mathrm{d}\upsigma}{\mathrm{d}\left|\Delta\Peta^{\ast}\right|}$
      \end{sideways}} 
     \put( -1, 97) { \small \begin{sideways} 
        $\frac{1}{\upsigma}\frac{\mathrm{d}\upsigma}{\mathrm{d}\mathcal{A}_{\mathrm{T}}}$
      \end{sideways}} 
    \put( 74,  91) { \small \begin{sideways} 
        $\frac{1}{\upsigma}\frac{\mathrm{d}\upsigma}{\mathrm{d} m^{\jpsi\jpsi} }$
      \end{sideways}} 
    \put( -3, 38) { \small \begin{sideways} 
        $\frac{1}{\upsigma}\frac{\mathrm{d}\upsigma}{\mathrm{d}p_{\mathrm{T}}^{\jpsi\jpsi}}$
      \end{sideways}} 
    \put( 74, 38) { \small \begin{sideways} 
        $\frac{1}{\upsigma}\frac{\mathrm{d}\upsigma}{\mathrm{d}y^{\jpsi\jpsi}}$
      \end{sideways}} 
    \put( 35,110) { $\left|\Delta\Pphi^{\ast}\right|/\pi$ } 
    \put(110,110) { $\left|\Delta\Peta^{\ast}\right|$     } 
    \put( 35, 55) { $\mathcal{A}_{\mathrm{T}}$ } 
    \put(110, 55) { $m^{\jpsi\jpsi}$ } 
    \put(132, 55) { $\left[\!\gevcc\right]$}
    \put( 35,  0) { $p_{\mathrm{T}}^{\jpsi\jpsi}$ } 
    \put( 58,  0) { $\left[\!\gevc\right]$}
    \put(110,  0) { $y^{\jpsi\jpsi}$ } 
    \put(115,152) { \small $\begin{array}{r}\mathrm{LHCb} \\
        \sqrt{s}=7,8\,\mathrm{TeV} \\
        p_{\mathrm{T}}^{\jpsi}>2\gevc \end{array}$}
    \put( 16,155) {a)}
    \put( 91,155) {b)}
    \put( 16,100) {c)}
    \put( 91,100) {d)}
    \put( 16, 45) {e)}
    \put( 91, 45) {f)}
    \put(20,155) { \color[rgb]{1,0.747,0}{\begin{tikzpicture}[x=1mm,y=1mm]\draw[thin,pattern=crosshatch, pattern color=Root92]  (0,0) rectangle (12,2.0);\end{tikzpicture}} }
    \put(33,155) { \sc{Powheg} } 
    \put(20,150) { \color[rgb]{0.35,0.83,0.33} {\rule{12mm}{2.2mm}}}
    \put(33,150) { \sc{Pythia} } 
    \put(20,145) { \color[rgb]{1,0,1} {\hdashrule[0.5ex][x]{1.2cm}{0.7pt}{3mm 0.5mm} } } 
    \put(33,145) { \small uncorrelated $\mathrm{b}\bar{\mathrm{b}}$}
  \end{picture}
  \caption { \small
    Normalized differential production cross\nobreakdash-sections\,(points with error bars)
    for 
    a)\,$\left|\Delta\Pphi^{\ast}\right|/\pi$,
    b)\,$\left|\Delta\Peta^{\ast}\right|$,
    c)\,$\mathcal{A}_{\mathrm{T}}$,
    d)\,$m^{\jpsi\jpsi}$,
    e)\,$p^{\jpsi\jpsi}_{\mathrm{T}}$ and 
    f)\,$y^{\jpsi\jpsi}$
    together with the~{\sc{Powheg}}\,(orange line)
    and \pythia\,(green band) predictions. 
    The~expectations for 
    uncorrelated $\bquark\bquarkbar$~production are shown 
    by the~dashed magenta line. 
    The~uncertainties  
    in the~{\sc{Powheg}}~and~$\pythia$
    predictions due to the~choice
    of factorization and renormalization scales
    are shown as orange cross\nobreakdash-hatched 
    and green solid areas, respectively.
  }
  \label{fig:gev2}
\end{figure}

\begin{figure}[t]
  \setlength{\unitlength}{1mm}
  \centering
  \begin{picture}(150,165)
    \put(  0,110){ 
      \includegraphics*[width=75mm,height=55mm,%
      ]{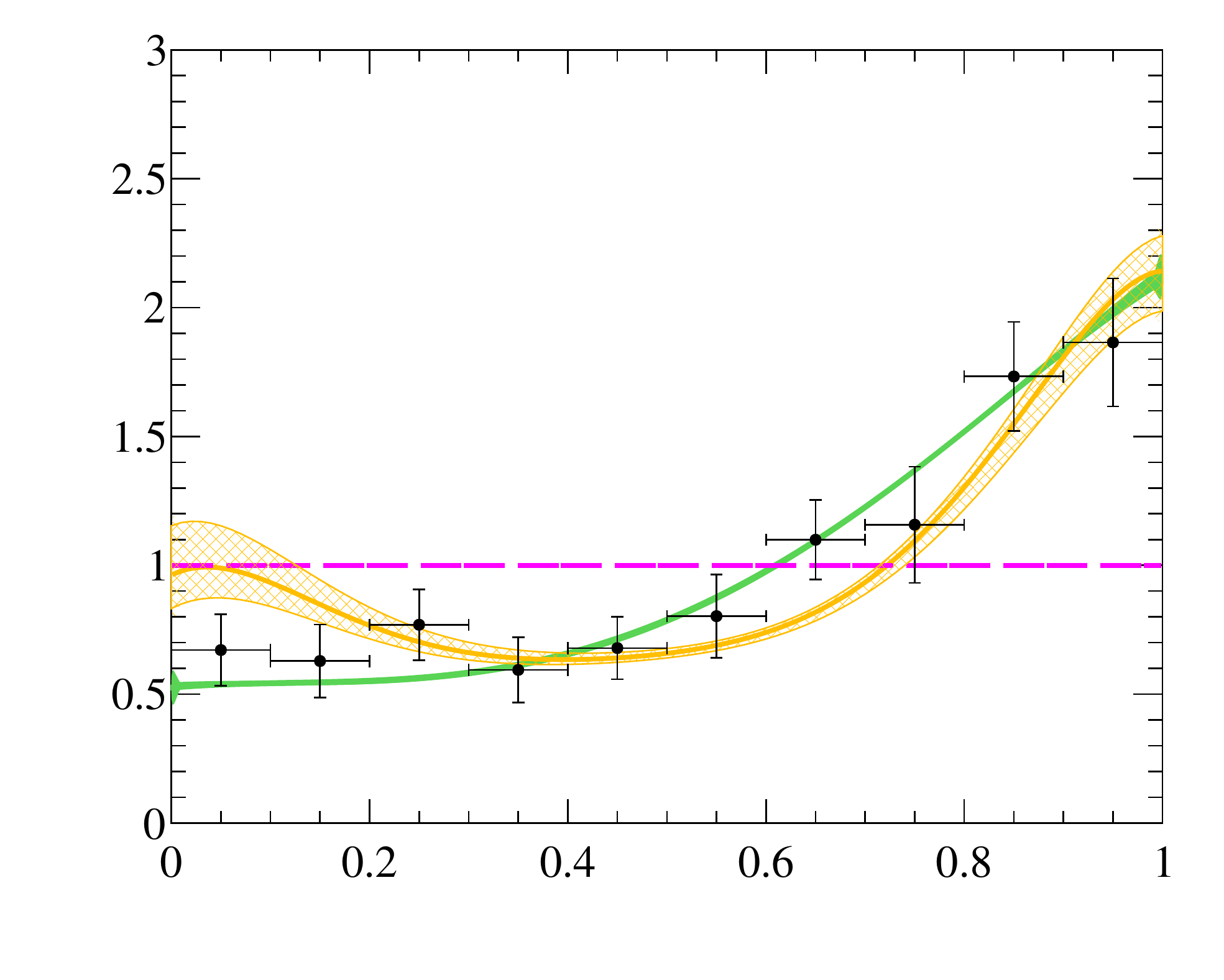}
    }
    \put( 75,110){ 
      \includegraphics*[width=75mm,height=55mm,%
      ]{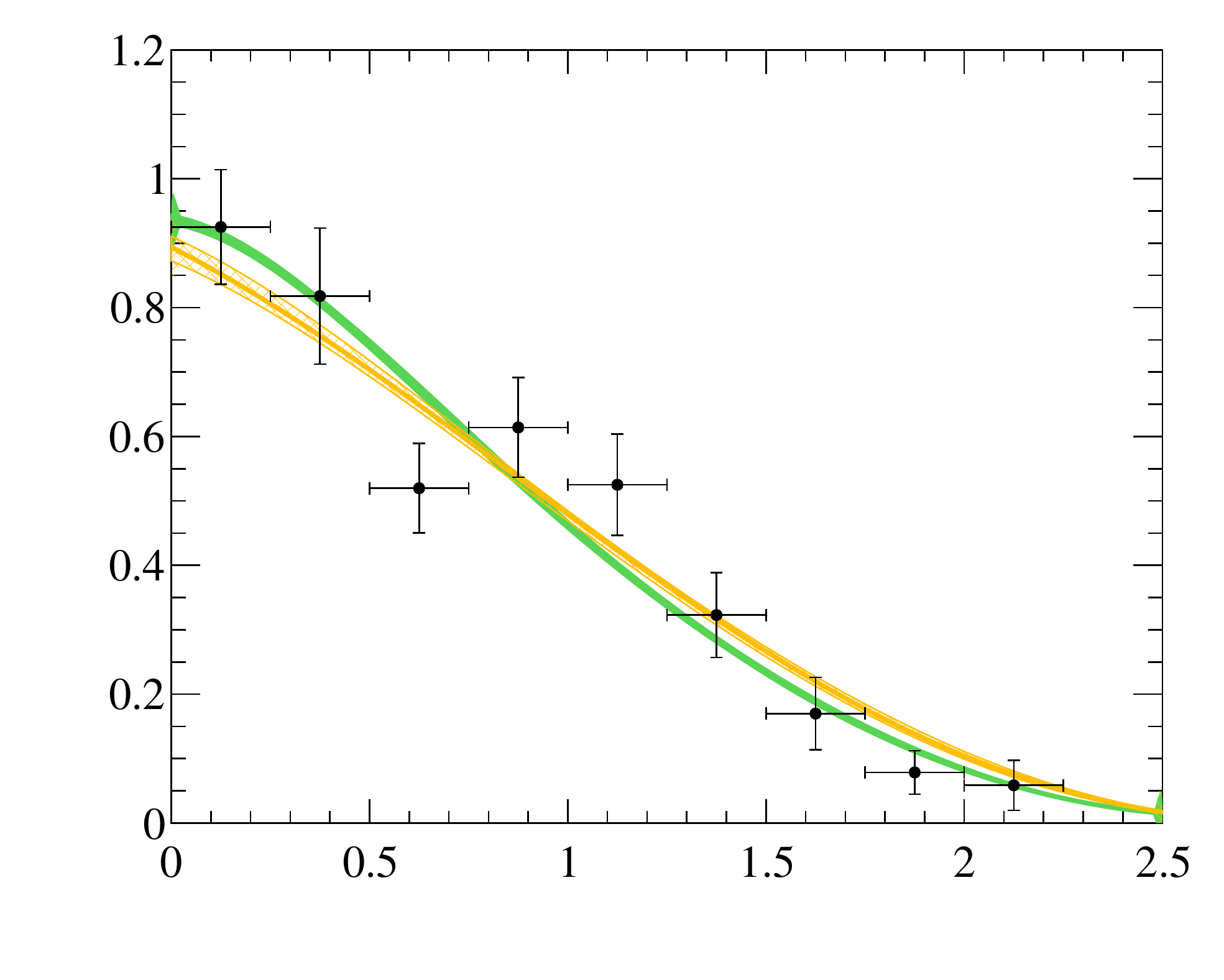}
    }
    \put(  0, 55){ 
      \includegraphics*[width=75mm,height=55mm,%
      ]{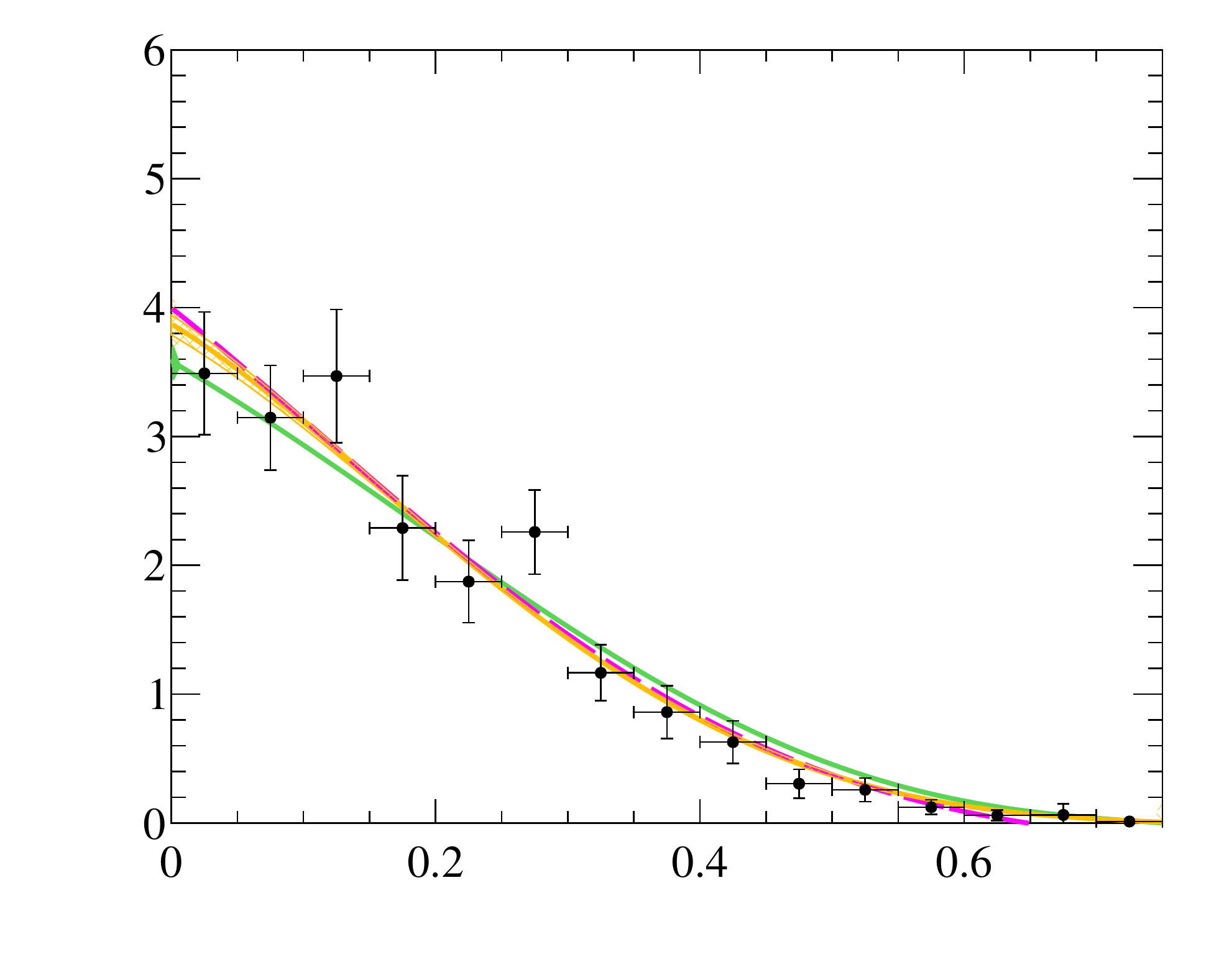}
    }
    \put( 75, 55){ 
      \includegraphics*[width=75mm,height=55mm,%
      ]{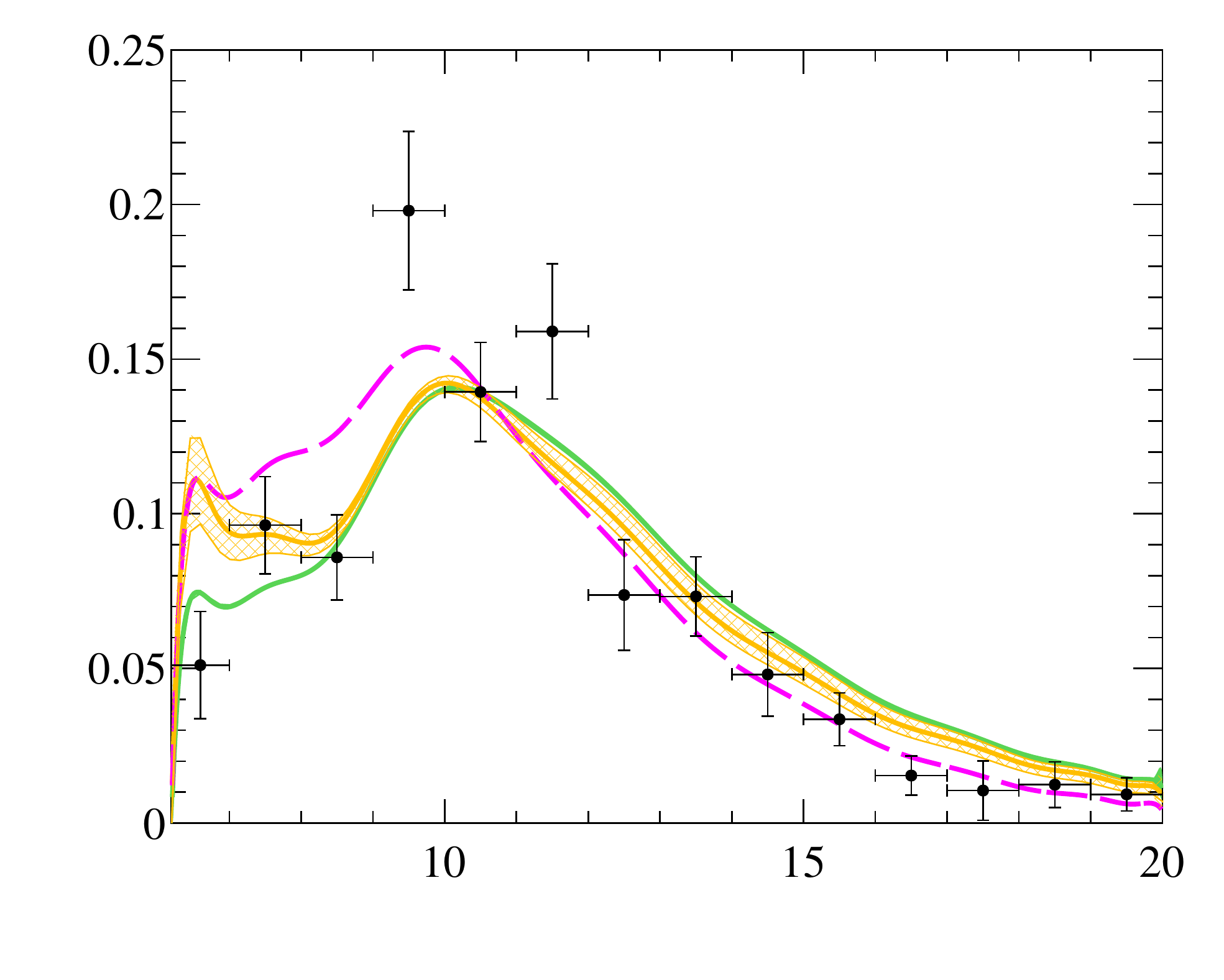}
    }
    \put(  0,  0){ 
      \includegraphics*[width=75mm,height=55mm,%
      ]{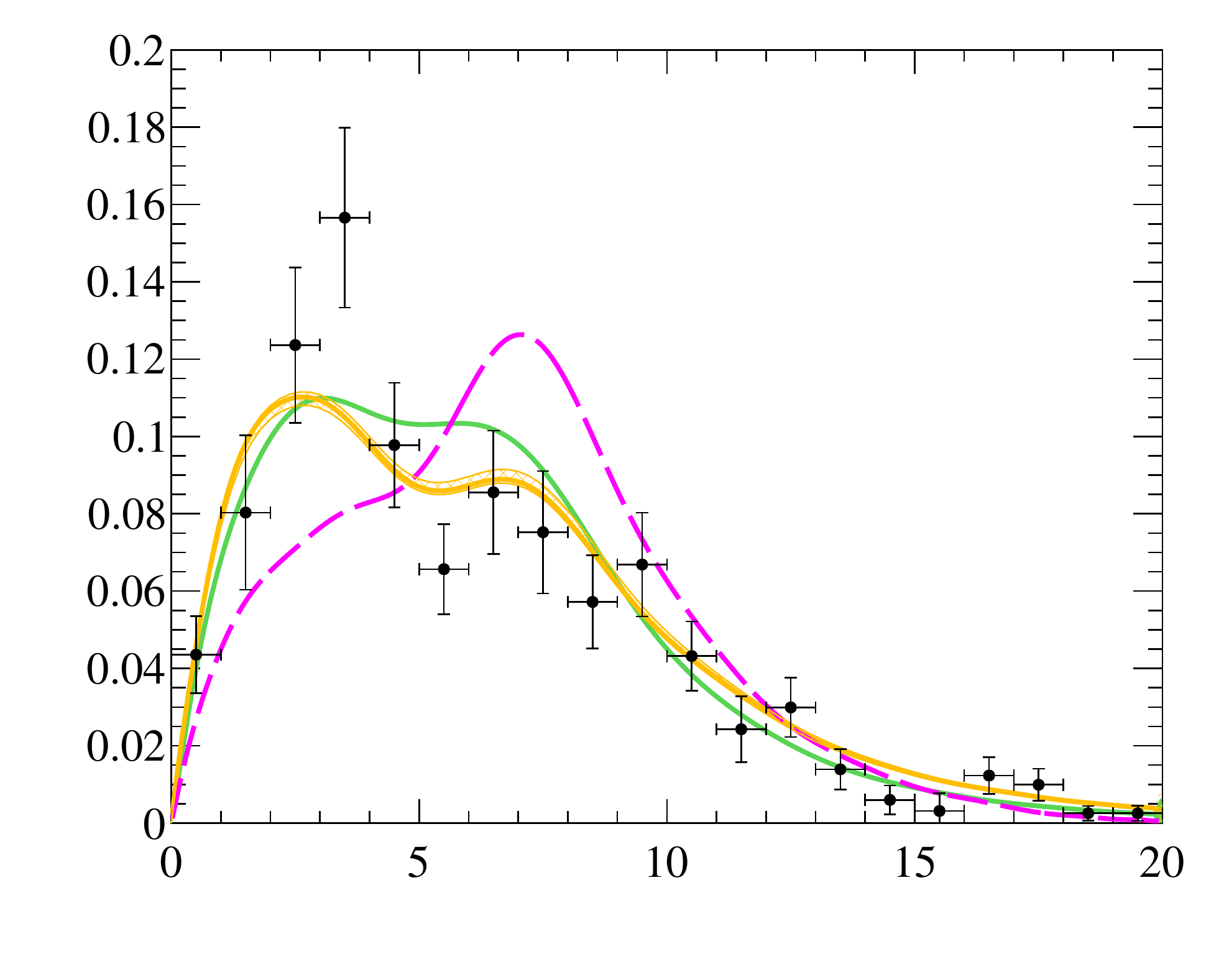}
    }
    \put( 75,  0){ 
      \includegraphics*[width=75mm,height=55mm,%
      ]{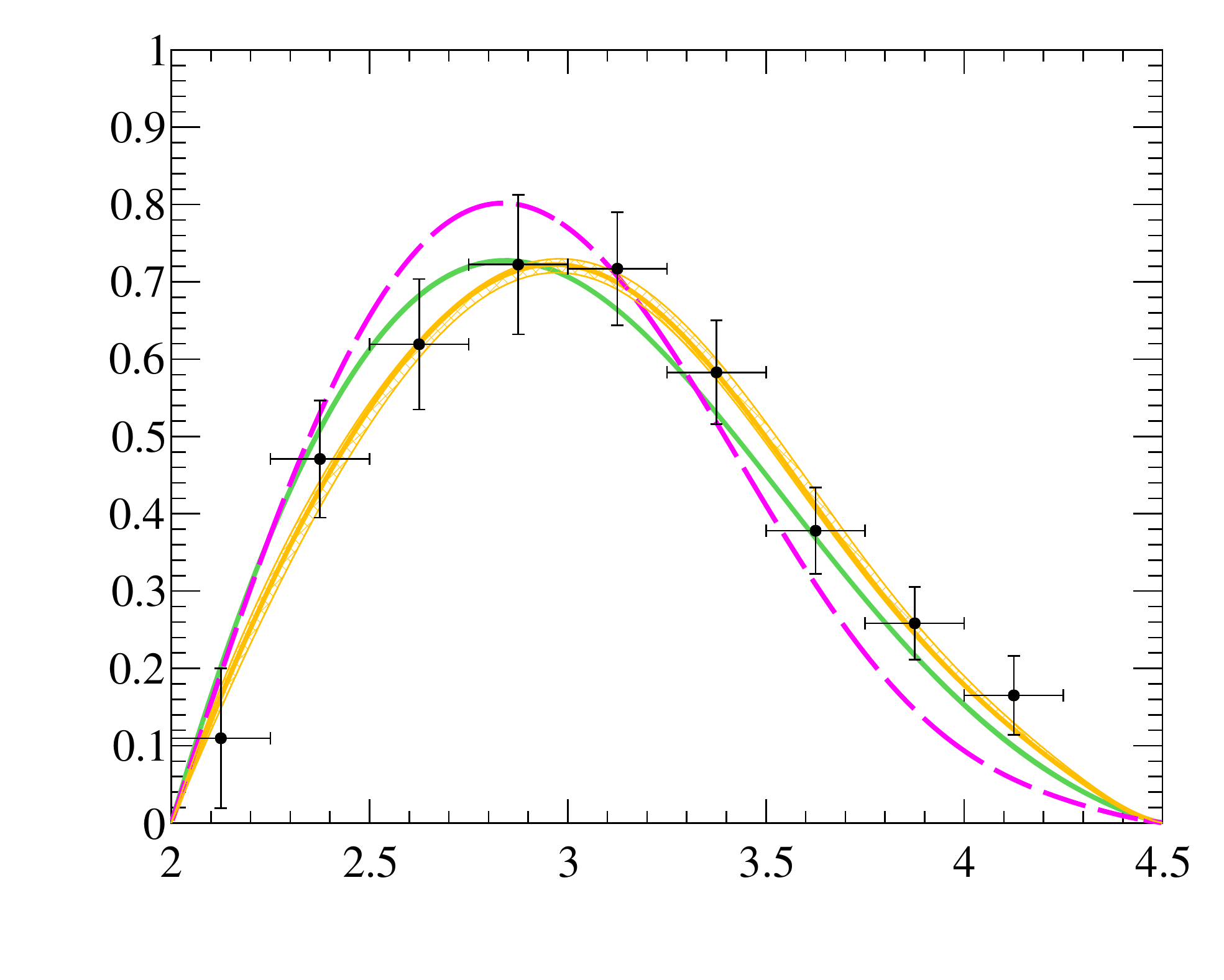}
    }
    \put( -1,150) { \small \begin{sideways} 
        $\frac{\pi}{\upsigma}\frac{\mathrm{d}\upsigma}{\mathrm{d}\left|\Delta\Pphi^{\ast}\right|}$
      \end{sideways}} 
    \put( 74,150) { \small \begin{sideways} 
        $\frac{1}{\upsigma}\frac{\mathrm{d}\upsigma}{\mathrm{d}\left|\Delta\Peta^{\ast}\right|}$
      \end{sideways}} 
     \put( -1, 97) { \small \begin{sideways} 
        $\frac{1}{\upsigma}\frac{\mathrm{d}\upsigma}{\mathrm{d}\mathcal{A}_{\mathrm{T}}}$
      \end{sideways}} 
    \put( 74,  91) { \small \begin{sideways} 
        $\frac{1}{\upsigma}\frac{\mathrm{d}\upsigma}{\mathrm{d} m^{\jpsi\jpsi} }$
      \end{sideways}} 
    \put( -3, 38) { \small \begin{sideways} 
        $\frac{1}{\upsigma}\frac{\mathrm{d}\upsigma}{\mathrm{d}p_{\mathrm{T}}^{\jpsi\jpsi}}$
      \end{sideways}} 
    \put( 74, 38) { \small \begin{sideways} 
        $\frac{1}{\upsigma}\frac{\mathrm{d}\upsigma}{\mathrm{d}y^{\jpsi\jpsi}}$
      \end{sideways}} 
    \put( 35,110) { $\left|\Delta\Pphi^{\ast}\right|/\pi$ } 
    \put(110,110) { $\left|\Delta\Peta^{\ast}\right|$     } 
    \put( 35, 55) { $\mathcal{A}_{\mathrm{T}}$ } 
    \put(110, 55) { $m^{\jpsi\jpsi}$ } 
    \put(132, 55) { $\left[\!\gevcc\right]$}
    \put( 35,  0) { $p_{\mathrm{T}}^{\jpsi\jpsi}$ } 
    \put( 58,  0) { $\left[\!\gevc\right]$}
    \put(110,  0) { $y^{\jpsi\jpsi}$ } 
    \put(115,152) { \small $\begin{array}{r}\mathrm{LHCb} \\
        \sqrt{s}=7,8\,\mathrm{TeV} \\
        p_{\mathrm{T}}^{\jpsi}>3\gevc \end{array}$}
    \put( 16,155) {a)}
    \put( 91,155) {b)}
    \put( 16,100) {c)}
    \put( 91,100) {d)}
    \put( 16, 45) {e)}
    \put( 91, 45) {f)}
    \put(20,155) { \color[rgb]{1,0.747,0}{\begin{tikzpicture}[x=1mm,y=1mm]\draw[thin,pattern=crosshatch, pattern color=Root92]  (0,0) rectangle (12,2.0);\end{tikzpicture}} }
    \put(33,155) { \sc{Powheg} } 
    \put(20,150) { \color[rgb]{0.35,0.83,0.33} {\rule{12mm}{2.2mm}}}
    \put(33,150) { \sc{Pythia} } 
    \put(20,145) { \color[rgb]{1,0,1} {\hdashrule[0.5ex][x]{1.2cm}{0.7pt}{3mm 0.5mm} } } 
    \put(33,145) { \small uncorrelated $\mathrm{b}\bar{\mathrm{b}}$}    
  \end{picture}
  \caption { \small
    Normalized differential production cross\nobreakdash-sections\,(points with error bars) 
    for 
    a)\,$\left|\Delta\Pphi^{\ast}\right|/\pi$,
    b)\,$\left|\Delta\Peta^{\ast}\right|$,
    c)\,$\mathcal{A}_{\mathrm{T}}$,
    d)\,$m^{\jpsi\jpsi}$,
    e)\,$p^{\jpsi\jpsi}_{\mathrm{T}}$ and 
    f)\,$y^{\jpsi\jpsi}$
    together with the~{\sc{Powheg}}\,(orange line)
    and \pythia\,(green band) predictions. 
    The~expectations for 
    uncorrelated $\bquark\bquarkbar$~production are shown by the~dashed magenta line. 
    The~uncertainties  
    in the~{\sc{Powheg}}~and~$\pythia$
    predictions due to the~choice
    of factorization and renormalization scales
    are shown as cross\nobreakdash-hatched 
    and green solid areas, respectively.
  }
  \label{fig:gev3}
\end{figure}

\begin{figure}[t]
  \setlength{\unitlength}{1mm}
  \centering
  \begin{picture}(150,165)
    \put(  0,110){ 
      \includegraphics*[width=75mm,height=55mm,%
      ]{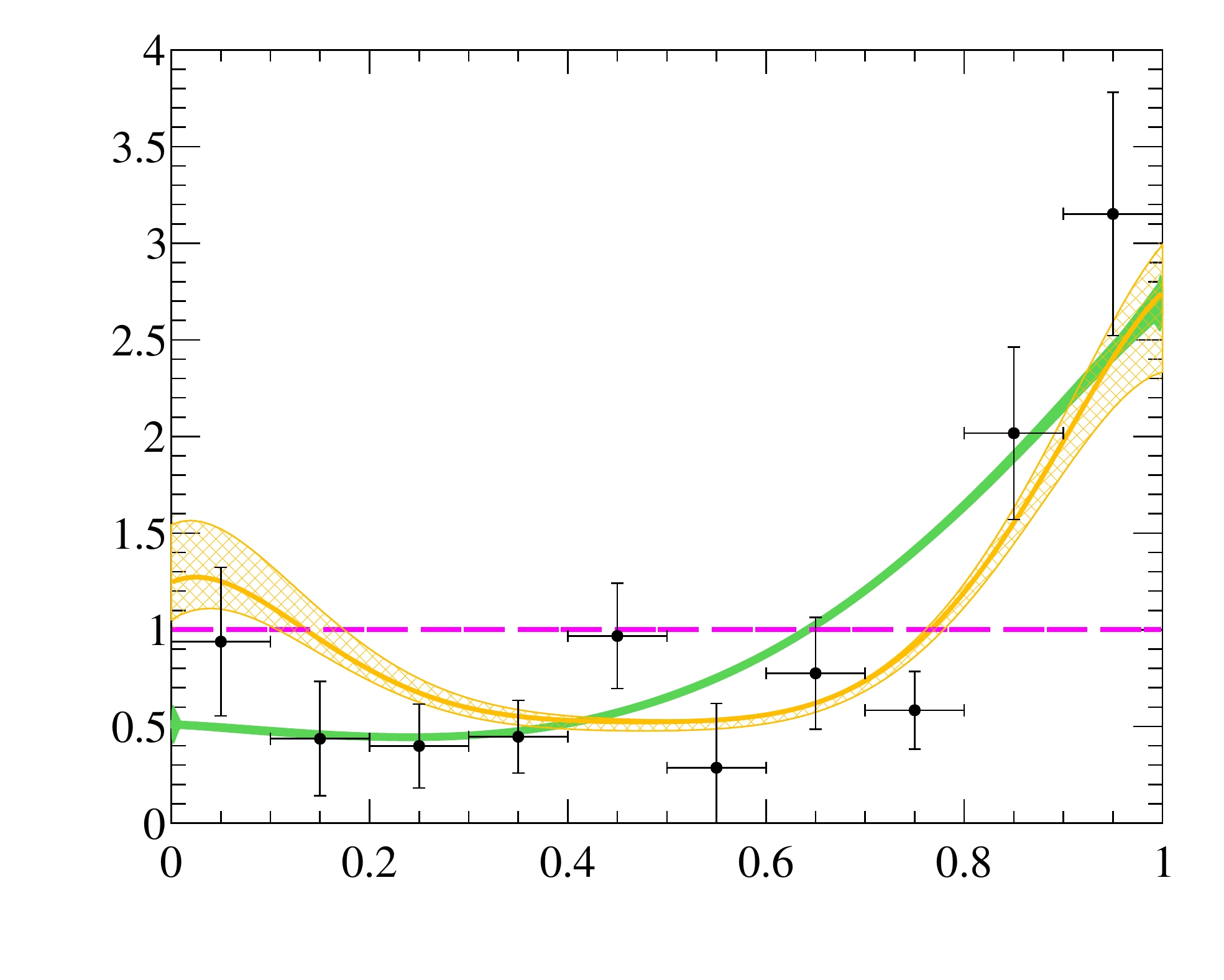}
    }
    \put( 75,110){ 
      \includegraphics*[width=75mm,height=55mm,%
      ]{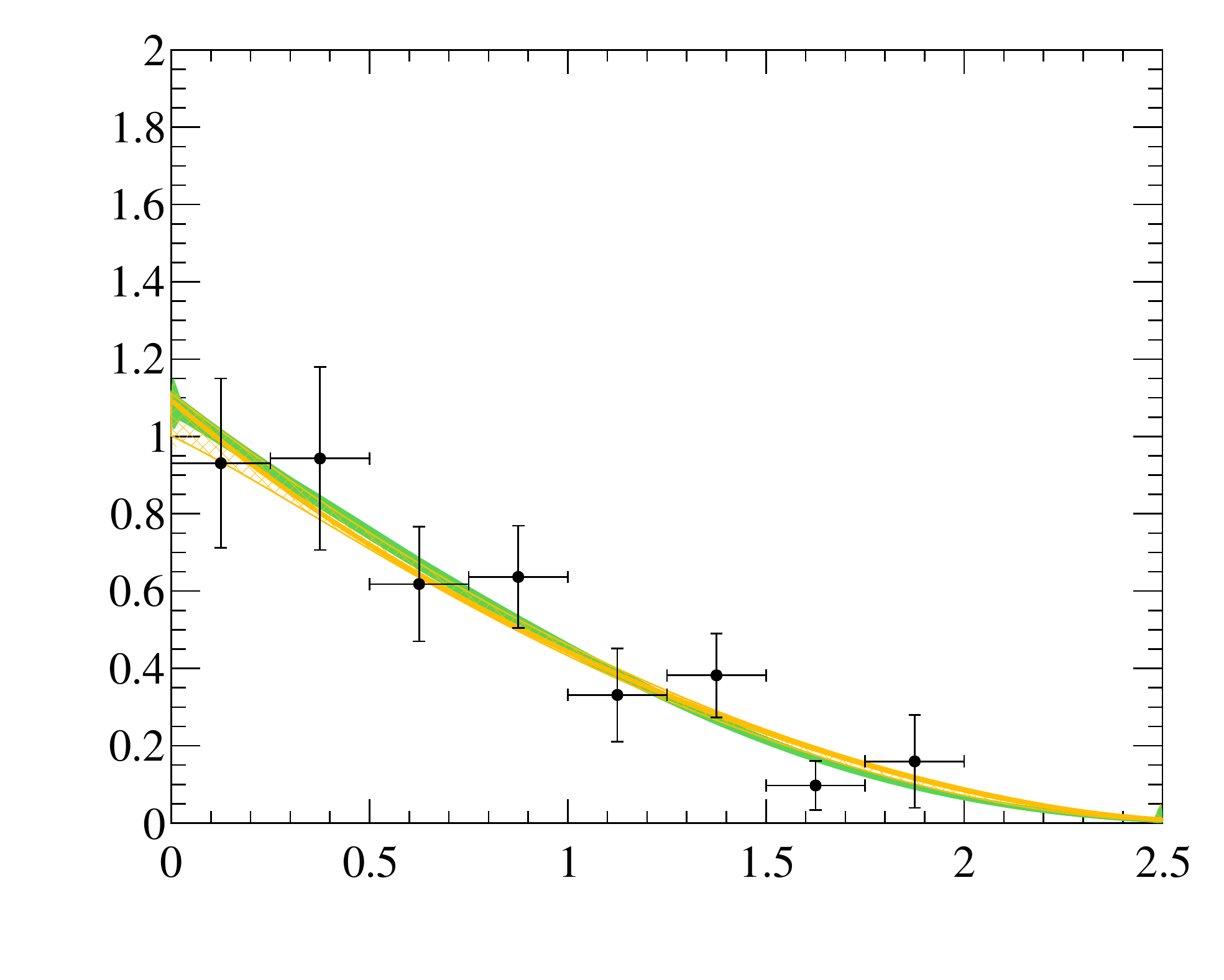}
    }
    \put(  0, 55){ 
      \includegraphics*[width=75mm,height=55mm,%
      ]{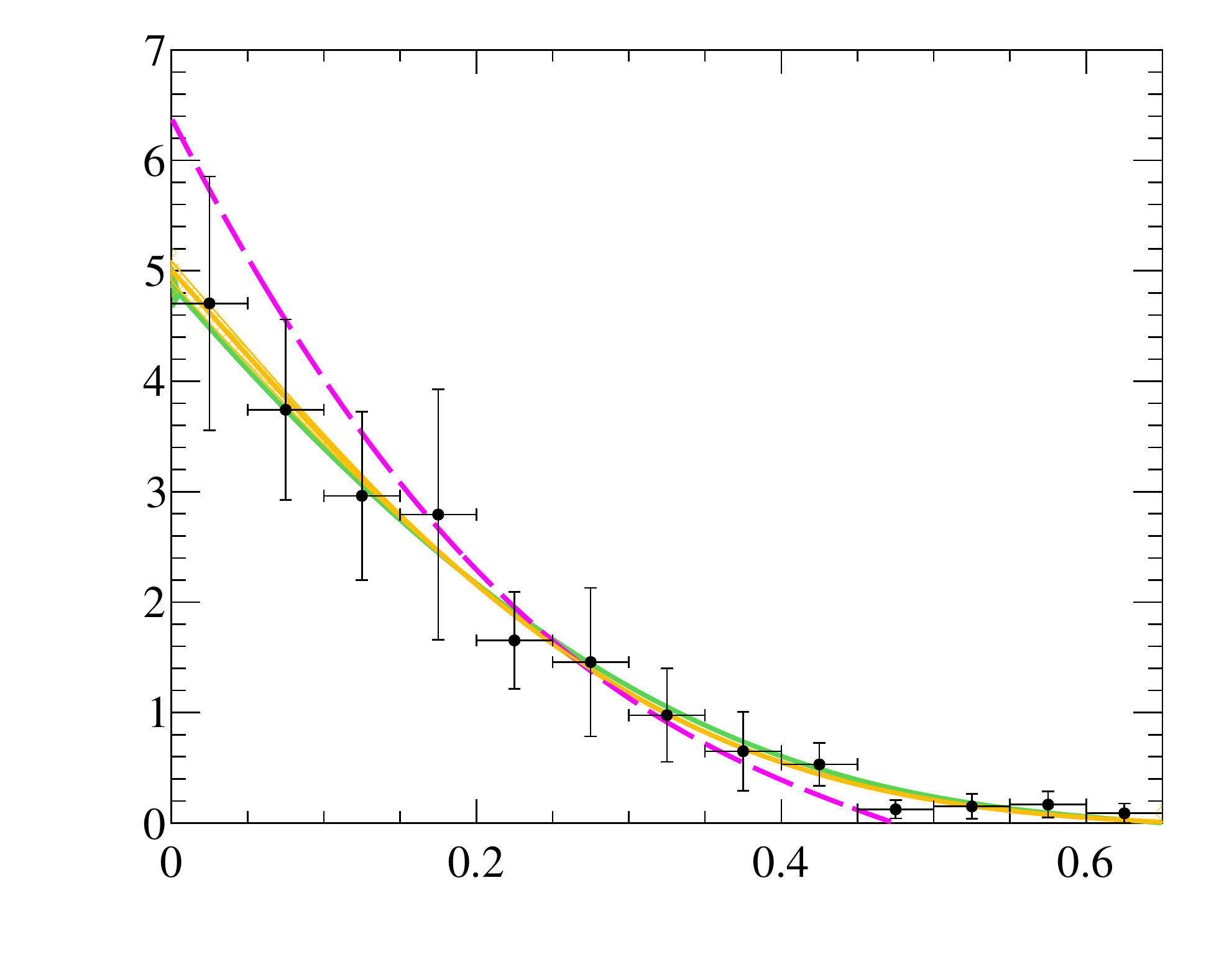}
    }
    \put( 75, 55){ 
      \includegraphics*[width=75mm,height=55mm,%
      ]{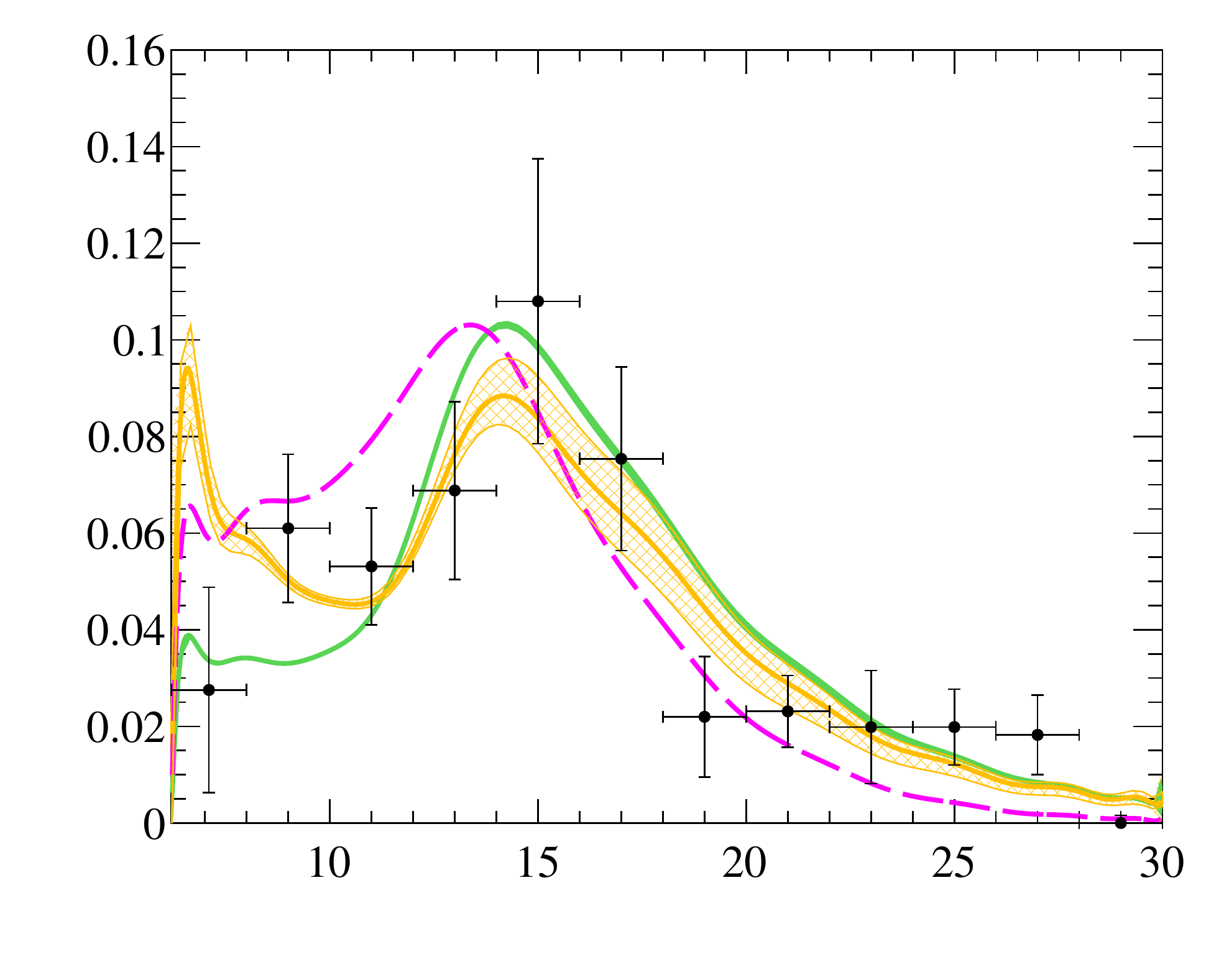}
    }
    \put(  0,  0){ 
      \includegraphics*[width=75mm,height=55mm,%
      ]{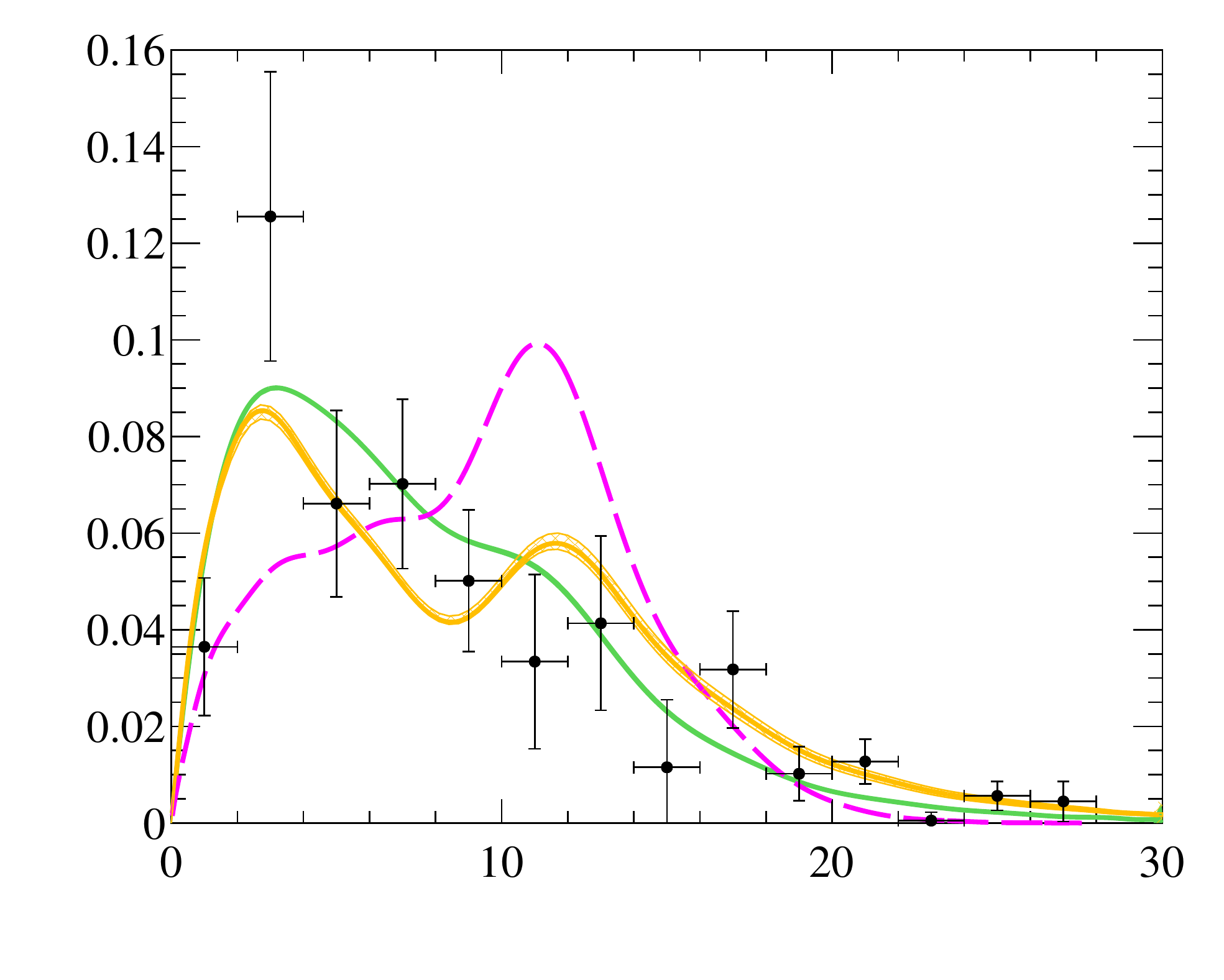}
    }
    \put( 75,  0){ 
      \includegraphics*[width=75mm,height=55mm,%
      ]{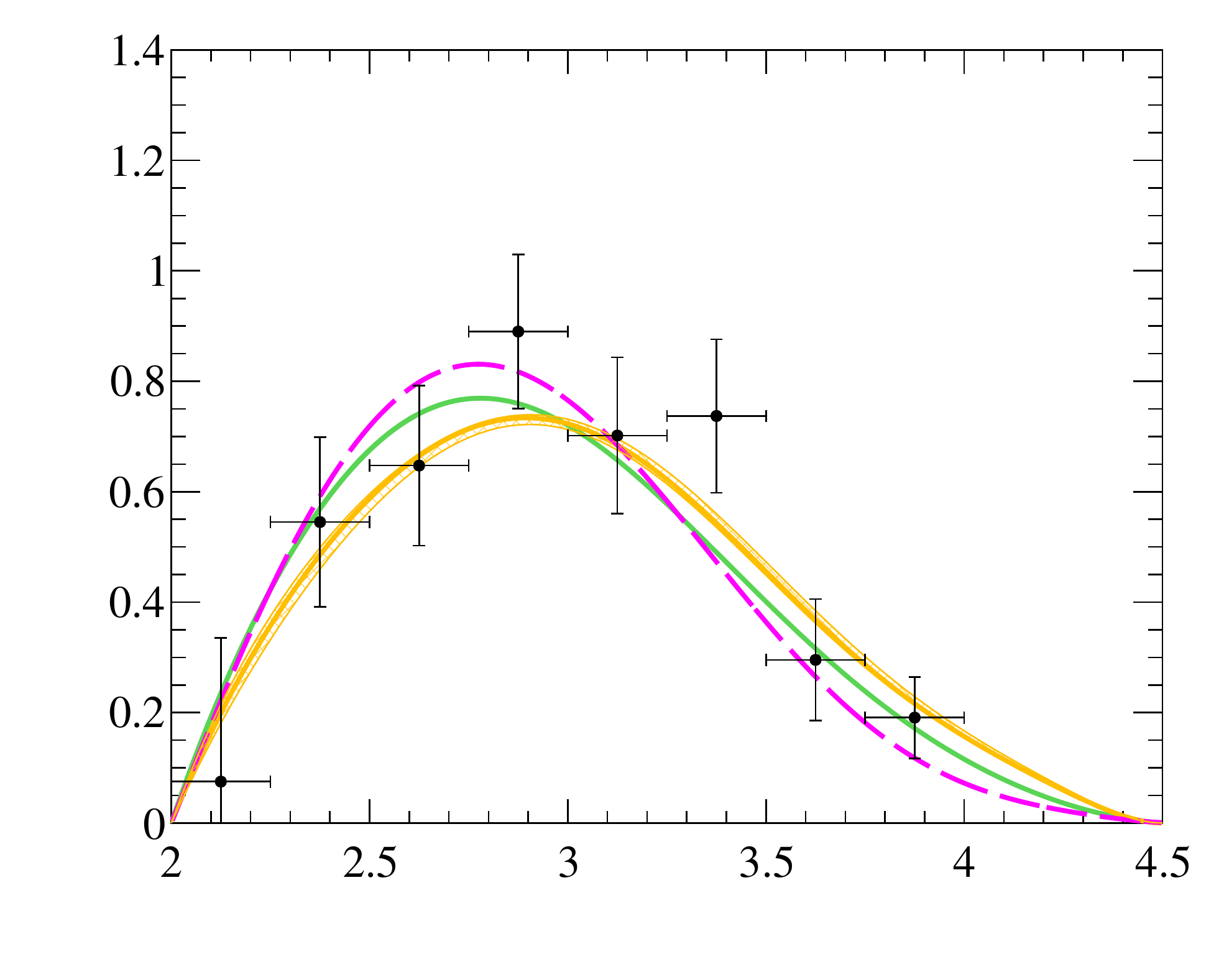}
    }
    \put( -1,150) { \small \begin{sideways} 
        $\frac{\pi}{\upsigma}\frac{\mathrm{d}\upsigma}{\mathrm{d}\left|\Delta\Pphi^{\ast}\right|}$
      \end{sideways}} 
    \put( 74,150) { \small \begin{sideways} 
        $\frac{1}{\upsigma}\frac{\mathrm{d}\upsigma}{\mathrm{d}\left|\Delta\Peta^{\ast}\right|}$
      \end{sideways}} 
     \put( -1, 97) { \small \begin{sideways} 
        $\frac{1}{\upsigma}\frac{\mathrm{d}\upsigma}{\mathrm{d}\mathcal{A}_{\mathrm{T}}}$
      \end{sideways}} 
    \put( 74,  91) { \small \begin{sideways} 
        $\frac{1}{\upsigma}\frac{\mathrm{d}\upsigma}{\mathrm{d} m^{\jpsi\jpsi} }$
      \end{sideways}} 
    \put( -3, 38) { \small \begin{sideways} 
        $\frac{1}{\upsigma}\frac{\mathrm{d}\upsigma}{\mathrm{d}p_{\mathrm{T}}^{\jpsi\jpsi}}$
      \end{sideways}} 
    \put( 74, 38) { \small \begin{sideways} 
        $\frac{1}{\upsigma}\frac{\mathrm{d}\upsigma}{\mathrm{d}y^{\jpsi\jpsi}}$
      \end{sideways}} 
    \put( 35,110) { $\left|\Delta\Pphi^{\ast}\right|/\pi$ } 
    \put(110,110) { $\left|\Delta\Peta^{\ast}\right|$     } 
    \put( 35, 55) { $\mathcal{A}_{\mathrm{T}}$ } 
    \put(110, 55) { $m^{\jpsi\jpsi}$ } 
    \put(132, 55) { $\left[\!\gevcc\right]$}
    \put( 35,  0) { $p_{\mathrm{T}}^{\jpsi\jpsi}$ } 
    \put( 58,  0) { $\left[\!\gevc\right]$}
    \put(110,  0) { $y^{\jpsi\jpsi}$ } 
    \put(115,152) { \small $\begin{array}{r}\mathrm{LHCb} \\
        \sqrt{s}=7,8\,\mathrm{TeV} \\
        p_{\mathrm{T}}^{\jpsi}>5\gevc \end{array}$}
    \put( 16,155) {a)}
    \put( 91,155) {b)}
    \put( 16,100) {c)}
    \put( 91,100) {d)}
    \put( 20, 45) {e)}
    \put( 91, 45) {f)}
    \put(20,155) { \color[rgb]{1,0.747,0}{\begin{tikzpicture}[x=1mm,y=1mm]\draw[thin,pattern=crosshatch, pattern color=Root92]  (0,0) rectangle (12,2.0);\end{tikzpicture}} }
    \put(33,155) { \sc{Powheg} } 
    \put(20,150) { \color[rgb]{0.35,0.83,0.33} {\rule{12mm}{2.2mm}}}
    \put(33,150) { \sc{Pythia} } 
    \put(20,145) { \color[rgb]{1,0,1} {\hdashrule[0.5ex][x]{1.2cm}{0.7pt}{3mm 0.5mm} } } 
    \put(33,145) { \small uncorrelated $\mathrm{b}\bar{\mathrm{b}}$}    
  \end{picture}
  \caption { \small
    Normalized differential production cross\nobreakdash-sections\,(points with error bars) 
    for 
    a)\,$\left|\Delta\Pphi^{\ast}\right|/\pi$,
    b)\,$\left|\Delta\Peta^{\ast}\right|$,
    c)\,$\mathcal{A}_{\mathrm{T}}$,
    d)\,$m^{\jpsi\jpsi}$,
    e)\,$p^{\jpsi\jpsi}_{\mathrm{T}}$ and 
    f)\,$y^{\jpsi\jpsi}$
    together with the~{\sc{Powheg}}\,(orange line)
    and \pythia\,(green band) predictions. 
    The~expectations for 
    uncorrelated $\bquark\bquarkbar$~production are shown by the~dashed magenta line. 
    The~uncertainties  
    in the~{\sc{Powheg}}~and~$\pythia$
    predictions due to the~choice
    of factorization and renormalization scales
    are shown as orange cross\nobreakdash-hatched 
    and green solid areas, respectively.
  }
  \label{fig:gev5}
\end{figure}

\begin{figure}[t]
  \setlength{\unitlength}{1mm}
  \centering
  \begin{picture}(150,165)
    \put(  0,110){ 
      \includegraphics*[width=75mm,height=55mm,%
      ]{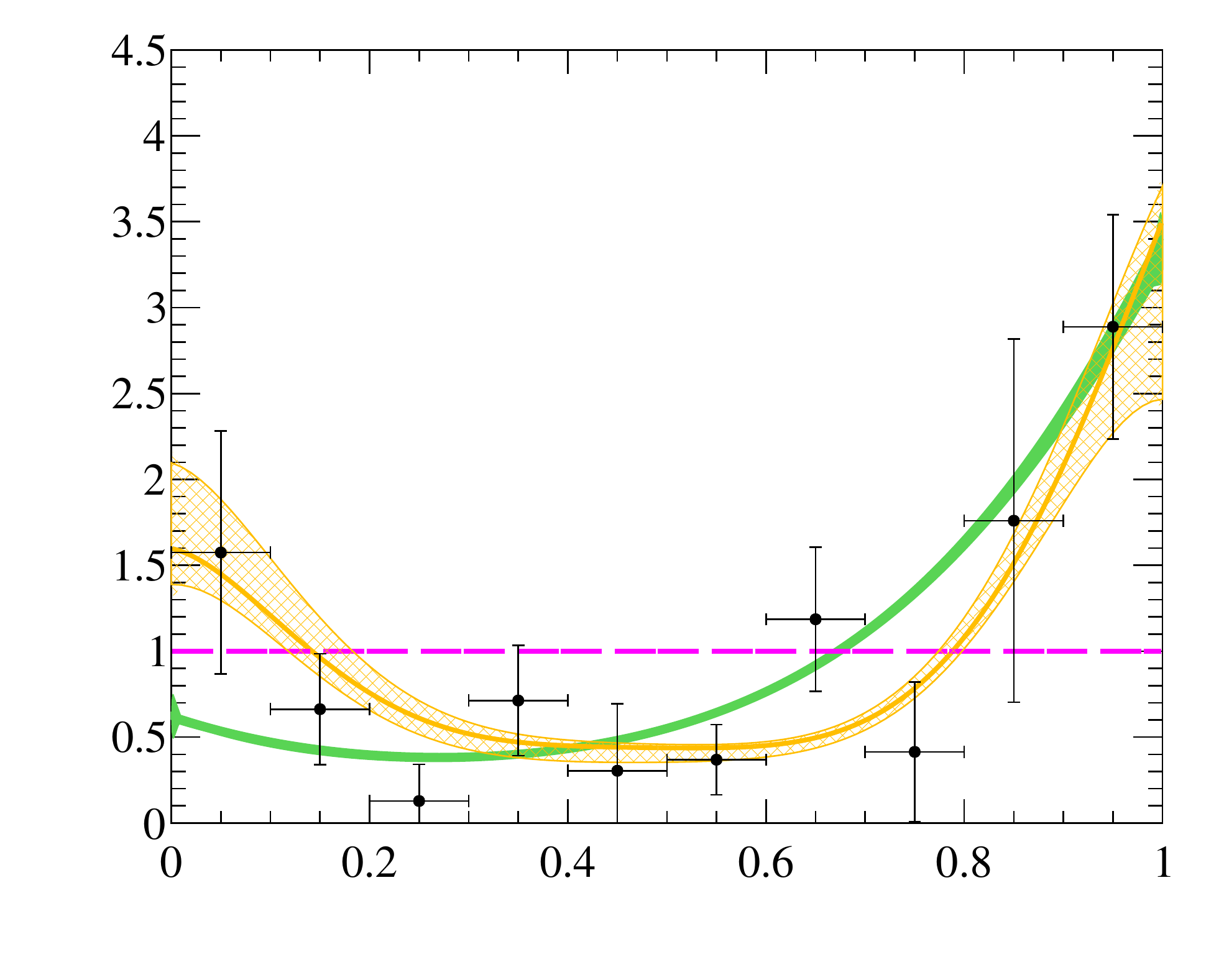}
    }
    \put( 75,110){ 
      \includegraphics*[width=75mm,height=55mm,%
      ]{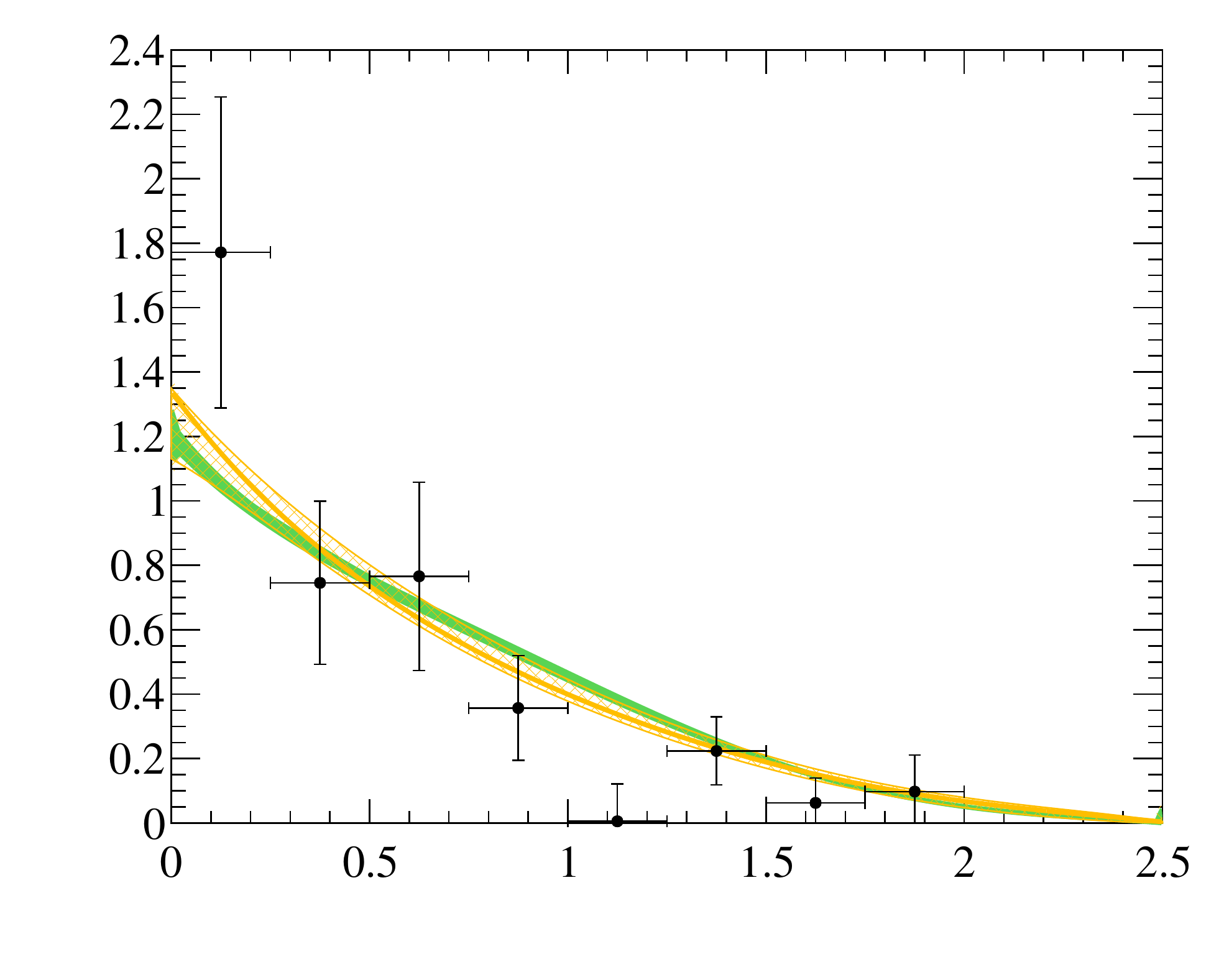}
    }
    \put(  0, 55){ 
      \includegraphics*[width=75mm,height=55mm,%
      ]{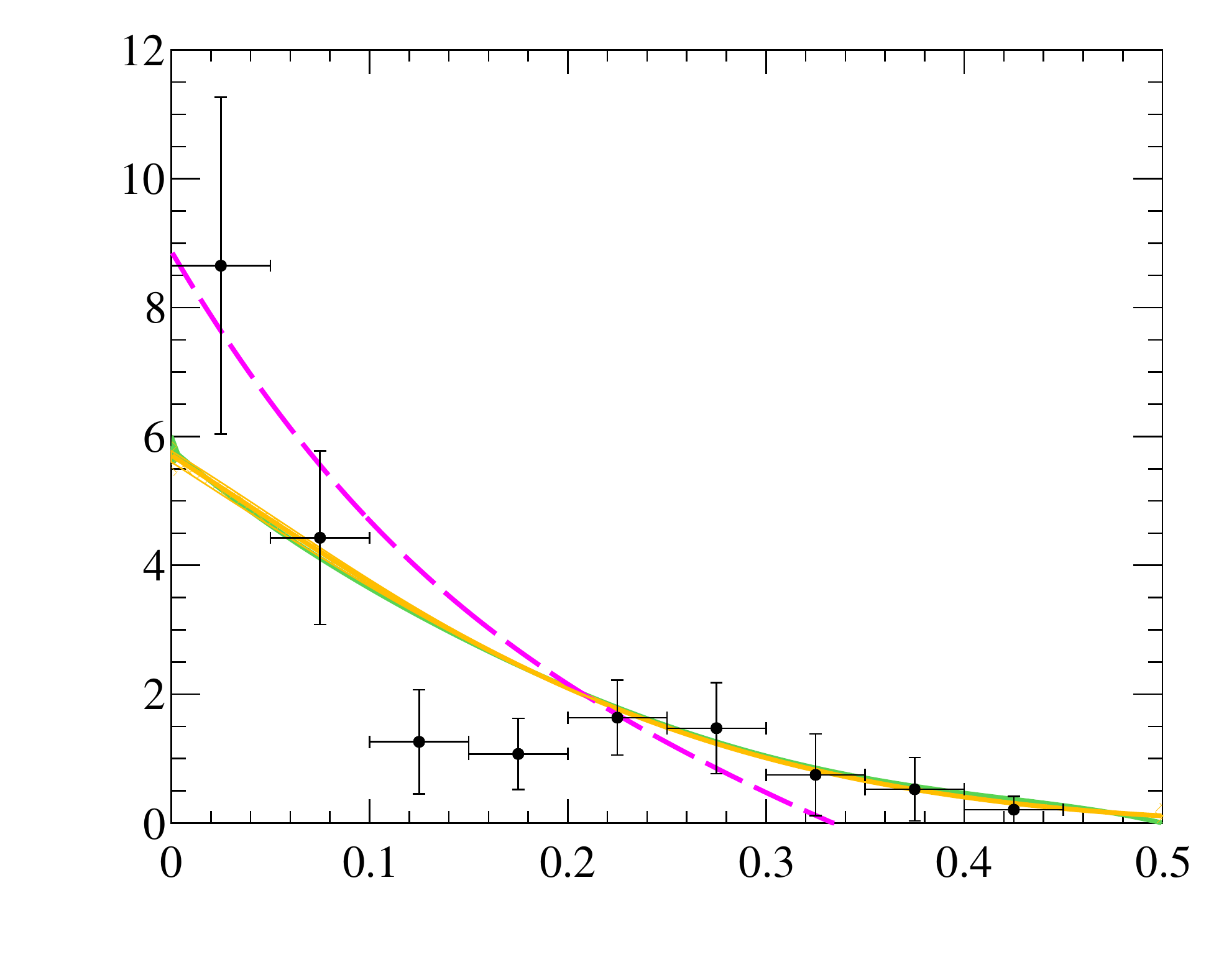}
    }
    \put( 75, 55){ 
      \includegraphics*[width=75mm,height=55mm,%
      ]{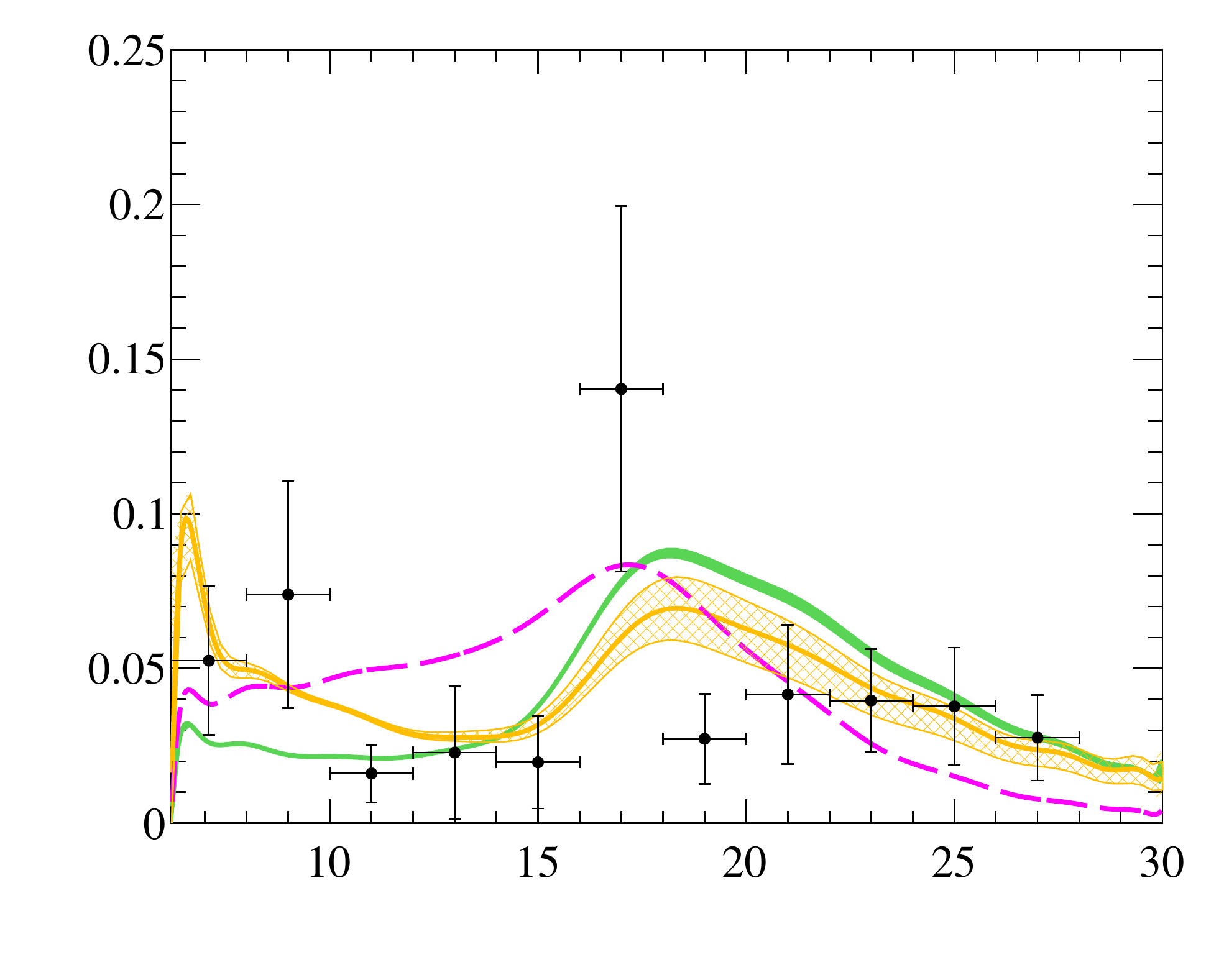}
    }
    \put(  0,  0){ 
      \includegraphics*[width=75mm,height=55mm,%
      ]{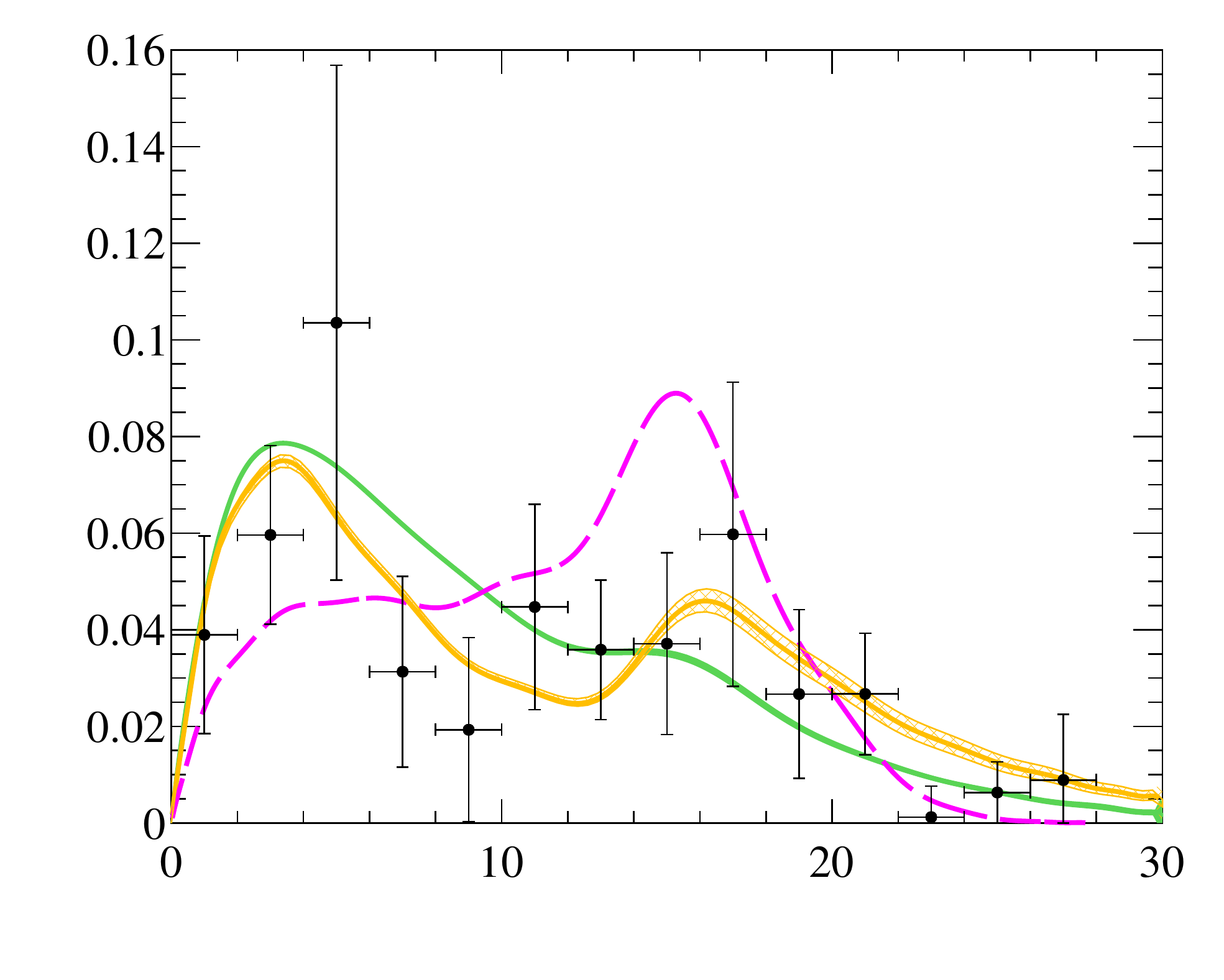}
    }
    \put( 75,  0){ 
      \includegraphics*[width=75mm,height=55mm,%
      ]{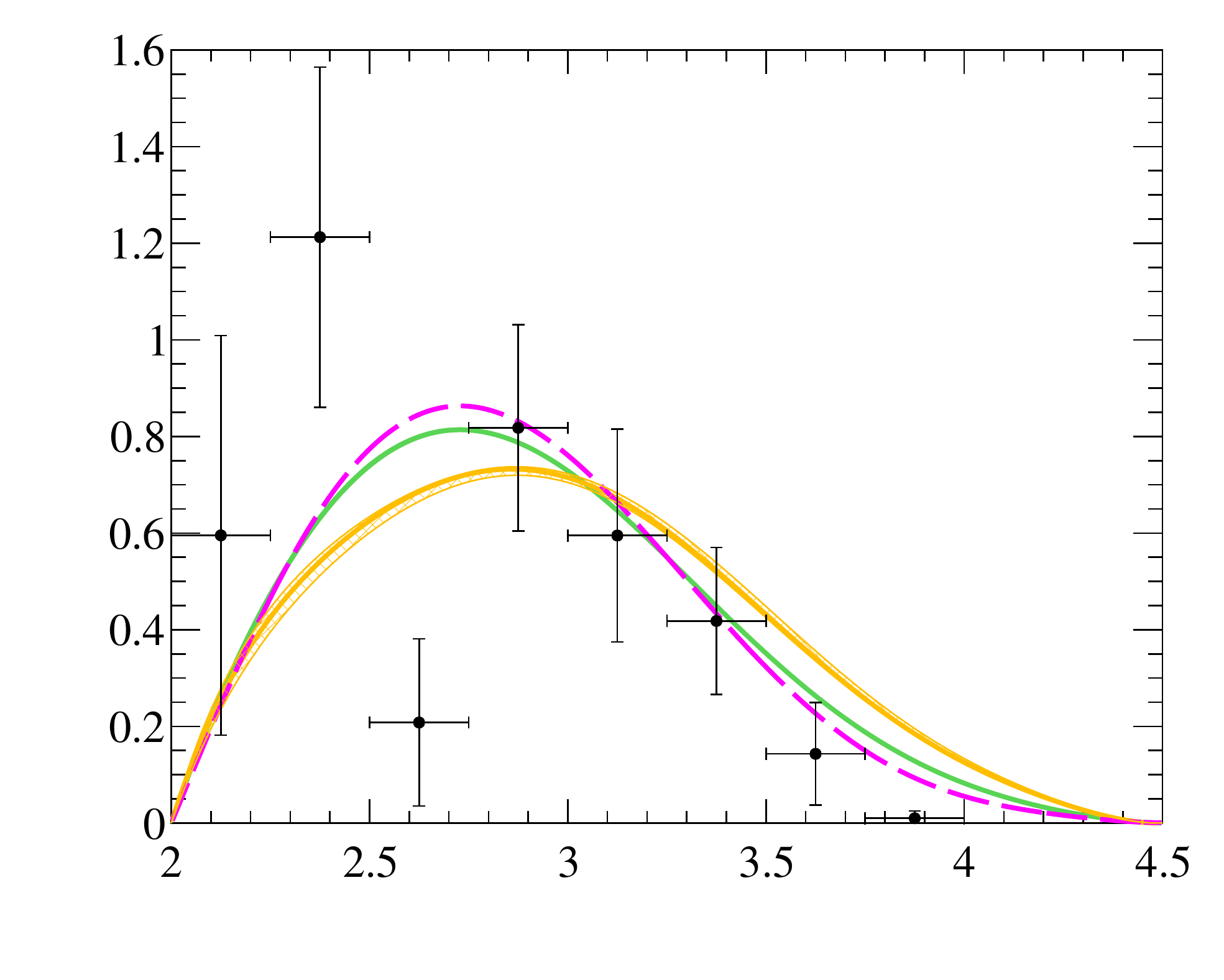}
    }
    \put( -1,150) { \small \begin{sideways} 
        $\frac{\pi}{\upsigma}\frac{\mathrm{d}\upsigma}{\mathrm{d}\left|\Delta\Pphi^{\ast}\right|}$
      \end{sideways}} 
    \put( 74,150) { \small \begin{sideways} 
        $\frac{1}{\upsigma}\frac{\mathrm{d}\upsigma}{\mathrm{d}\left|\Delta\Peta^{\ast}\right|}$
      \end{sideways}} 
     \put( -1, 97) { \small \begin{sideways} 
        $\frac{1}{\upsigma}\frac{\mathrm{d}\upsigma}{\mathrm{d}\mathcal{A}_{\mathrm{T}}}$
      \end{sideways}} 
    \put( 74,  91) { \small \begin{sideways} 
        $\frac{1}{\upsigma}\frac{\mathrm{d}\upsigma}{\mathrm{d} m^{\jpsi\jpsi} }$
      \end{sideways}} 
    \put( -3, 38) { \small \begin{sideways} 
        $\frac{1}{\upsigma}\frac{\mathrm{d}\upsigma}{\mathrm{d}p_{\mathrm{T}}^{\jpsi\jpsi}}$
      \end{sideways}} 
    \put( 74, 38) { \small \begin{sideways} 
        $\frac{1}{\upsigma}\frac{\mathrm{d}\upsigma}{\mathrm{d}y^{\jpsi\jpsi}}$
      \end{sideways}} 
    \put( 35,110) { $\left|\Delta\Pphi^{\ast}\right|/\pi$ } 
    \put(110,110) { $\left|\Delta\Peta^{\ast}\right|$     } 
    \put( 35, 55) { $\mathcal{A}_{\mathrm{T}}$ } 
    \put(110, 55) { $m^{\jpsi\jpsi}$ } 
    \put(132, 55) { $\left[\!\gevcc\right]$}
    \put( 35,  0) { $p_{\mathrm{T}}^{\jpsi\jpsi}$ } 
    \put( 58,  0) { $\left[\!\gevc\right]$}
    \put(110,  0) { $y^{\jpsi\jpsi}$ } 
    \put(115,152) { \small $\begin{array}{r}\mathrm{LHCb} \\
        \sqrt{s}=7,8\,\mathrm{TeV} \\
        p_{\mathrm{T}}^{\jpsi}>7\gevc \end{array}$}
    \put( 16,155) {a)}
    \put( 91,155) {b)}
    \put( 16,100) {c)}
    \put( 91,100) {d)}
    \put( 16, 45) {e)}
    \put( 91, 45) {f)}
    \put(20,155) { \color[rgb]{1,0.747,0}{\begin{tikzpicture}[x=1mm,y=1mm]\draw[thin,pattern=crosshatch, pattern color=Root92]  (0,0) rectangle (12,2.0);\end{tikzpicture}} }
    \put(33,155) { \sc{Powheg} } 
    \put(20,150) { \color[rgb]{0.35,0.83,0.33} {\rule{12mm}{2.2mm}}}
    \put(33,150) { \sc{Pythia} } 
    \put(20,145) { \color[rgb]{1,0,1} {\hdashrule[0.5ex][x]{1.2cm}{0.7pt}{3mm 0.5mm} } } 
    \put(33,145) { \small uncorrelated $\mathrm{b}\bar{\mathrm{b}}$}    
  \end{picture}
  \caption { \small
    Normalized differential production cross\nobreakdash-sections\,(points with error bars)
    for 
    a)\,$\left|\Delta\Pphi^{\ast}\right|/\pi$,
    b)\,$\left|\Delta\Peta^{\ast}\right|$,
    c)\,$\mathcal{A}_{\mathrm{T}}$,
    d)\,$m^{\jpsi\jpsi}$,
    e)\,$p^{\jpsi\jpsi}_{\mathrm{T}}$ and 
    f)\,$y^{\jpsi\jpsi}$
    together with the~{\sc{Powheg}}\,(orange line)
    and \pythia\,(green band) predictions. 
    The~expectations for 
    uncorrelated $\bquark\bquarkbar$~production are shown by the~dashed magenta line. 
    The~uncertainties  
    in the~{\sc{Powheg}}~and~$\pythia$
    predictions due to the~choice
    of factorization and renormalization scales
    are shown as orange cross\nobreakdash-hatched 
    and green solid areas, respectively.
  }
  \label{fig:gev7}
\end{figure}

The~normalized differential production cross\nobreakdash-sections
are shown in Figs.~\ref{fig:gev2},
\ref{fig:gev3}, \ref{fig:gev5} and~\ref{fig:gev7} for 
different requirements on the~minimum transverse momentum 
of the~\jpsi~mesons. 
Since the~distributions obtained for data accumulated 
at $\sqs=7$~and~$8\tev$ are very similar, they are treated together.
In general, the~width of the~resolution function is much smaller 
than the~bin width, \ie the~results are not 
affected by bin\nobreakdash-to\nobreakdash-bin migration.
The~exception to this is a~small fraction of events with
\mbox{$2.0<\ptpsi<2.5\gevc$}, 
where the~resolution for
$\left| \Delta \Pphi^{\ast} \right|$ and 
$\left| \Delta \Peta^{\ast} \right|$ 
is close to half of the~bin\nobreakdash-width.

The~normalized differential production cross\nobreakdash-sections 
are compared with  expectations from 
{\sc{Powheg}}~\cite{Nason:2004rx,Frixione:2007vw,Frixione:2007nw,Alioli:2010xd} 
and~\pythia~\cite{Sjostrand:2006za,Sjostrand:2007gs,LHCb-PROC-2010-056}
using
the~parton distribution functions from 
    {\sc{CT09MCS}}~\cite{Lai:2009ne},
    {\sc{CTEQ6L1}}~\cite{Pumplin:2002vw} and 
    {\sc{CTEQ6.6}}~\cite{Nadolsky:2008zw} 
    for the~samples produced with {\sc{Powheg}},
    {\sc{Pythia\,6}} and 
    {\sc{Pythia\,8}}, respectively.
    Since no visible difference between 
    {\sc{Pythia\,6}} and 
    {\sc{Pythia\,8}} samples are  found, they are combined.
    For the~{\sc{Powheg}} samples the~default configuration is used except
    for the~$\bquark$\nobreakdash-quark mass, which is set to~\mbox{$4.75\gevcc$}.
To~illustrate the~size of 
the~correlations
between the~two $\bquark$~quarks, 
predictions from an~artificial data\nobreakdash-driven model of uncorrelated 
$\bquark\bquarkbar$~production are also presented. 
This~model is based on the~measured transverse momenta and rapidity 
spectra for 
$\bquark\to\jpsi\mathrm{X}$~decays~\cite{LHCb-PAPER-2011-003,LHCb-PAPER-2013-016}, 
assuming uncorrelated production of \bquark~and \bquarkbar~quarks.
The~momenta of the~two $\jpsi$~mesons are sampled according to 
the~measured $(\ptpsi,\ypsi)$~spectra, assuming a~uniform 
distribution in the~azimuthal angle, $\Pphi^{\jpsi}$. 
This~allows
the~distributions for all variables 
except for $\left|\Delta\Peta^{\ast}\right|$ to be predicted.
This~model is considered as an~extreme case 
that corresponds to uncorrelated $\bquark\bquarkbar$~production;
in contrast, 
the~leading\nobreakdash-order collinear approximation, 
where the~transverse momentum of the~$\bquark\bquarkbar$~system 
from the~$\mathrm{gg}\to\bquark\bquarkbar$~process is zero, 
results in maximum correlation.
The~smearing of the~transverse momenta of 
the~initial gluons 
could result in significant decorrelations of the~initially 
highly correlated heavy\nobreakdash-flavour quarks. 
It~should be noted that the~model
using uncorrelated $\bquark\bquarkbar$~pairs
also mimics a~possible 
small contribution of double parton  scattering 
to $\bquark\bquarkbar$~pair production.

In general, both {\sc{Powheg}} and \pythia~describe 
the~data well for all distributions, suggesting 
that NLO effects 
in $\bquark\bquarkbar$~production
in the~studied kinematic region 
are small
compared with the~experimental precision.
Unlike the~measurements  with open\nobreakdash-charm 
mesons~\cite{Reisert:2007zz,Reisert:2007zza,LHCb-PAPER-2012-003},
no significant contribution from 
gluon splitting is observed at small~\mbox{$\left|\Delta \Pphi^{\ast}\right|$}.
This~observation is in agreement with expectations,  
since the~contribution from 
gluon splitting
is suppressed due to the~large mass of 
the~beauty quark. 
For~\mbox{$\ptpsi>5$} and~\mbox{$7\gev$},
there is a~hint of a~small enhancement
at small~\mbox{$\left|\Delta \Pphi^{\ast}\right|$}.
This~also agrees 
with the~expectation of a~larger contribution of 
gluon splitting at higher~\pt.
Another~large enhancement towards the~threshold in $m^{\jpsi\jpsi}$
is predicted by {\sc{Powheg}} for $\ptpsi>5$ and $7\gev$, 
due to large leading\nobreakdash-logarithm corrections~\cite{Nason:1987xz}.
No~evidence for  this enhancement is observed in the~LHCb data, as can be seen in Figs.~\ref{fig:gev5}d 
and~\ref{fig:gev7}d.
The~data~agree well with the~model 
of uncorrelated $\bquark\bquarkbar$~production for
$y^{\jpsi\jpsi}$ and $\mathcal{A}_{\mathrm{T}}$,
and also for $p^{\jpsi\jpsi}_{\mathrm{T}}$ and $m^{\jpsi\jpsi}$
in the~\mbox{$\ptpsi>2\gevc$}~region.
This~suggests gluon emission from the~initial and/or final state, 
or large effective smearing of the~transverse momenta of the~gluons, 
$\mathcal{O}(3\gevc)$, resulting in large decorrelation
of the~produced heavy~quarks.   

%% file: summary.tex
\section{Summary and conclusions}

 Kinematic correlations for pairs of beauty hadrons, 
 produced in high energy proton\nobreakdash-proton collisions, 
 are studied.
 The~data sample used was collected with the~\lhcb
 experiment at centre\nobreakdash-of\nobreakdash-mass energies of 
 7~and~8\tev and corresponds to an~integrated luminosity of 3\invfb.
 The~measurement is performed using $\bquark\to\jpsi\mathrm{X}$~decays
 in the~kinematic range \mbox{$2<y^{\jpsi}<4.5$}, \mbox{$2<\ptpsi<25\gevc$}.
 The~observed correlations agree with 
 \mbox{$\pythia\,$(LO)} and \mbox{{\sc{Powheg}}\,(NLO)}~predictions, 
 suggesting NLO~effects  in $\bquark\bquarkbar$~production are small.
 In~particular, 
 no~large contribution from gluon splitting 
 is observed.
 The~present data do not allow 
 discrimination of theory predictions 
 in the~region of large \pt of the~\jpsi~mesons, 
 where the~difference between {\sc{Powheg}} and
 \pythia~predictions is larger.
 Such~discrimination will be possible 
 with future~measurements with larger data samples 
 at~higher energy.

%% file: acknowledgements.tex
\section*{Acknowledgements}

 
\noindent 
We would like to thank P.~Nason and A.K.~Likhoded for interesting 
and stimulating discussions on production of heavy\nobreakdash-flavours.
We~express our gratitude to our colleagues in the~CERN accelerator
departments for the~excellent performance of the~LHC. 
We~thank the~technical and administrative staff at the~LHCb institutes. 
We~acknowledge support from CERN and from the~national agencies: 
CAPES, CNPq, FAPERJ and FINEP\,(Brazil); 
MOST and NSFC\,(China); 
CNRS/IN2P3\,(France); 
BMBF, DFG and MPG\,(Germany); 
INFN\,(Italy); 
NWO\,(The~Netherlands); 
MNiSW and NCN\,(Poland); 
MEN/IFA\,(Romania); 
MinES and FASO\,(Russia); 
MinECo\,(Spain); 
SNSF and SER\,(Switzerland); 
NASU\,(Ukraine); 
STFC\,(United Kingdom); 
NSF\,(USA).  
We~acknowledge the~computing resources that are provided by CERN, 
IN2P3\,(France), KIT and DESY\,(Germany), 
INFN\,(Italy), 
SURF\,(The~Netherlands), 
PIC\,(Spain),
GridPP\,(United~Kingdom), 
RRCKI and Yandex~LLC\,(Russia), 
CSCS\,(Switzerland), 
IFIN\nobreakdash-HH\,(Romania), 
CBPF\,(Brazil), 
PL\nobreakdash-GRID\,(Poland) and
OSC\,(USA). 
We~are indebted to the~communities behind the~multiple open
source software packages on which we depend. 
Individual groups or members have received support from 
AvH Foundation\,(Germany), 
EPLANET, Marie Sk\l{}odowska\nobreakdash-Curie Actions and ERC\,(European Union), 
ANR, Labex~P2IO, \mbox{ENIGMASS} and OCEVU, and R\'{e}gion Auvergne\nobreakdash-Rh\^{o}ne\nobreakdash-Alpes\,(France), 
RFBR and Yandex~LLC\,(Russia), 
GVA, XuntaGal and GENCAT\,(Spain), 
Herchel Smith Fund, 
the~Royal Society, 
the~English\nobreakdash-Speaking Union and 
the~Leverhulme Trust\,(United Kingdom).

%% file: appendix.tex

\clearpage

{\noindent\normalfont\bfseries\Large Appendices}
\appendix

\section{Additional variables}\label{app:more}

In this appendix the~normalized differential 
production cross\nobreakdash-sections are studied for additional variables, namely 
\begin{itemize}
\item $\left|\Delta\Pphi^{\jpsi}\right|$, the~difference in 
  the~azimuthal angle $\Pphi^{\jpsi}$ between the~momentum directions of two \jpsi~mesons;
\item $\left|\Delta\Peta^{\jpsi}\right|$, the~difference in 
  the~pseudorapidity $\Peta^{\jpsi}$ between the~momentum directions of two \jpsi~mesons;
\item $\left|\Delta y^{\jpsi}\right|$, the~difference in 
  the~rapidity $y^{\jpsi}$ between the~two \jpsi~mesons.
\end{itemize}
Unlike 
$\left|\Delta\Pphi^{\ast}\right|/\pi$ and 
$\left|\Delta\Peta^{\ast}\right|$, 
which are largely independent on the~decays of beauty hadrons,
all these variables 
have a~minor dependence both on the~branching 
fractions of different beauty hadrons,  
as well as on the~\mbox{$\bquark\to\jpsi\mathrm{X}$} decay kinematics. 

\begin{figure}[t]
  \setlength{\unitlength}{1mm}
  \centering
  \begin{picture}(150,165)
    \put(  0,110){ 
      \includegraphics*[width=75mm,height=55mm,%
      ]{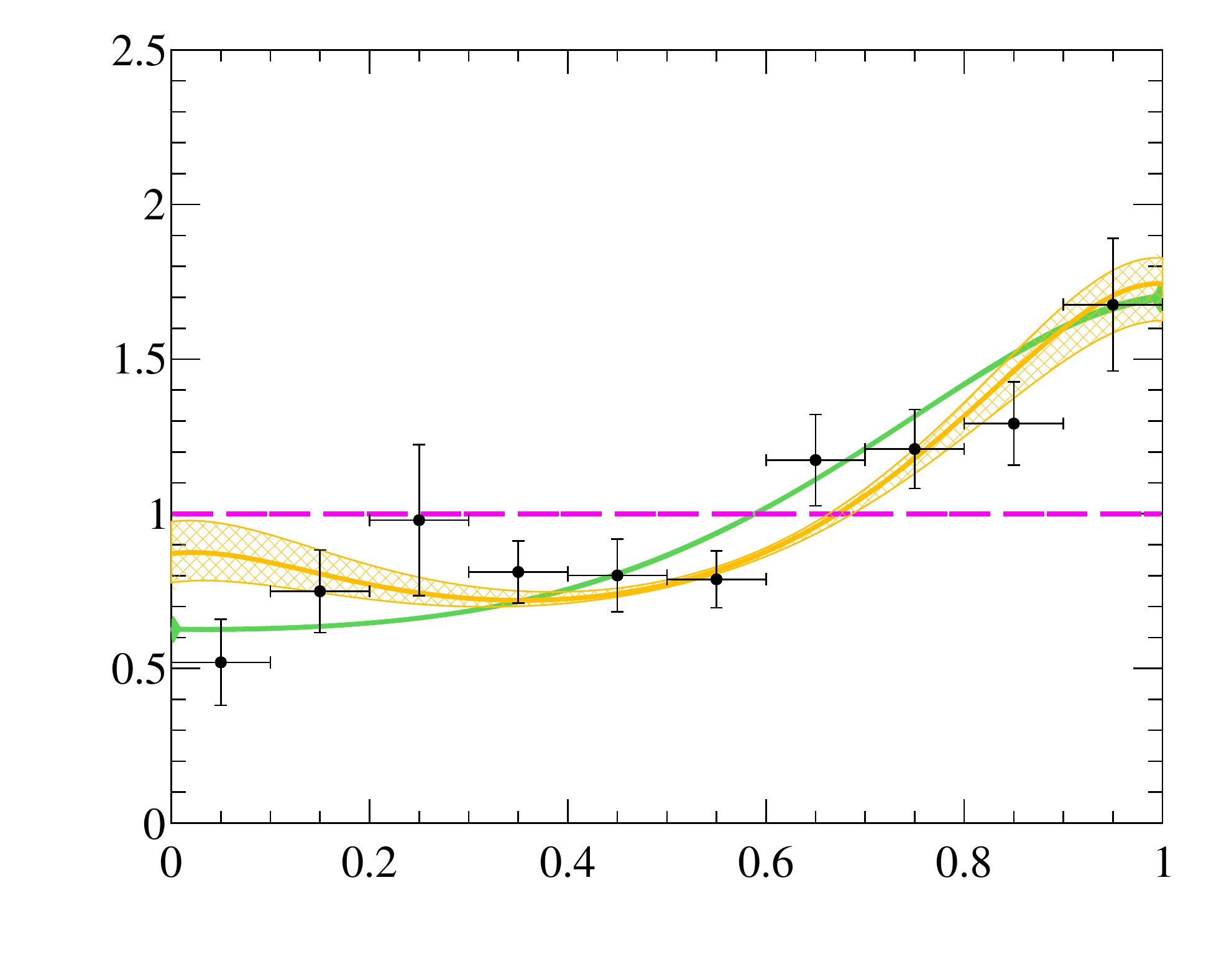}
    }
    \put( 75,110){ 
      \includegraphics*[width=75mm,height=55mm,%
      ]{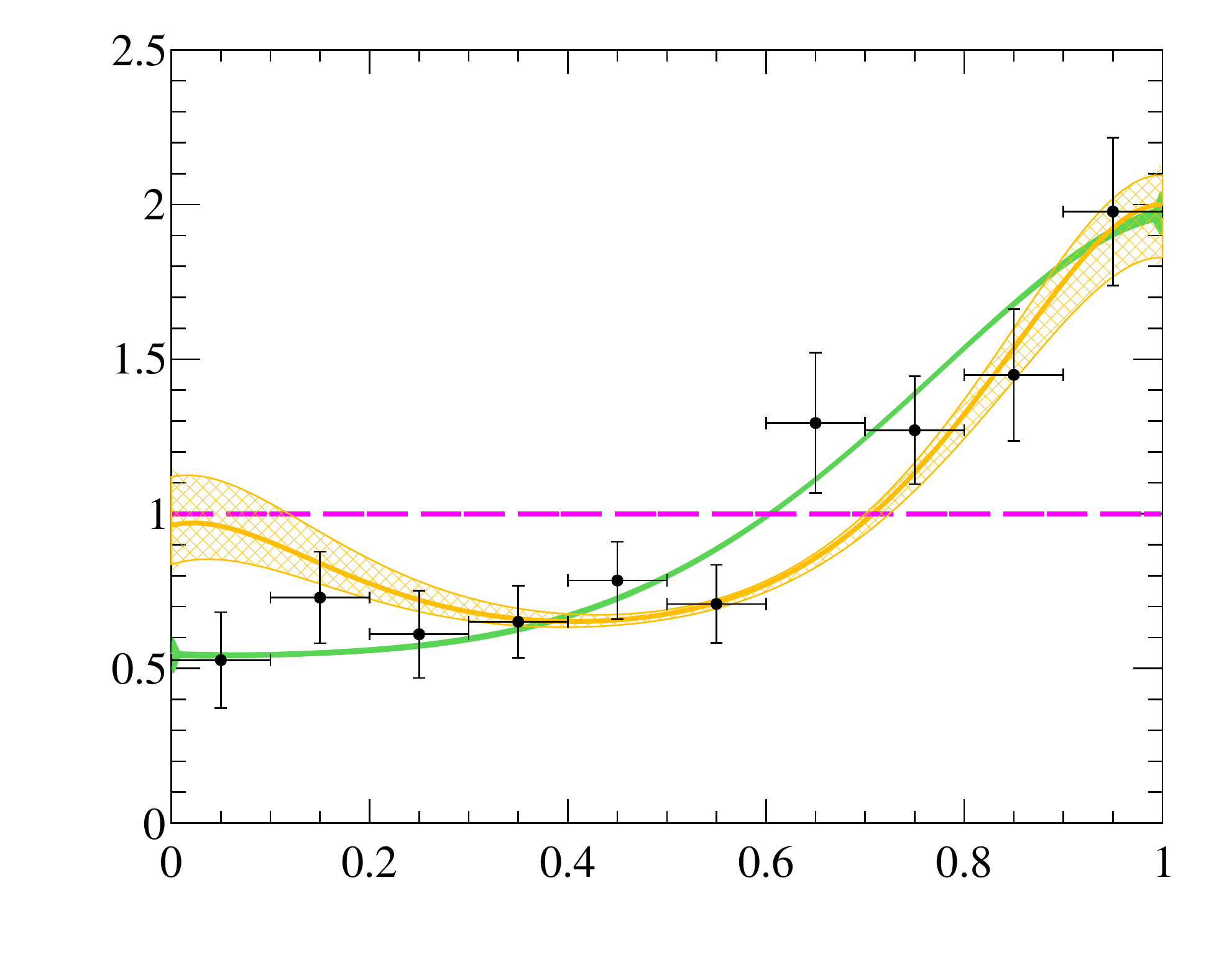}
    }
    \put(  0, 55){ 
      \includegraphics*[width=75mm,height=55mm,%
      ]{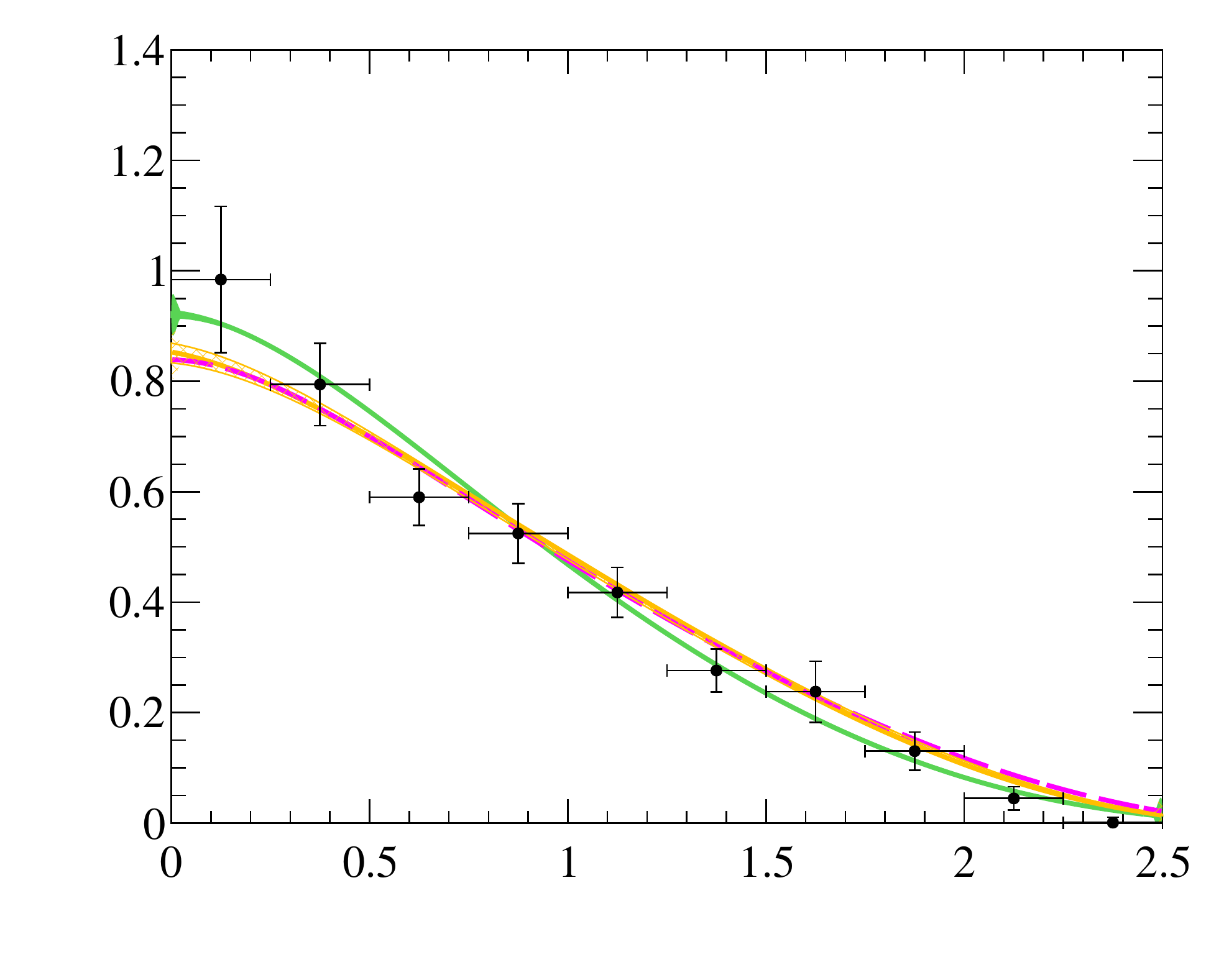}
    }
    \put( 75, 55){ 
      \includegraphics*[width=75mm,height=55mm,%
      ]{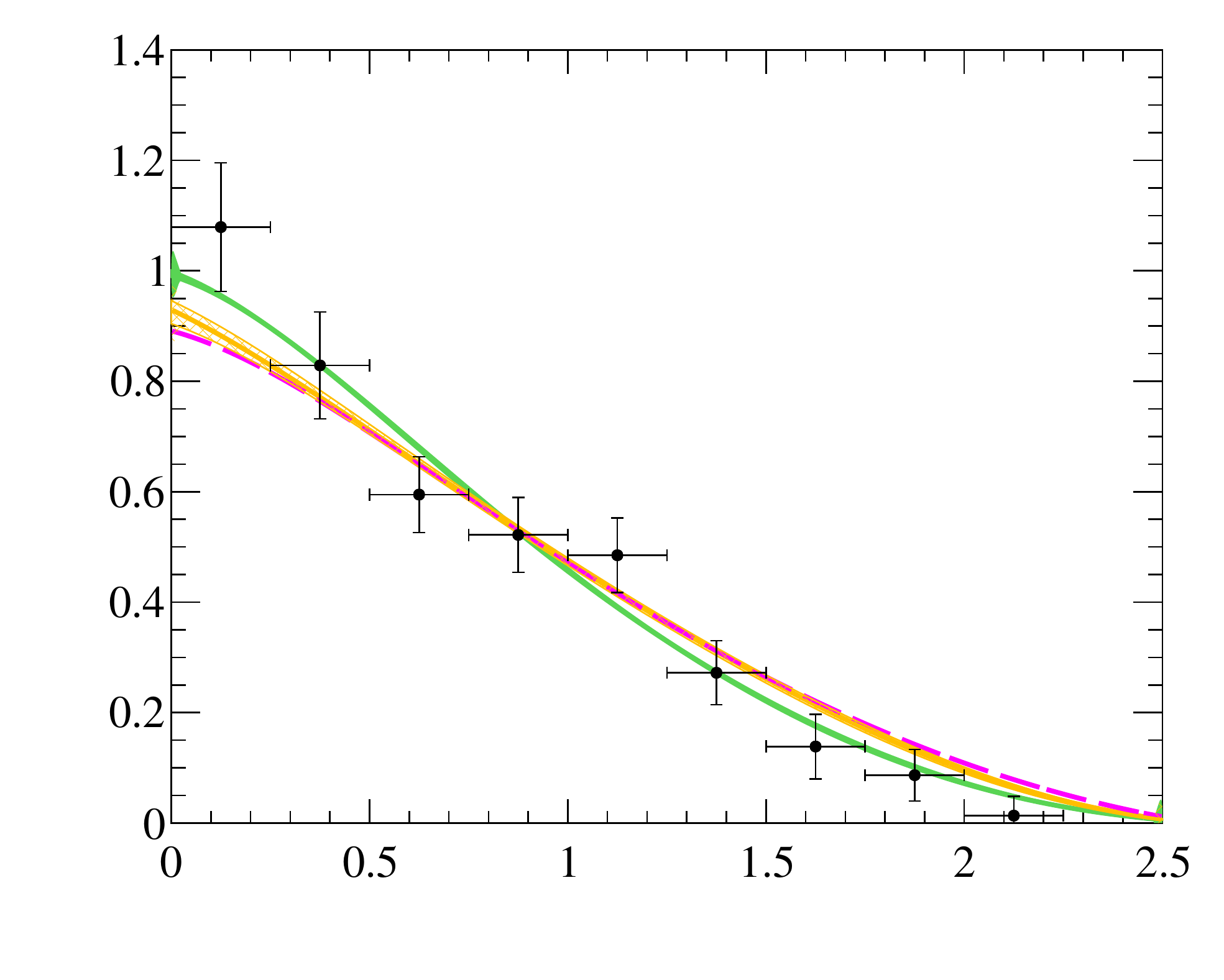}
    }
    \put(  0,  0){ 
      \includegraphics*[width=75mm,height=55mm,%
      ]{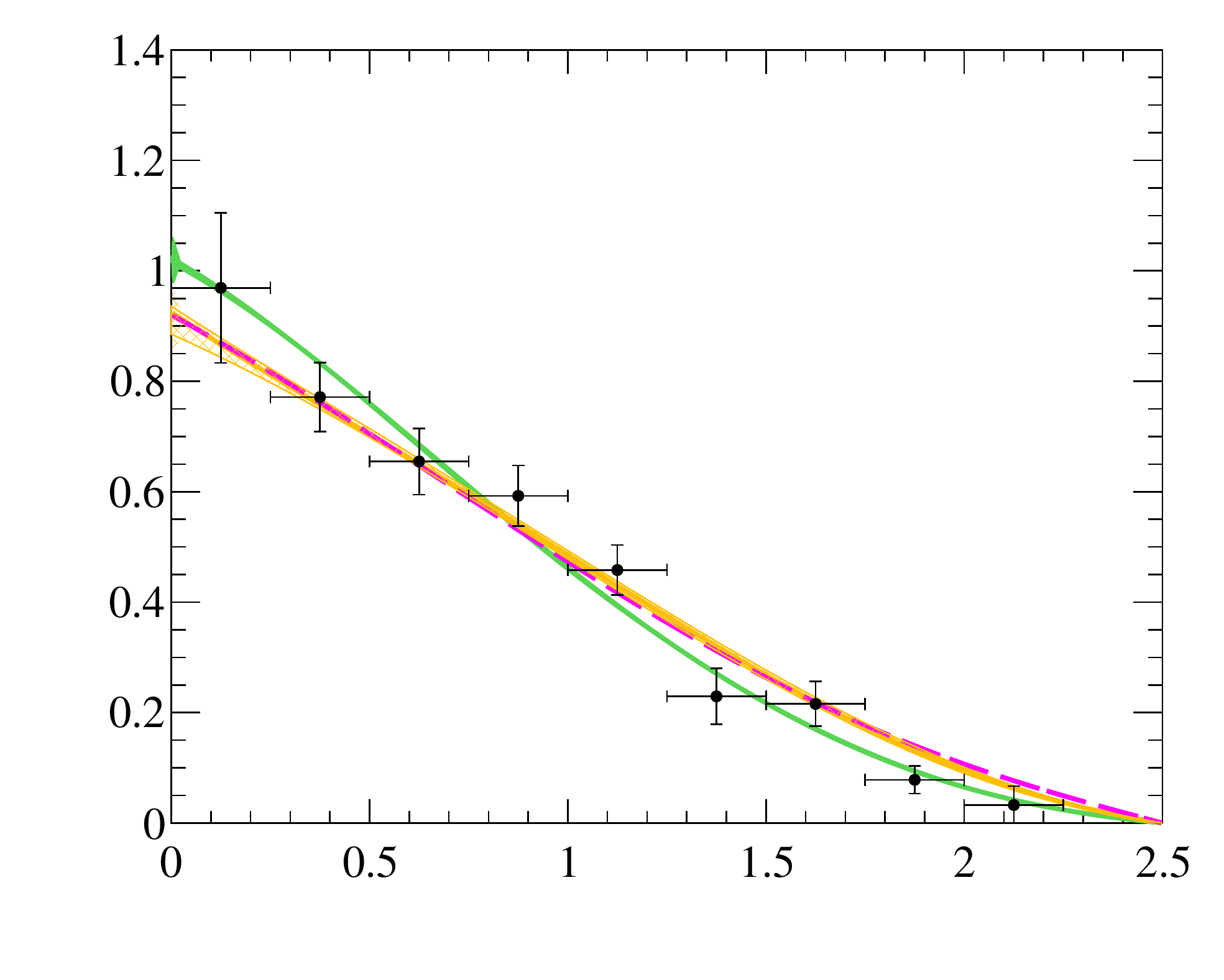}
    }
    \put( 75,  0){ 
      \includegraphics*[width=75mm,height=55mm,%
      ]{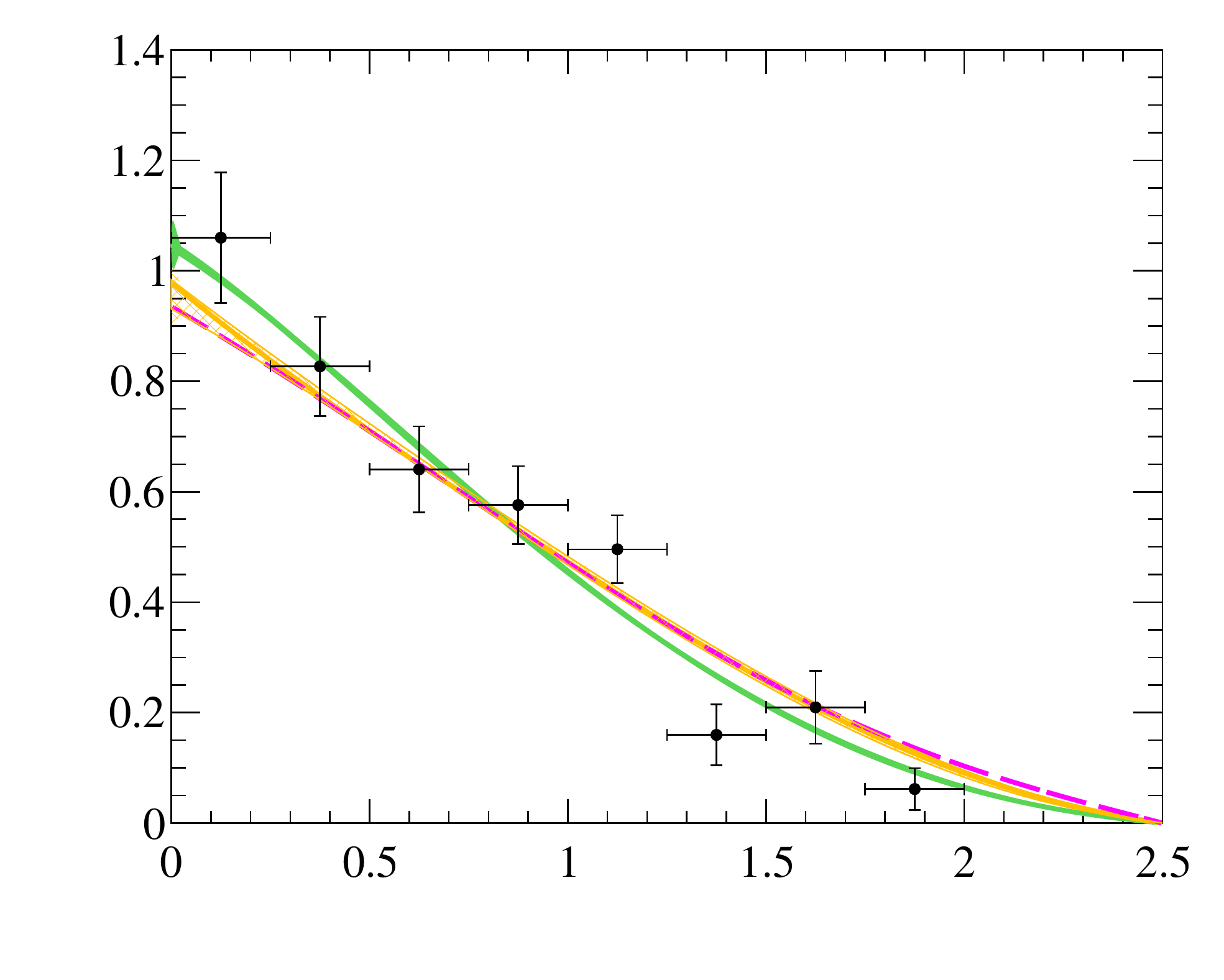}
    }
    \put( -1,146) { \small \begin{sideways} 
        $\frac{\pi}{\upsigma}\frac{\mathrm{d}\upsigma}{\mathrm{d}\left|\Delta\Pphi^{\jpsi}\right|}$
      \end{sideways}} 
    \put( 74,146) { \small \begin{sideways} 
        $\frac{\pi}{\upsigma}\frac{\mathrm{d}\upsigma}{\mathrm{d}\left|\Delta\Pphi^{\jpsi}\right|}$
      \end{sideways}} 
    \put( -1, 92) { \small \begin{sideways} 
        $\frac{1}{\upsigma}\frac{\mathrm{d}\upsigma}{\mathrm{d}\left|\Delta\Peta^{\jpsi}\right|}$
      \end{sideways}} 
    \put( 74, 92) { \small \begin{sideways} 
        $\frac{1}{\upsigma}\frac{\mathrm{d}\upsigma}{\mathrm{d}\left|\Delta\Peta^{\jpsi}\right|}$
      \end{sideways}} 
    \put( -1, 37) { \small \begin{sideways} 
        $\frac{1}{\upsigma}\frac{\mathrm{d}\upsigma}{\mathrm{d}\left|\Delta y^{\jpsi}\right|}$
      \end{sideways}} 
    \put( 74, 37) { \small \begin{sideways} 
        $\frac{1}{\upsigma}\frac{\mathrm{d}\upsigma}{\mathrm{d}\left|\Delta y^{\jpsi}\right|}$
      \end{sideways}} 
    \put( 35,110) { $\left|\Delta\Pphi^{\jpsi}\right|/\pi$ } 
    \put(110,110) { $\left|\Delta\Pphi^{\jpsi}\right|/\pi$ } 
    \put( 35, 55) { $\left|\Delta\Peta^{\jpsi}\right|$ } 
    \put(110, 55) { $\left|\Delta\Peta^{\jpsi}\right|$ } 
    \put( 35,  0) { $\left|\Delta y^{\jpsi}\right|$ } 
    \put(110,  0) { $\left|\Delta y^{\jpsi}\right|$ } 
    \put(113,155) { \small $\begin{array}{r}\mathrm{LHCb}\\
        \sqrt{s}=7,8\,\mathrm{TeV} \end{array}$}
    \put( 16,155) {a)}
    \put( 91,155) {b)}
    \put( 16,100) {c)}
    \put( 91,100) {d)}
    \put( 16, 45) {e)}
    \put( 91, 45) {f)}
    \put(20,155) { \color[rgb]{1,0.747,0}{\begin{tikzpicture}[x=1mm,y=1mm]\draw[thin,pattern=crosshatch, pattern color=Root92]  (0,0) rectangle (12,2.0);\end{tikzpicture}} }
    \put(33,155) { \sc{Powheg} } 
    \put(20,150) { \color[rgb]{0.35,0.83,0.33} {\rule{12mm}{2.2mm}}}
    \put(33,150) { \sc{Pythia} } 
    \put(20,145) { \color[rgb]{1,0,1} {\hdashrule[0.5ex][x]{1.2cm}{0.7pt}{3mm 0.5mm} } } 
    \put(33,145) { \small uncorrelated $\mathrm{b}\bar{\mathrm{b}}$}
  \end{picture}
  \caption { \small     
    Normalized differential production cross\nobreakdash-sections\,(points with error bars) 
    for~\mbox{$\ptpsi>2\gevc$}\,(left)     
    and~\mbox{$\ptpsi>3\gevc$}\,(right) data for 
    a,b)\,$\left|\Delta\Pphi^{\jpsi}\right|/\pi$,
    c,d)\,$\left|\Delta\Peta^{\jpsi}\right|$, and 
    e,f)\,$\left|\Delta y^{\jpsi}\right|$,
    together with the~{\sc{Powheg}}\,(orange line)
    and \pythia\,(green band) predictions. 
    The~expectations for 
    uncorrelated $\bquark\bquarkbar$~production are shown by the~dashed magenta line. 
    The~uncertainties  
    in the~{\sc{Powheg}}~and~$\pythia$
    predictions due to the~choice
    of factorization and renormalization scales
    are shown 
    as orange cross\nobreakdash-hatched and 
    and green solid areas, respectively.
  }
  \label{fig:a1}
\end{figure}

\begin{figure}[t]
  \setlength{\unitlength}{1mm}
  \centering
  \begin{picture}(150,165)
    \put(  0,110){ 
      \includegraphics*[width=75mm,height=55mm,%
      ]{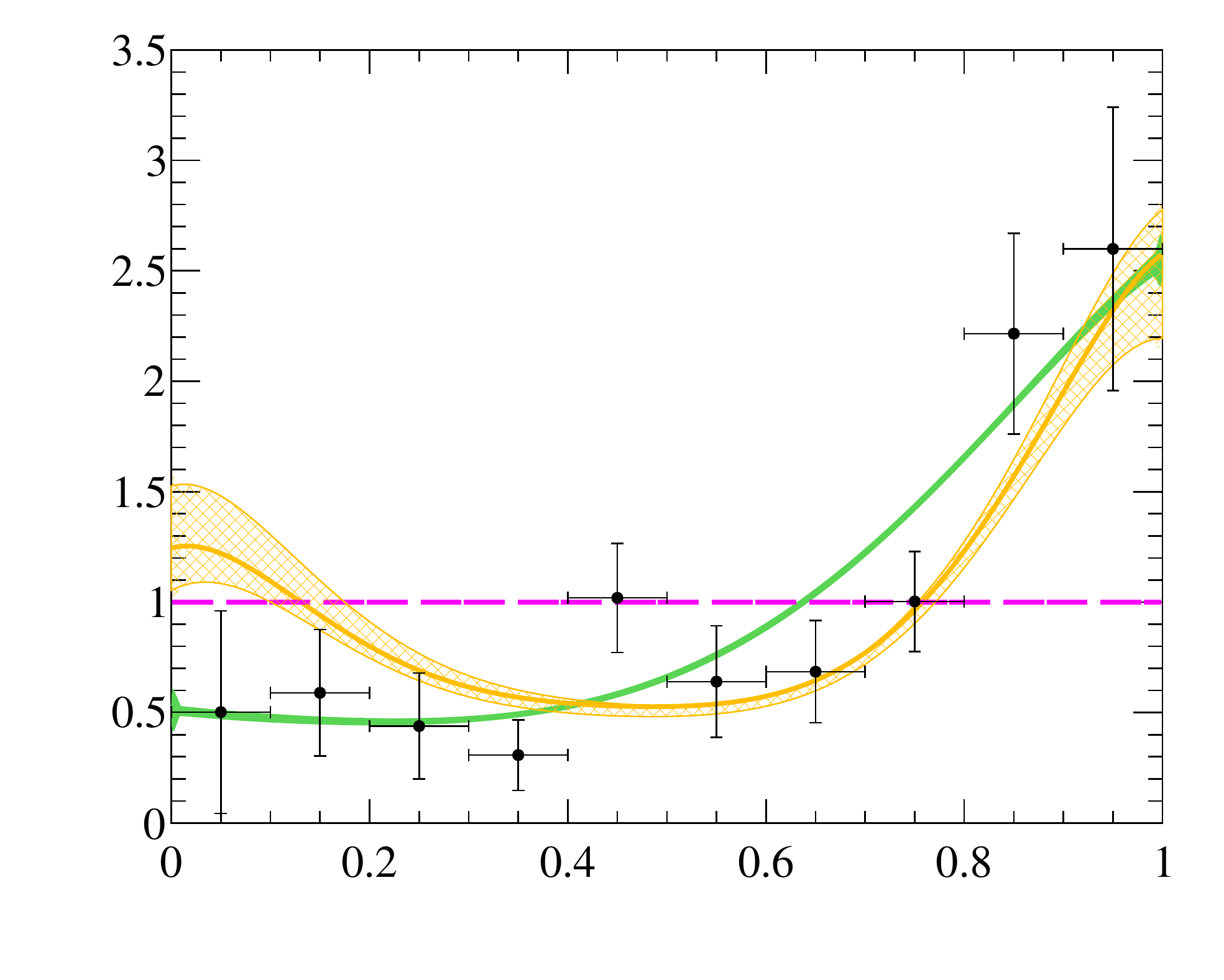}
    }
    \put( 75,110){ 
      \includegraphics*[width=75mm,height=55mm,%
      ]{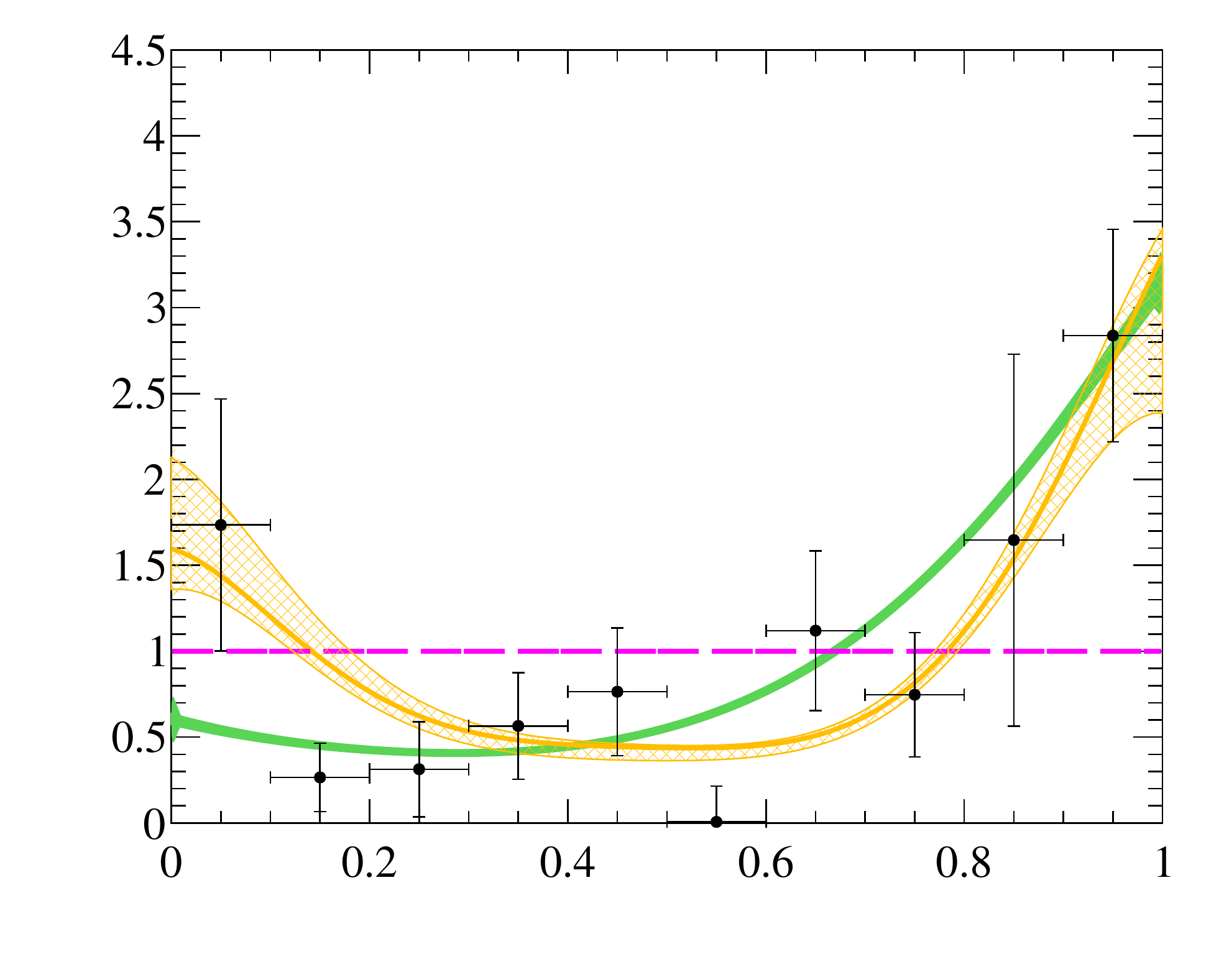}
    }
    \put(  0, 55){ 
      \includegraphics*[width=75mm,height=55mm,%
      ]{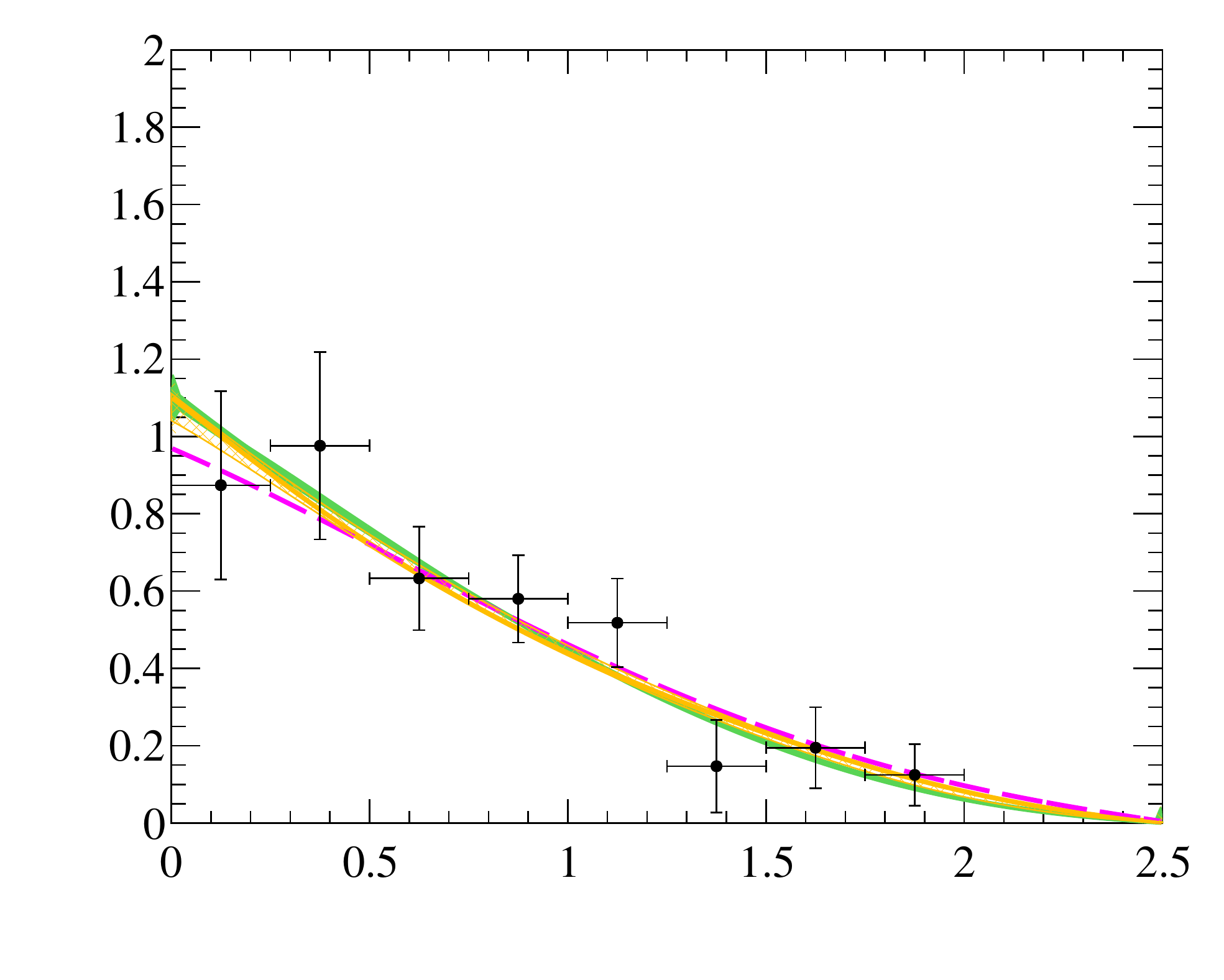}
    }
    \put( 75, 55){ 
      \includegraphics*[width=75mm,height=55mm,%
      ]{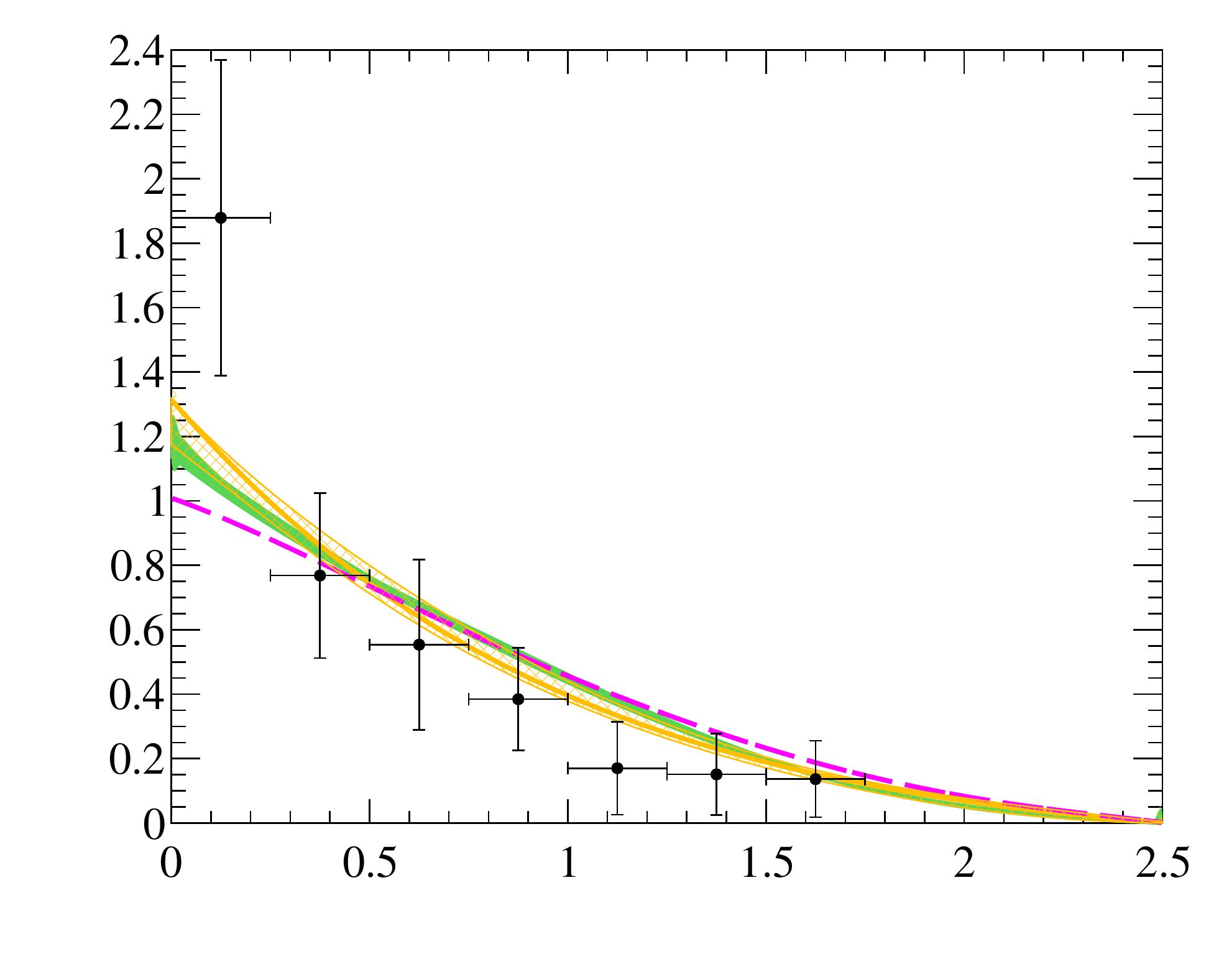}
    }
    \put(  0,  0){ 
      \includegraphics*[width=75mm,height=55mm,%
      ]{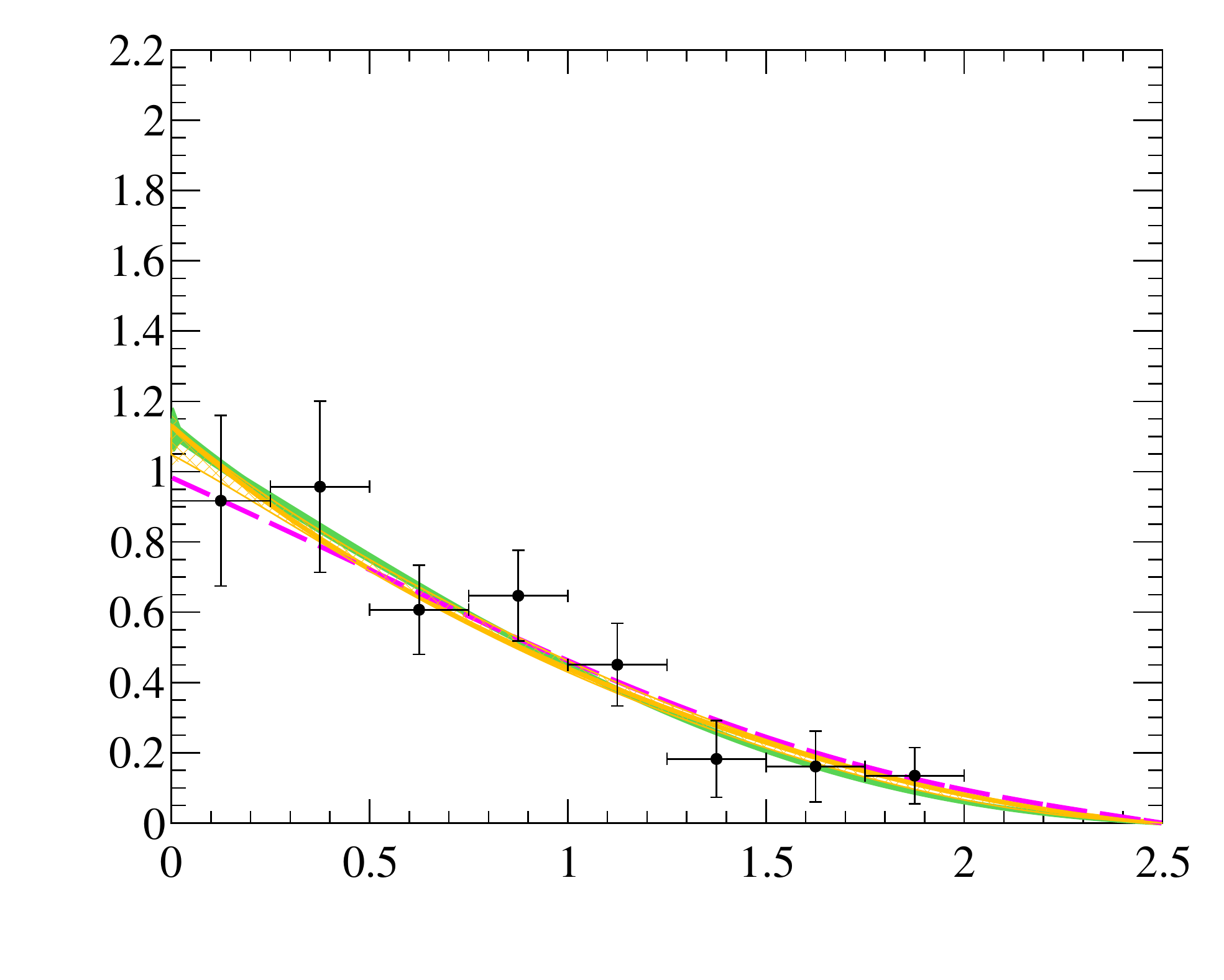}
    }
    \put( 75,  0){ 
      \includegraphics*[width=75mm,height=55mm,%
      ]{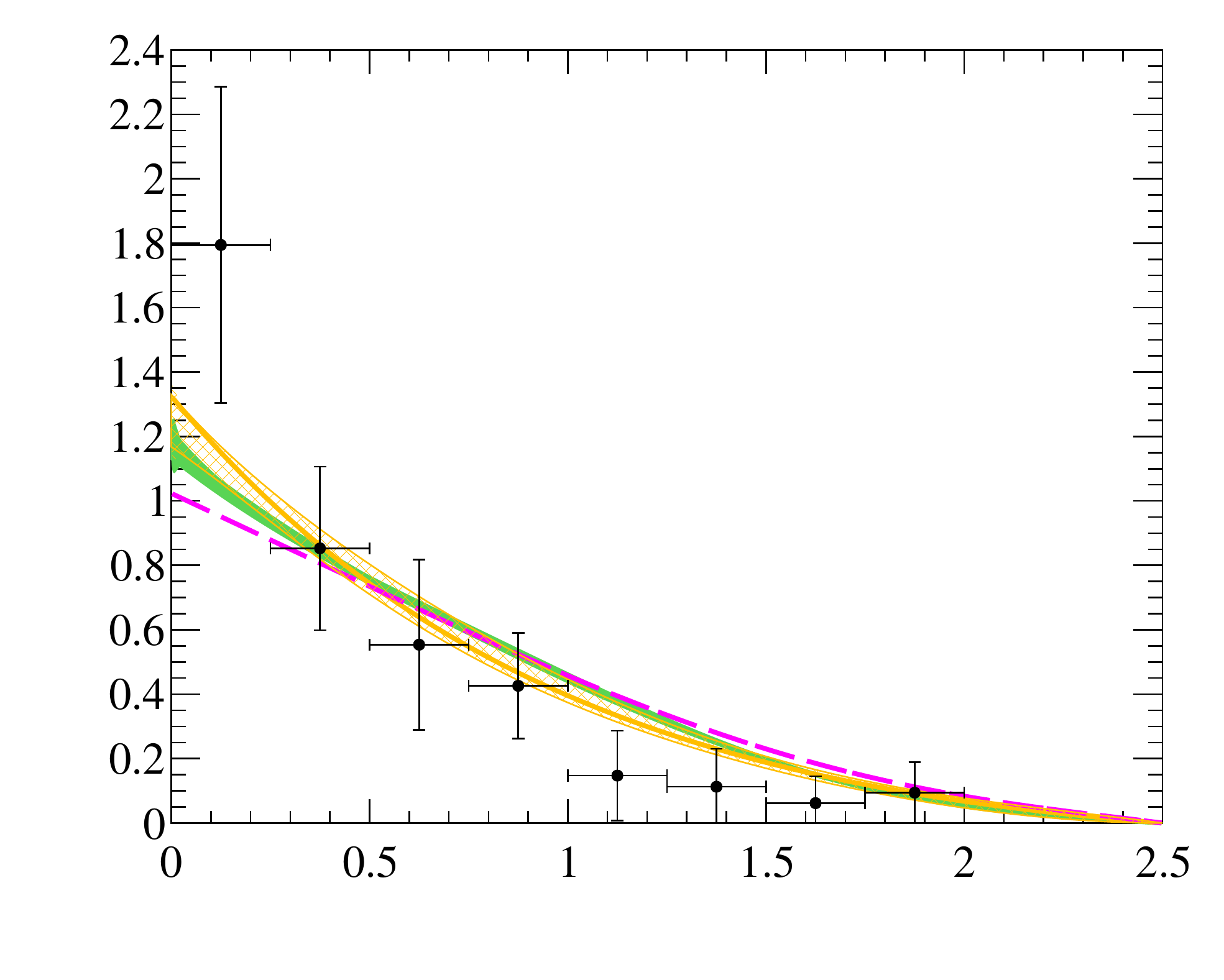}
    }
    \put( -1,146) { \small \begin{sideways} 
        $\frac{\pi}{\upsigma}\frac{\mathrm{d}\upsigma}{\mathrm{d}\left|\Delta\Pphi^{\jpsi}\right|}$
      \end{sideways}} 
    \put( 74,146) { \small \begin{sideways} 
        $\frac{\pi}{\upsigma}\frac{\mathrm{d}\upsigma}{\mathrm{d}\left|\Delta\Pphi^{\jpsi}\right|}$
      \end{sideways}} 
    \put( -1, 92) { \small \begin{sideways} 
        $\frac{1}{\upsigma}\frac{\mathrm{d}\upsigma}{\mathrm{d}\left|\Delta\Peta^{\jpsi}\right|}$
      \end{sideways}} 
    \put( 74, 92) { \small \begin{sideways} 
        $\frac{1}{\upsigma}\frac{\mathrm{d}\upsigma}{\mathrm{d}\left|\Delta\Peta^{\jpsi}\right|}$
      \end{sideways}} 
    \put( -1, 37) { \small \begin{sideways} 
        $\frac{1}{\upsigma}\frac{\mathrm{d}\upsigma}{\mathrm{d}\left|\Delta y^{\jpsi}\right|}$
      \end{sideways}} 
    \put( 74, 37) { \small \begin{sideways} 
        $\frac{1}{\upsigma}\frac{\mathrm{d}\upsigma}{\mathrm{d}\left|\Delta y^{\jpsi}\right|}$
      \end{sideways}} 
    \put( 35,110) { $\left|\Delta\Pphi^{\jpsi}\right|/\pi$ } 
    \put(110,110) { $\left|\Delta\Pphi^{\jpsi}\right|/\pi$ } 
    \put( 35, 55) { $\left|\Delta\Peta^{\jpsi}\right|$ } 
    \put(110, 55) { $\left|\Delta\Peta^{\jpsi}\right|$ } 
    \put( 35,  0) { $\left|\Delta y^{\jpsi}\right|$ } 
    \put(110,  0) { $\left|\Delta y^{\jpsi}\right|$ } 
    \put(113,155) { \small $\begin{array}{r}\mathrm{LHCb}\\
        \sqrt{s}=7,8\,\mathrm{TeV} \end{array}$}
    \put( 16,155) {a)}
    \put( 91,155) {b)}
    \put( 16,100) {c)}
    \put( 91,100) {d)}
    \put( 16, 45) {e)}
    \put( 91, 45) {f)}
    \put(20,155) { \color[rgb]{1,0.747,0}{\begin{tikzpicture}[x=1mm,y=1mm]\draw[thin,pattern=crosshatch, pattern color=Root92]  (0,0) rectangle (12,2.0);\end{tikzpicture}} }
    \put(33,155) { \sc{Powheg} } 
    \put(20,150) { \color[rgb]{0.35,0.83,0.33} {\rule{12mm}{2.2mm}}}
    \put(33,150) { \sc{Pythia} } 
    \put(20,145) { \color[rgb]{1,0,1} {\hdashrule[0.5ex][x]{1.2cm}{0.7pt}{3mm 0.5mm} } } 
    \put(33,145) { \small uncorrelated $\mathrm{b}\bar{\mathrm{b}}$}
  \end{picture}
  \caption { \small
    Normalized differential production cross\nobreakdash-sections\,(points with error bars)
    for~\mbox{$\ptpsi>5\gevc$}\,(left) 
    and~\mbox{$\ptpsi>7\gevc$}\,(right) data for 
    a,b)\,$\left|\Delta\Pphi^{\jpsi}\right|/\pi$,
    c,d)\,$\left|\Delta\Peta^{\jpsi}\right|$, and 
    e,f)\,$\left|\Delta y^{\jpsi}\right|$,
    together with the~{\sc{Powheg}}\,(orange line)
    and \pythia\,(green band) predictions. 
    The~expectations for 
    uncorrelated $\bquark\bquarkbar$~production are shown by the~dashed magenta line. 
    The~uncertainties  
    in the~{\sc{Powheg}} and \pythia
    predictions due to the~choice
    of factorization and renormalization scales
    are shown as orange cross\nobreakdash-hatched and 
    and green solid areas, respectively.
  }
  \label{fig:a2}
\end{figure}

The~corresponding differential cross\nobreakdash-sections 
are presented in Figs.~\ref{fig:a1} and~\ref{fig:a2}.
They~are compared with  expectations from 
the~{\sc{Powheg}}~\cite{Nason:2004rx,Frixione:2007vw,Frixione:2007nw,Alioli:2010xd}
and~\pythia~\cite{Sjostrand:2006za,Sjostrand:2007gs,LHCb-PROC-2010-056} generators 
and with expectations 
from the data\nobreakdash-driven model  of uncorrelated 
$\bquark\bquarkbar$~production, described in Sect.~\ref{sec:correlations}.
Also~in this case both {\sc{Powheg}}~and \pythia~describe 
the~data well for all distributions, 
suggesting a~small role 
of next\nobreakdash-to\nobreakdash-leading order effects 
in $\bquark\bquarkbar$~production
in the~studied kinematical range
compared to the~experimental precision.
The~data~agree well with the~model 
of uncorrelated $\bquark\bquarkbar$~production for 
$\left|\Delta\Peta^{\jpsi}\right|$  and 
$\left|\Delta y^{\jpsi}\right|$,
supporting the~hypothesis of large effective decorrelation
of the~produced heavy~quarks. 

%% file: LHCb_Authorship_flat_06-Jun-2017.tex
\centerline{\large\bf LHCb collaboration}
\begin{flushleft}
\small
R.~Aaij$^{40}$,
B.~Adeva$^{39}$,
M.~Adinolfi$^{48}$,
Z.~Ajaltouni$^{5}$,
S.~Akar$^{59}$,
J.~Albrecht$^{10}$,
F.~Alessio$^{40}$,
M.~Alexander$^{53}$,
A.~Alfonso~Albero$^{38}$,
S.~Ali$^{43}$
% ...

{\footnotesize \it
% ...
$ ^{5}$Clermont Universit{\'e}, Universit{\'e} Blaise Pascal, CNRS/IN2P3, LPC, Clermont-Ferrand, France\\
% ...
$ ^{10}$Fakult{\"a}t Physik, Technische Universit{\"a}t Dortmund, Dortmund, Germany\\
% ...
$ ^{38}$ICCUB, Universitat de Barcelona, Barcelona, Spain\\
$ ^{39}$Universidad de Santiago de Compostela, Santiago de Compostela, Spain\\
$ ^{40}$European Organization for Nuclear Research (CERN), Geneva, Switzerland\\
% ...
$ ^{43}$Nikhef National Institute for Subatomic Physics, Amsterdam, The Netherlands\\
% ...
$ ^{48}$H.H. Wills Physics Laboratory, University of Bristol, Bristol, United Kingdom\\
% ...
$ ^{53}$School of Physics and Astronomy, University of Glasgow, Glasgow, United Kingdom\\
% ...
$ ^{59}$University of Cincinnati, Cincinnati, OH, United States\\
% ...
}
\end{flushleft}